\documentclass[12pt]{report}

\usepackage{xcolor}
\usepackage[normalem]{ulem}
\usepackage{physics}
\usepackage[utf8]{inputenc}
\usepackage{graphicx}
\usepackage[export]{adjustbox}
\usepackage{subfigure}
\usepackage{caption}
\usepackage{newtxmath}
\usepackage{multirow}
\usepackage{tabularx}
\usepackage{longtable}
\usepackage{ltablex}
\usepackage[autostyle]{csquotes}

\usepackage{setspace}
\usepackage{graphicx}
\graphicspath{{Images/}}

\usepackage[Lenny]{fncychap}
\usepackage[a4paper,width=150mm,top=25mm,bottom=25mm,bindingoffset=6mm]{geometry}
\usepackage{fancyhdr}
\usepackage{nomencl}
\makenomenclature

\usepackage[authoryear]{natbib}
\begin{document}

\begin{titlepage}

\begin{minipage}{0.3\textwidth}%
\includegraphics[width=0.8\textwidth]{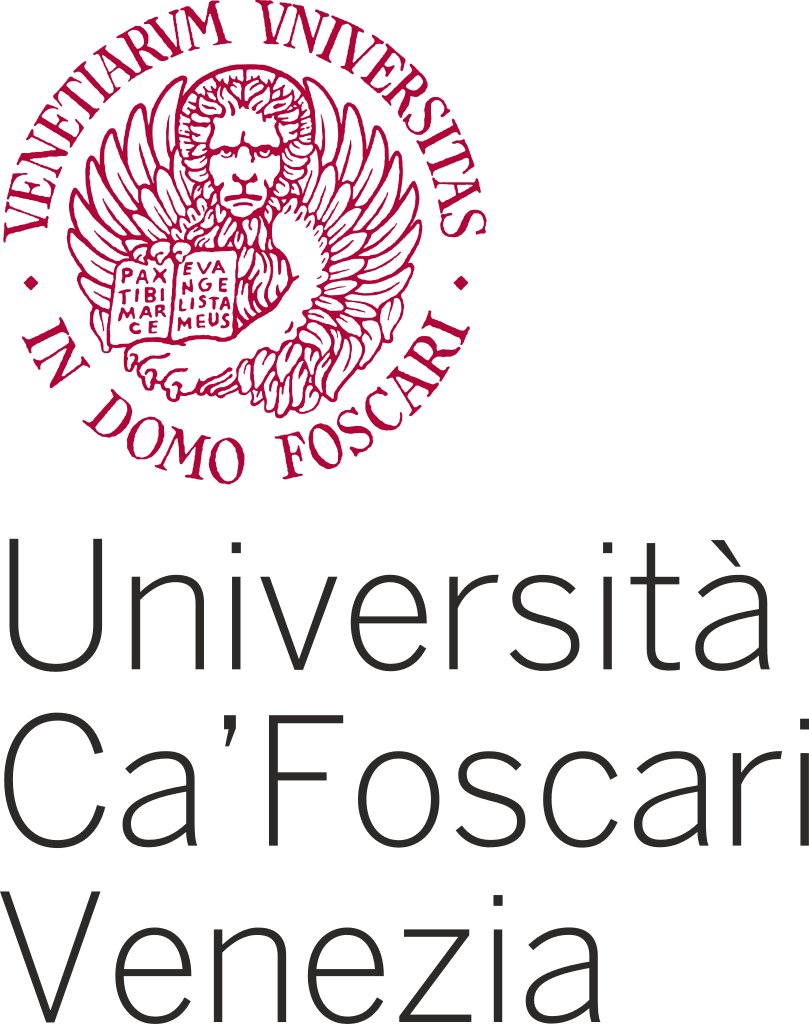}%
\end{minipage}\hspace{155pt}
\begin{minipage}{0.3\textwidth}%
\includegraphics[width=0.8\textwidth]{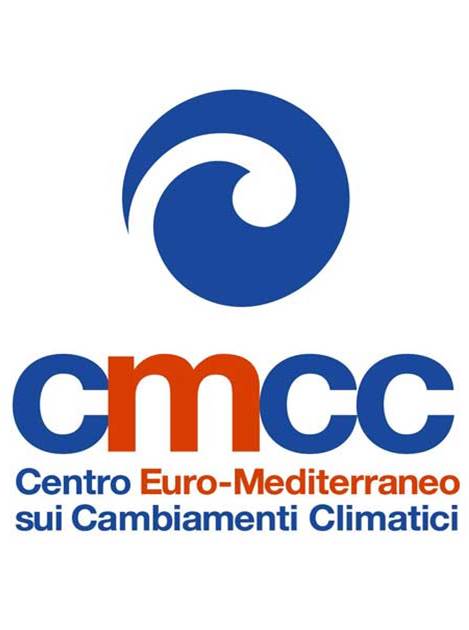}%
\end{minipage}

\begin{center}

\vspace{1cm}

\large
PhD course in\\
"Science and Management of Climate Change"\\
34$^{th}$ cycle

\vspace{1cm}

\large
Thesis title\\

\Large
\textbf{Ocean--atmosphere interaction at the Gulf Stream sea surface temperature front: variability and impacts on midlatitude atmospheric circulation}
            
\vspace{1.5cm}

\large
\textbf{\underline{Author:}}
\hfill
\textbf{\underline{Supervisor:}} \\
Luca Famooss Paolini
\hfill
Alessio Bellucci\\
\hfill
\textbf{\underline{Co-supervisors:}} \\
\hfill
Panos Athanasiadis\\
\hfill
Paolo Ruggieri

\large
Note: This PhD thesis was defended on \textbf{May 26, 2023},\\
and includes work that has been published; see the \textbf{Declaration}.

\end{center}
\end{titlepage}

\chapter*{\textbf{Abstract}}
Recent observational and modelling studies show that the strong sea surface temperature (SST) gradients (oceanic fronts) associated with the western boundary currents affect the atmospheric circulation both on local and large spatial scale and on different timescales. Despite these advancements, the nature of the ocean--atmosphere interactions associated with oceanic fronts variability is still not fully understood. Indeed, the character of the atmospheric response to oceanic fronts variability has not been established yet. Furthermore, the effective role of the atmospheric forcing on the oceanic fronts variability is unclear.

The present PhD thesis analyses the ocean--atmosphere interactions associated with the Gulf Stream SST front (GSF) variability. Specifically, the first part of the thesis assesses the character of the atmospheric response to the interannual GSF meridional shifts and its dependence on the model horizontal resolution. The second part of the thesis assesses the spectral features of the North Atlantic Oscillation (NAO)--GSF interaction and the mechanims through which the NAO forces the GSF meridional shifts on the decadal timescale.

The character of the atmospheric response to the interannual GSF meridional shifts was assessed in the ERA5 reanalysis dataset and in an ensemble of multi-member atmosphere-only historical simulations forced with observed SSTs (1950--2014). The role of the model horizontal resolution was assessed analyzing models with the horizontal resolution from 25 km to 100 km. Results show that the atmospheric response is strongly resolution dependent, with the response in the high-resolution simulations (resolution finer than 50 km) resembling the observed anomalies. More specifically: (i) analysis of the atmospheric thermodynamic balance close to the GSF showed that the anomalous diabatic heating associated with the GSF displacement is mainly balanced in the atmosphere by vertical motion and by meridional transient eddy heat transport (not the case for low-resolution models), while (ii) the large-scale response includes a meridional shift of the North Atlantic eddy-driven jet and stormtrack homo-directional to the GSF displacement. This atmospheric response is accompanied by changes in low-level baroclinicity close to and north of the GSF, resulting from the oceanic forcing and the zonal atmospheric circulation anomalies respectively. The low-level baroclinicity anomalies lead to changes in baroclinic eddy activity and, ultimately, in the jet via eddy--mean flow interaction. 

The spectral features of the NAO--GSF interaction in the past decades and the mechanisms underlying this interaction were assessed in a set of global atmosphere
and ocean reanalyses. Results show that the NAO and the GSF indices covary on the decadal timescales but only during 1972--2018. A secondary peak in the NAO--GSF covariability emerges on multiannual timescales but only for a limited period of time (2005--2015). The non-stationarity in the decadal NAO--GSF covariability is also manifested through the dependency of their lead--lag relationship on the analyzed time period. Indeed, the NAO leads the GSF shifts by 3 years during 1972--1990 and by 2 years during 1990--2018. Results show that the lag between the NAO forcing and the GSF response on decadal timescales can be interpreted as the joint effect of the fast response of wind-driven oceanic circulation, the lagged response of deep oceanic circulation, and the propagation of Rossby waves. However, not all the mechanisms are stationary. There is evidence of Rossby wave propagation only before 1990. Here it is suggested that the non-stationarity of the Rossby wave propagation causes the time lag between the NAO and the GSF latitudinal position on decadal timescales to differ before and after 1990. 

Considering the impact that the GSF variability has on the North Atlantic variability, this PhD thesis helps to improve our understanding of extratropical climate variability.


\chapter*{\textbf{Acknowledgements}}
I decided to start this PhD at the age of 28 years old, after some years of work in private companies in a completely different field. An hard decision. Especially in a cultural context like the Italian one, where leaving a permanent, well-paying but unsatisfying job for the unstable, poorly paid yet passionate scientific research is simply and hastily considered crazy. At least, this was the most common reaction around me at that time.

After 4 years of \enquote{studio matto e disperatissimo}, sleepless nights spent in front of the black screen of Python, moments of stellar enthusiasm where one thinks to be able to publish in Nature journal followed by moments of dark and gloomy discouragement where one wonders what sense it makes to \enquote{play scientist}, I can say that choosing that unstable, poorly paid yet passionate scientific research was one of the most important and right decisions of my life.

I would like to thank my supervisor Alessio Bellucci and my co-supervisors Paolo Ruggieri and Panos Athanasiadis, who made it possible to make this choice a reality. The tense moments faced in the past years within the group certainly cannot erase the invaluable help they have given me with scientific and non-scientific discussions, lunches, beers, phone calls, \enquote{psicological support}, comments, corrections and advice. All of this has allowed me to grow not only as a scientist but also, and more importantly, as a person.

Furthermore, I thank Noureddine Omrani at the University of Bergen. In him, I have found not only one of the brightest and most passionate scientists of today but also a person with whom to share a friendship. At the University of Bergen, I also thank Noel Keenlyside. It is especially thanks to him that I managed to spend my surprising period abroad in Bergen. Few words shared with him were able to help me in my research more than reading tens of scientific papers.

In addition, it is essential to thank external reviewers Gerard McCarthy and Fumiaki Ogawa who, with their punctual and valuable comments, have made this thesis a significantly better work.

Finally, I want to thank all the other people (scientists and non-scientists) that I have met, and in some cases \enquote{lost}, along this amazing journey. Rome (yes, it is like a person) and all the people that represent my roots, friends and colleagues at CMCC in Bologna, my PhD classmates, Venice (again, other human identity) and my new Venetian family, Duegradi's friends, Bergen community, etc. With their presence and their passing, they have given me the necessary energy to complete this experience.

\chapter*{\textbf{Declaration}}
I confirm that this is my own work and the use of all material from other sources has been properly and fully acknowledged. Part of the work presented in Chapter 3 has been published prior to the submission of this thesis\footnote{Famooss Paolini, L., Athanasiadis, P. J., Ruggieri, P., \& Bellucci, A. (2022). The Atmospheric Response to Meridional Shifts of the Gulf Stream SST Front and Its Dependence on Model Resolution, \textit{Journal of Climate}, 35(18), 6007-6030, https://doi.org/10.1175/JCLI-D-21-0530.1}, whereas part of the work presented in Chapter 4 was published after the thesis defence\footnote{Famooss Paolini, L., Omrani, N.-E., Bellucci, A., Athanasiadis, P. J., Ruggieri, P., Patrizio, C.R., \& Keenlyside, N. (2024). Nonstationarity in the NAO-Gulf Stream SST Front Interaction, \textit{Journal of Climate}, 37, 1629-1650, https://doi.org/10.1175/JCLI-D-23-0476.1}.

\tableofcontents{\textbf{}}

\printnomenclature[3cm]

\addcontentsline{toc}{chapter}{Abbreviations}

\chapter{\textit{\textbf{Motivation of work}}}
\label{motivation_of_work}
\fancyhead[R]{\textit{Chapter \ref{motivation_of_work} - Motivation of work}}

The seminal work by \citet{frankignoul1977} has shown that the large-scale, low-frequency upper ocean variability in the extratropics can be largely interpreted as the oceanic integral response to short timescale randomic atmospheric disturbances. These results have led to the emergence of the well-known paradigm of white-noise atmospheric forcing and red-noise oceanic passive response, which has been an important paradigm to interpret the low-frequency sea surface temperature (SST)\nomenclature{SST}{sea surface temperature} variability in middle latitudes during the last 40 years. As an example, this paradigm has been used to explain the SST tripole variability as the oceanic response to the North Atlantic Oscillation \citep[NAO;][]{hurrell1995}\nomenclature{NAO}{North Atlantic Oscillation} forcing via surface heat fluxes (SHF)\nomenclature{SHF}{surface heat fluxes} and Ekman currents \citep{cayan1992,deser2010}. On the other hand, the ocean has been shown to exert a weak effect on the short timescale atmospheric variability, with the ocean-induced atmospheric response projecting onto intrinsic atmospheric variability modes \citep{rodwell1999,czaja2002,kushnir2002,peng2003,cassou2007}. 

However, several observational and modelling studies have shown that this interpretation of air--sea interaction in the extratropics is altered by the presence of meso-scale oceanic features, such as the western boundary currents \citep[WBCs;][]{chelton2001, chelton2004, xie2004,bellucci2021}\nomenclature{WBC}{western boundary current}. In these areas, a great part of the oceanic variability is induced by intrinsic oceanic processes, such as advection and diffusion \citep{kelly2010,kwon2010,minobe2010,patrizio2021}. Furthermore, the WBCs are characterized by strong SST gradients (oceanic fronts), which strongly influence the atmospheric circulation within the marine atmospheric boundary layer \citep[MABL;][]{chelton2001,small2008}\nomenclature{MABL}{marine atmospheric boundary layer}. The impact of the oceanic fronts extend also beyond the MABL, influencing the atmospheric circulation on the large-scale and on different timescales \citep{oreilly2015,sato2014,wills2016,joyce2019}. These results show that the oceanic fronts may be acting as a source of atmospheric predictability on a wide range of timescales \citep{joyce2019,athanasiadis2020}. The oceanic fronts are also important because their presence shapes many features of the time-mean atmospheric circulation. In particular, they anchor zones of intense time-mean upward motion thanks to different pressure conditions across the fronts themselves \citep{minobe2008a}. In addition, they shape the extratropical stormtracks and eddy-driven jets maintaining the near-surface baroclinicity necessary for baroclinic instability growth \citep{sampe2010}. Finally, they affect the time-mean stratospheric circulation enhancing the upward wave propagation \citep{omrani2019}.

Despite these advancements, the nature of the ocean--atmosphere interactions associated with oceanic fronts variability is still not fully understood. Indeed, the character of the atmospheric response to oceanic fronts variability has not been established yet. Furthermore, the effective role of the atmospheric forcing on the oceanic fronts variability is unclear. This is especially true for the SST front associated with the Gulf Stream (GS)\nomenclature{GS}{Gulf Stream}, on which this PhD thesis focuses. As one of the major WBCs, the GS plays a key role in driving the northern hemisphere climate and variability \citep{minobe2008a,omrani2019}. Therefore, improving our understanding of its variability and interaction with the atmosphere is highly relevant for the comprehension of extratropical climate variability on a wide range of timescales.

Regarding the ocean-to-atmosphere forcing, there are observational studies showing that positive SST anomalies in the GS region induce positive NAO-like anomalies in the atmospheric circulation \citep{sato2014,joyce2019}, whereas other studies show atmospheric circulation anomalies of opposite sign \citep{kwon2013,wills2016}. Taking into account that a positive (negative) NAO phase is associated with positive (negative) SST anomalies in the GS region \citep{frankignoul2001}, the first results shed light on the possible existence of a positive feedback between the NAO and the GS SST front (GSF)\nomenclature{GSF}{Gulf Stream sea surface temperature front} variability, whereas the second results show an opposite relationship between them. These differences may be induced either by discrepancies in the methodological approach  or by the chaotic nature of the atmospheric variability, especially on timescale shorter than the interannual one. 

The use of general circulation models (GCMs)\nomenclature{GCM}{general circulation model} could help to disentangle this issue, increasing the signal-to-noise ratio associated with the phenomena of interest. For example, ad-hoc simulations can be performed with observed SSTs in the area of interest and climatological SSTs elsewhere, thus reducing possible impact from remote SST variability. Using multi-member simulations also helps in letting the forced signal emerge from the chaotic atmospheric variability. Furthemore, simulations performed with atmosphere-only models allow to focus on the single ocean-to-atmosphere direction of the two-way interaction between the two realms. However, important inter-model discrepancies still emerge in studies analysing the impact of the oceanic fronts variability on the atmospheric circulation \citep{czaja2019}. In this context, there are indications that the horizontal resolution may play a role \citep{smirnov2014}, highlighting the need for models with grids adequately resolving the oceanic fronts.

Regarding the atmosphere-to-ocean forcing, the variability of the GSF has been mainly linked with the NAO \citep{taylor1998b,wolfe2019}, which is the dominant mode of variability of the surface atmospheric circulation in the North Atlantic region \citep{hurrell1995}. Interestingly, the discrepancies in the lead--lag relationship between the NAO forcing and the GSF response suggested in the literature show the possible existence of non-stationarity in their interaction. Such non-stationarity may explain why there is also a lack of consensus on the mechanisms through which the atmosphere influences the GSF latitudinal position \citep{rossby2000,zhang2007,sasaki2011,gangopadhyay2016}.

The present PhD thesis analyses the ocean--atmosphere interactions associated with the GSF variability, addressing the following scientific questions:
\begin{enumerate}
\item What is the character of the atmospheric response to the interannual GSF meridional shifts and its dependence on the model horizontal resolution?
\item What are the spectral features of the NAO--GSF interaction and the mechanisms through which the NAO forces the GSF meridional shifts on the decadal timescale?
\end{enumerate}

Specifically, Chapter \ref{introduction} gives a general introduction on the GS system and the North Atlantic circulation. Then, it provides a brief description of the NAO and an overview about the role of the GSF on the time-mean atmospheric circulation and the atmospheric variability. Chapter \ref{atmospheric_response} responds to the first question above, assessing the atmospheric response to the interannual meridional shift of the GSF and its dependence on model resolution. Chapter \ref{decadal_variability} responds to the second question above, assessing the spectral features of the NAO--GSF interaction and the mechanisms through which the NAO may be forcing changes in the GSF latitudinal position on the decadal timescale. Finally, Chapter \ref{conclusions} summarizes the most important findings of the PhD thesis.

\chapter{\textit{\textbf{Introduction}}}
\label{introduction}

\fancyhead[R]{\textit{Chapter \ref{introduction} - Introduction}}

\section{The Gulf Stream system and the North Atlantic circulation}
\label{GS}

The GS is the WBC located in the North Atlantic basin and it is an essential component of northern hemisphere climate. Indeed, it constitutes the upper limb of the Atlantic Meridional Overturning Circulation (AMOC)\nomenclature{AMOC}{Atlantic Meridional Overturning Circulation} and it transports huge amounts of warm and salty water poleward, partly balancing the heat disequilibrium between the tropics and the higher latitudes \citep{peixoto1992}. Once the GS reaches higher latitudes, the heat gets partly released into the atmosphere, which becomes predominant in redistributing the heat over extra-tropical and polar areas \citep{fasullo2008,yang2015}. The ocean-to-atmosphere heat exchanges are particularly intense in the winter season between Cape Hatteras (75$^{\circ}$W--37$^{\circ}$N) and the area south of the Grand Banks, where the GS enters in contact with cold, dry air masses coming from the North America continent \citep{kallberg2005,yu2007}.

The intense ocean-to-atmosphere heat exchanges in the western North Atlantic determine large-scale asymmetry in the horizontal distribution of surface diabatic heating within the subtropical gyre (STG)\nomenclature{STG}{subtropical gyre}. The STG is the large-scale anticyclonic oceanic circulation mainly driven by the wind forcing between the equator and mid-latitudes in the North Atlantic \citep{sverdrup1947}. This aspect is fundamental in setting the mid-latitude stationary and quasi-stationary atmospheric eddies \citep{smagorinsky1951,ambaum2007}, thus affecting the climate at extratropics. Furthermore, the GS plays an important role in northern hemisphere climate because it is characterized by a strong SST gradient (oceanic front), which can affect the atmospheric circulation both on local and large spatial scale and on different timescales \citep{minobe2010,sampe2010,brayshaw2011,joyce2019}. Refer to section \ref{GSF_impact_on_atm_circulation} for more details about the impact of GS front on the atmospheric circulation. 

The GS originates in the Gulf of Mexico and it flows northward along the eastern coast of North America until Cape Hatteras. Then, it detaches from the coast and turns northeastward, flowing along the continental shelf break until the Grand Banks. The location of the GS detachment from the coast has been explained through several mechanisms, such as the position of zero wind-stress curl \citep[strictly linked with the zero Sverdrup transport;][]{stommel1948,munk1950}, the outcropping of isopycnal surfaces \citep{parsons1969,veronis1973} and the interaction with adjacent currents \citep{ezer1992}. Furthermore, the position of the separation point has been shown to be controlled by local factors such as the coastline geometry, the bottom topography and the interaction with the deep western boundary current\nomenclature{DWBC}{deep western boundary current} \citep[DWBC;][]{thompson1989,spall1996a,tansley2000,zhang2007}.

Once detached from the coast, the GS starts meandering around 70$^{\circ}$W and an intense eddy activity develops up to the Grand Banks \citep{ducet2000}. Some of the large-amplitude, propagating meanders can give rise to warm (cold) core rings north (south) of the time-mean GS path \citep{brown1986,cornillon1986,cornillon1987}.

Near 50$^{\circ}$W, the GS decreases its transport and branches into several current bands. The main ones are the northern recirculation gyre\nomenclature{NRG}{northern recirculation gyre} \citep[NRG;][]{hogg1986}, the southern recirculation gyre\nomenclature{SRG}{southern recirculation gyre} \citep[SRG;][]{worthington1976}, the Azores current\nomenclature{AC}{Azores current} \citep[AC;][]{krauss1990} and the  North  Atlantic  current\nomenclature{NAC}{North Atlantic current} \citep[NAC;][]{krauss1986}. The NRG (SRG) is a barotropic cyclonic (anticyclonic) recirculation pattern, located north (south) of the GS \citep{hogg1992}. The NRG and SRG recirculate part of the GS water westward, rejoining the main stream at different longitudes downstream Cape Hatteras \citep{halkin1985,johns1995}. Because of these two recirculation gyres, the GS transport almost doubles between the separation point and the Grand Banks, going from 88 Sv at 73$^{\circ}$W \citep{halkin1985} to 150 Sv at 60$^{\circ}$W \citep{hogg1992}. The AC is the branch of the GS that flows eastward after the Grand Banks and it represents the eastern side of the STG \citep{sverdrup1947}. Due to its position, this current separates the subtropical water to the south from the subpolar water to the north \citep{krauss1990}. The GS water that is not recirculated westward by the NRG and SRG and does not proceed eastward along the AC, flows northward off the eastern flank of the Grand Banks and then turns abruptly eastward around 52$^{\circ}$N. This branch of the GS is the NAC and it transports warm and saline water to much higher latitudes than currents in any other ocean basin. The transport of warm water reaches the Nordic Seas, thus playing a fundamental role in determining the mild climate over northern Europe \citep{krauss1986}. The NAC flows poleward via three main channels: the Irminger Current and the subpolar front west and east of Iceland respectively, and the eastern NAC flowing in the eastern side of North Atlantic. The rest of the NAC water interacts with Arctic Sea water and turns westward, flowing along the coast of Greenland, as Eastern and Western Greenland currents, and then along the coast of eastern North America, as Labrador Current (LC)\nomenclature{LC}{Labrador Current}. These currents form a large-scale cyclonic circulation pattern called subpolar gyre (SPG)\nomenclature{SPG}{subpolar gyre}, which is also driven by the prevailing winds \citep{sverdrup1947}.

Within the SPG, the heat loss from the ocean to the atmosphere and the interaction between regional oceanic currents facilitate vertical mixing and convection in the ocean \citep{marshall1999,lazier2001}. Once the deep water is formed, it is partly recirculated within the Labrador Sea (LS)\nomenclature{LS}{Labrador Sea} and along the NAC and partly travels southward. The deep water is transported southward via the DWBC \citep{stommel1958}, flowing from the LS along the continental shelf of the North American sea-board and then passing underneath the GS at the crossover region \citep[near Cape Hatteras;][]{spall1996a,spall1996b}. This current has been shown to strongly interact with the GS, affecting the position of the separation point and the GS path downstream Cape Hatteras \citep{zhang2007,chassinet2008}. Furthermore the deep water is transported southward via interior pathways which pass underneath the NAC close to the Grand Banks \citep{bower2009,bower2019}. Differently from what was thought before, these interior pathways have been shown to be the principal conduits through which the ventilated water is transported from the SPG to the STG \citep{getzlaff2006,bower2011}. The DWBC and the interior pathways constitute the deep limb of the AMOC and then they constitute the deep branch of the meridional--vertical thermohaline circulation of the Atlantic ocean \citep{lozier2010}.

\begin{figure}[ht!]
    \centering
    \includegraphics[max size={\textwidth}{\textheight}]{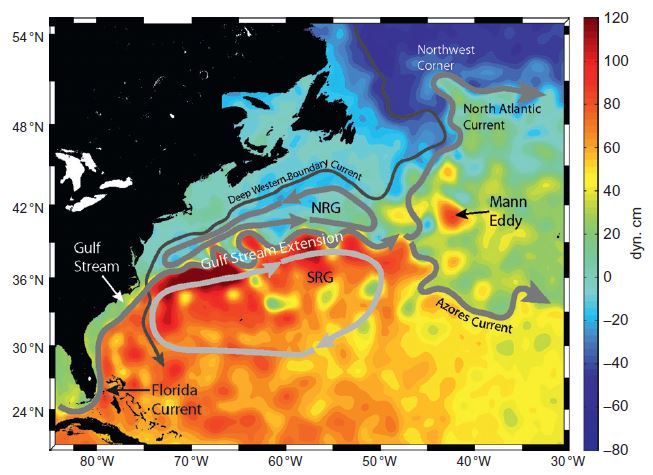}
    \caption{Maps of the main oceanic currents connected with the GS. The figure is reproduced from \citet{imawaki2013}.}
    \label{fig:imawaki2013.JPG}
\end{figure}

\section{The western intensification}
The GS is the western portion of STG. The STG is a large-scale oceanic circulation pattern mainly driven by the stress exerted by the prevailing winds on the sea surface \citep{sverdrup1947}. The same is true for other ocean gyres in the globe. In the North Atlantic, the prevailing winds are westerlies at mid-latitudes and easterlies (the trades) at tropics. Then, the ocean is forced to have an anticyclonic circulation between the equator and around 40$^{\circ}$N because of the negative wind-stress torque.

Wind-driven ocean gyres are asymmetric in the zonal direction, showing narrow, deep and strong poleward currents in their western side (i.e. the GS in the North Atlantic) and broad, shallow and weaker equatorward currents in their eastern side. The intensification of currents along the western side of the ocean gyres has been referred to as \enquote{western intensification} and it is caused by the variation of Coriolis force with latitude \citep{stommel1948} and the friction exerted by the coasts \citep{munk1950}. This section provides a brief description of the theory explaining the western intensification.

Under the assumption of barotropic flow in a rotating system, starting from the momentum and the continuity equations it is possible to obtain the following potential vorticity (PV)\nomenclature{PV}{potential vorticity} equation: 

\begin{equation}
\label{PV_equation}
\dv{t}(\frac{\xi +f}{h}) = \frac{F}{h}
\end{equation}

Here $\xi$ represents the relative vorticity; f is the planetary vorticity; h is the height of the water column; F is the forcing and the dissipation of vorticity, while the quantity inside the parenthesis is the PV. The equation \ref{PV_equation} states that the PV of a column of water is conserved following the motion in the absence of PV sources and sinks on the fluid (i.e. F is equal to zero).

In the subtropical area, the midlatitude westerlies and the tropical easterlies induce southward and northward Ekman transport, respectively. Here, the Ekman transport is the net motion of ocean water within the Ekman layer as the result of a balance between Coriolis and turbulent drag forces exerted by the wind-stress. In the northern hemisphere, the Ekman transport is oriented about 90$^{\circ}$ from wind direction. For the balance of water mass, the convergence in the interior of the SPG determines downward motion, i.e. the squeezing of the water column (Ekman pumping). Being the forcing and the dissipation of PV zero (or almost) in the interior of the STG, the PV must be conserved (equation \ref{PV_equation}). Thus, the squashing of the water column is balanced by changes in planetary vorticity as the relative vorticity is small in the interior of the STG. This means that the water column moves southward. The reverse is true over the SPG, where the wind-stress curl induces divergence in the interior of the gyre, stretching of the water column and poleward motion. Generally speaking, this means that the large-scale wind-stress forcing tends to induce changes in the height of column water in the interior of ocean gyres, which are balanced by meridional displacement. This relation is referred to as Sverdrup balance and it is generally expressed as follows: 

\begin{equation}
\label{Sverdrup_equation}
\beta\overline{v} = f\pdv{\omega}{z} = f \pdv{(curl(\frac{\tau}{f\rho}))}{z}
\end{equation}

Here $\beta$ is the linear variation of the Coriolis parameter f; $\overline{v}$ is the meridional velocity vertically integrated over the ocean depth; $\tau$ is the wind-stress exerted on the sea surface; $\rho$ is the water density.

\begin{figure}[ht!]
    \centering
    \includegraphics[max size={\textwidth}{\textheight}]{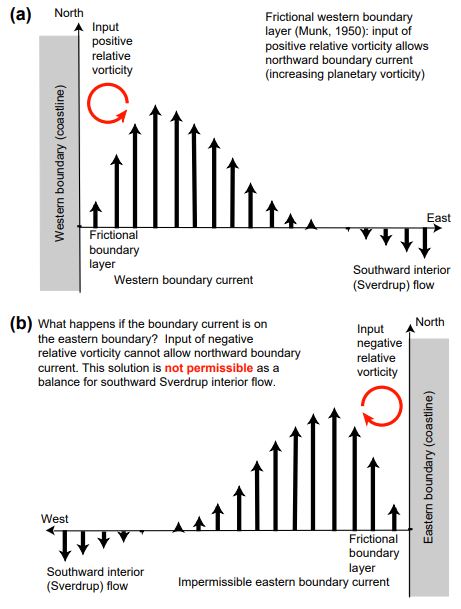}
    \caption{Vorticity balance at the western and eastern boundary because of the sidewalls friction (Munk's model). The figure is reproduced from \citet{talley2011}.}
    \label{fig:talley2011.JPG}
\end{figure}

Once the water column has reached a lower latitude in the SPG, it must be recirculated back in order to balance the total mass. Given the Sverdrup balance is valid for the large-scale oceanic circulation, the return flow must be in a narrow poleward current. From the view of mass conservation, the poleward current can be either in the form of a western boundary current or an eastern boundary current. However, only the western boundary current can close the vorticity balance over the oceanic basin. Indeed, while the water column travels southward within the SPG, it acquires negative relative vorticity because of the large-scale wind forcing. Then, positive relative vorticity must be injected into the fluid in order to counterbalance the negative PV due to the wind-stress curl. The injection of positive relative vorticity is provided by the friction that acts as a dissipation term in the equation \ref{PV_equation} (F different from zero). Specifically, \citet{stommel1948} has added friction to the bottom in the form of linear drag to balance the PV close to the boundaries. However, it is hard for the wind-driven oceanic circulation to reach the ocean bottom. For this reason, \citet{munk1950} has added friction along the side walls using a harmonic Newtonian viscosity. Because of the friction along the western sidewalls, the velocity parallel to the wall is zero along the coastlines and increases offshore. Then, an input of positive relative vorticity is provided to the western boundary of the basin, thus balancing the PV and recirculating the water column poleward. On the other hand, such a closure of the circulation could not be possible on the eastern boundary, because in this case the friction provides a further input of negative PV. Refer to Figure \ref{fig:talley2011.JPG} for a schematic of the vorticity balance at the western and eastern boundary because of the sidewall friction, as in the Munk’s model.

As specified above, the western boundary current intensification is present in all the main ocean gyres over the globe. This means that the oceanic currents intensify along the western portion of the ocean independently from the wind-stress curl sign. This is due to the Earth’s sphericity (beta effect). Indeed, \citet{stommel1948} has shown that, in the case of a no-rotating or uniformly rotating ocean, the western intensification disappears and the ocean currents show an east-west symmetric circulation centered respect the oceanic basin.

\section{The North Atlantic Oscillation}

As a result of intrinsic oceanic variability as well as atmospheric forcing, the GS undergoes meridional shifts, meandering and changes in mass/heat transport \citep{kelly2010}. Among other atmospheric drivers, the NAO has been shown to play a fundamental role in forcing GS variability on different timescales \citep{taylor1998a,joyce2000,gangopadhyay2016}. Being this PhD thesis focused on the ocean--atmosphere interaction associated with the GSF variability, the NAO represents a key atmospheric process to properly understand the results described in following sections. For this reason, a brief description of the NAO is provided in this chapter.

The NAO is a large-scale seesaw of pressure between the high-latitude and subtropical North Atlantic and represents the dominant mode of atmospheric low-frequency variability over the Euro-Atlantic region as it explains about the 30\% of the annual atmospheric variability over that region \citep{jianping2003,hurrell2010}. This percentage increases up to about 36\% during winter season, whereas it decreases to about 22\% during summer \citep{barnston1987,hurrell2003}.

\begin{figure}[ht!]
    \centering
    \includegraphics[max size={\textwidth}{\textheight}]{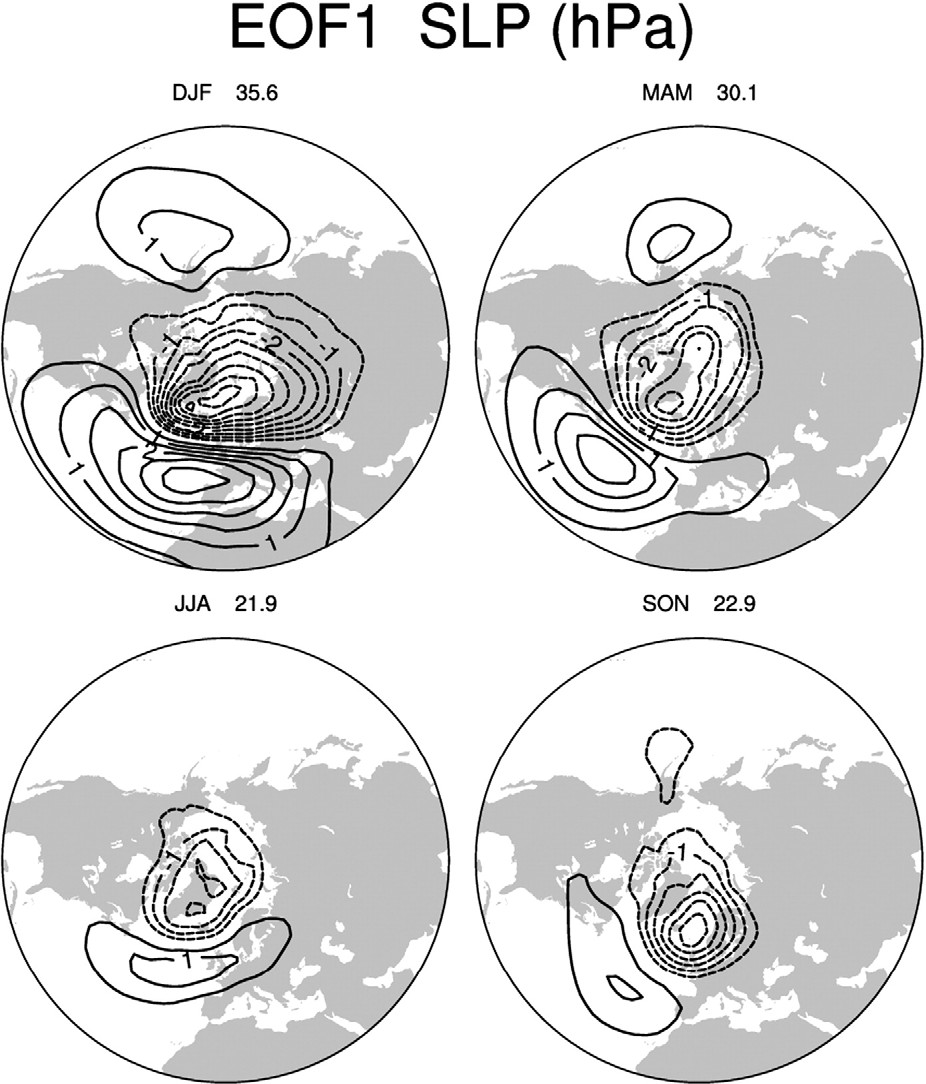}
    \caption{Leading empirical orthogonal function of the seasonal SLP anomalies in the North Atlantic sector (20$^{\circ}$--70$^{\circ}$N, 90$^{\circ}$W--40$^{\circ}$E) during winter (DJF), spring (MAM), summer (JJA) and autumn (SON). The percentage of the total variability the SLP anomalies explain in North Atlantic sector is shown on the top of each subplot. The figure is reproduced from \citet{hurrell2003}.}
    \label{fig:hurrell2010.JPG}
\end{figure}

The NAO is considered to be in its positive (negative) phase when the pressure gradient between the subtropical and the northern region is increased (reduced). Consistently, the positive NAO phase is associated with negative and positive mean sea-level pressure (SLP)\nomenclature{SLP}{mean sea-level pressure} anomalies in respect to the atmospheric mean state over the subpolar and subtropical North Atlantic, respectively (Figure \ref{fig:hurrell2010.JPG}). The reverse is true for the negative NAO phase. The pressure anomalies associated with the NAO are equivalent barotropic (high coherence along the vertical) and extend from the surface up to the lower stratosphere with amplitude increasing with height.

\begin{figure}[ht!]
    \centering
    \includegraphics[max size={\textwidth}{\textheight}]{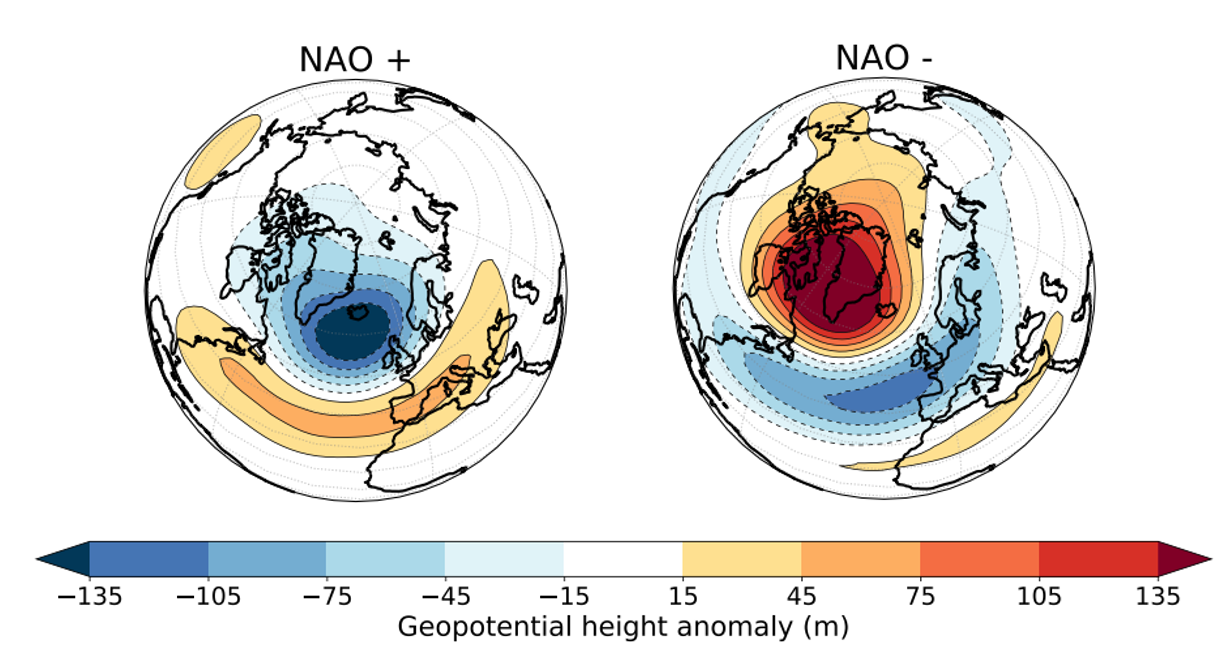}
    \caption{Main patterns of the positive (left) and negative (right) NAO phase. These patterns have been extracted through a K-means clustering algorithm applied to winter (DJF) geopotential height anomalies at 500\,hPa, using ERA40 (1957--1978) and ERAInterim (1979--2014 data. The figure is a partial reproduction of Figure 1 in \citet{fabiano2020}.}
    \label{fig:fabiano2020.png}
\end{figure}

The spatial pattern of the NAO has been shown to be non-stationary throughout the year, with different tilting and position of the centers of action (i.e. where the pressure anomalies are the most intense) during the four seasons \citep[Figure \ref{fig:hurrell2010.JPG};][]{hurrell2010}. Furthermore, the spatial pattern of the NAO has been shown to be asymmetrical between its positive and negative phase \citep{hurrell2003,fabiano2020}. Specifically, the positive NAO phase exhibits a northwest--southeast tilting, whereas the negative NAO phase exhibits a northeast--southwest tilting (Figure \ref{fig:fabiano2020.png}).

In describing the spatial pattern of the NAO, it is highlighted that the pressure anomalies extend also over the Aleutian Low (North Pacific). Because of that, it has been (and it is still) debated whether the NAO should be considered a specific mode of variability or a regional manifestation of the hemispheric mode of atmospheric variability known as the Arctic Oscillation (AO)\nomenclature{AO}{Arctic Oscillation} or Northern Hemisphere Annular Mode \citep[NAM;][]{thompson2000,deser2000,ambaum2001}\nomenclature{NAM}{Northern Hemisphere Annular Mode}. Apart from some subtle differences over the Euro-Atlantic region, the main difference between the AO/NAM and the NAO is the greatest amplitude of the SLP anomalies that characterise the AO/NAM over the North Pacific. These anomalies make the AO/NAM a (almost) zonally-symmetric mode of variability in the northern hemisphere. Some authors have suggested that the NAO should be considered as part of the AO/NAM, regionally modified by zonally-asymmetric forcings such as topography and land--sea contrast \citep{thompson1998,thompson2000}. Differently, other authors have suggested that the AO/NAM cannot be considered as a teleconnection pattern \citep{deser2000,ambaum2001}. For example, \citet{deser2000} has reached this conclusion showing that the Atlantic and Pacific centers of action are weakly correlated with each other and that the interannual SLP anomalies over the Arctic and Pacific sectors are not significantly correlated as they are between the Arctic and the North Atlantic.

\begin{figure}[ht!]
    \centering
    \includegraphics[max size={\textwidth}{\textheight}]{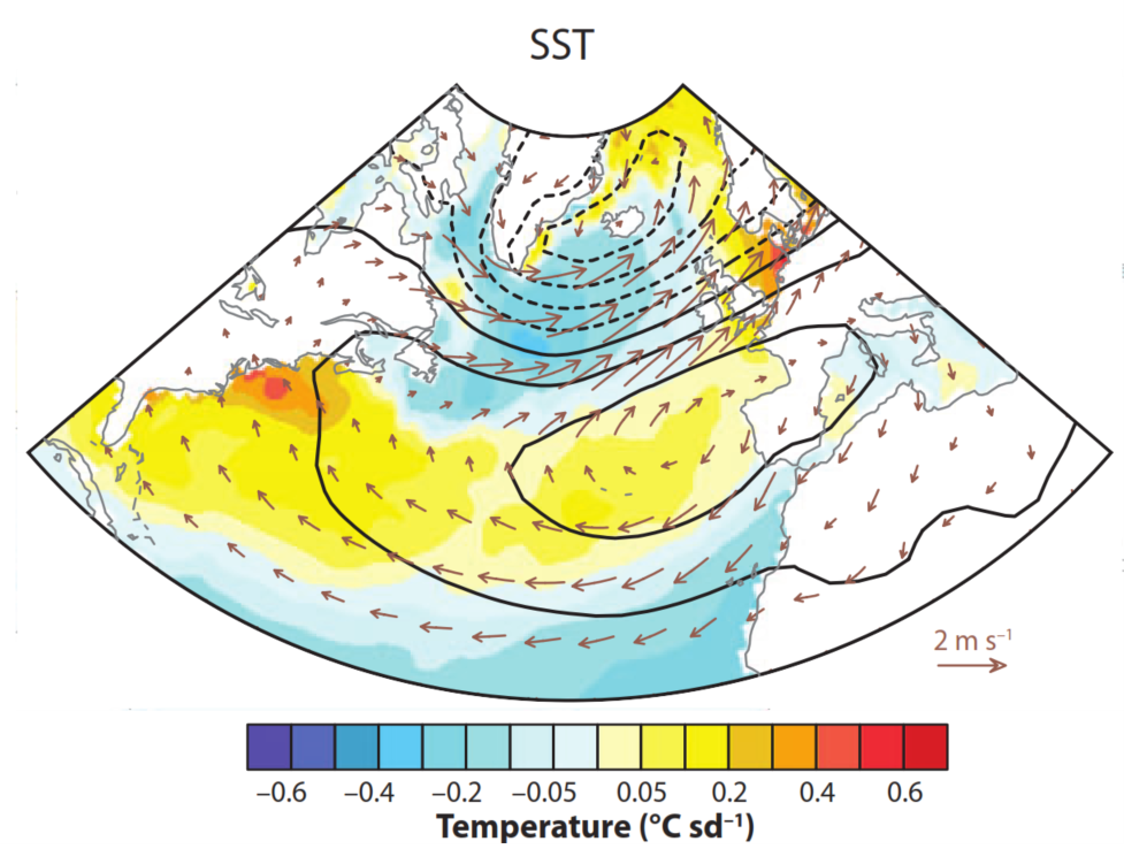}
    \caption{SST ($^{\circ}$C; colour shadings), SLP (hPa; contours; contour interval is 1\,hPa) and surface wind (m s$^{-1}$; arrows) anomalies associated with one positive standard deviation departure of the NAO index during winter. The NAO index has been defined as the station index of \citet{hurrell2003}. The figure is a reproduction of Figure 1a in \citet{deser2010}.}
    \label{fig:deser2010.png}
\end{figure}

The pressure anomalies associated with the NAO have a large impact on the Euro-Atlantic weather and climate, with important consequences for socio-economic and ecological systems \citep{drinkwater2003,mysterud2003}. For example, relevant in the context of this thesis are the NAO impacts on the SST, stormtrack and eddy-driven jet in North Atlantic region. Specifically, the NAO is associated with an SST tripolar pattern that extends to the entire North Atlantic (Figure \ref{fig:deser2010.png}). The SST tripole is characterized by negative (positive) SST anomalies in the SPG (STG), interpreted as the oceanic response to anomalous SHF and wind-driven oceanic circulation associated with NAO forcing. This SST tripole pattern represents an important portion of the extratropical SST variability on seasonal and interannual timescale \citep{cayan1992,deser2010}.

\begin{figure}[ht!]
    \centering
    \includegraphics[max size={\textwidth}{\textheight}]{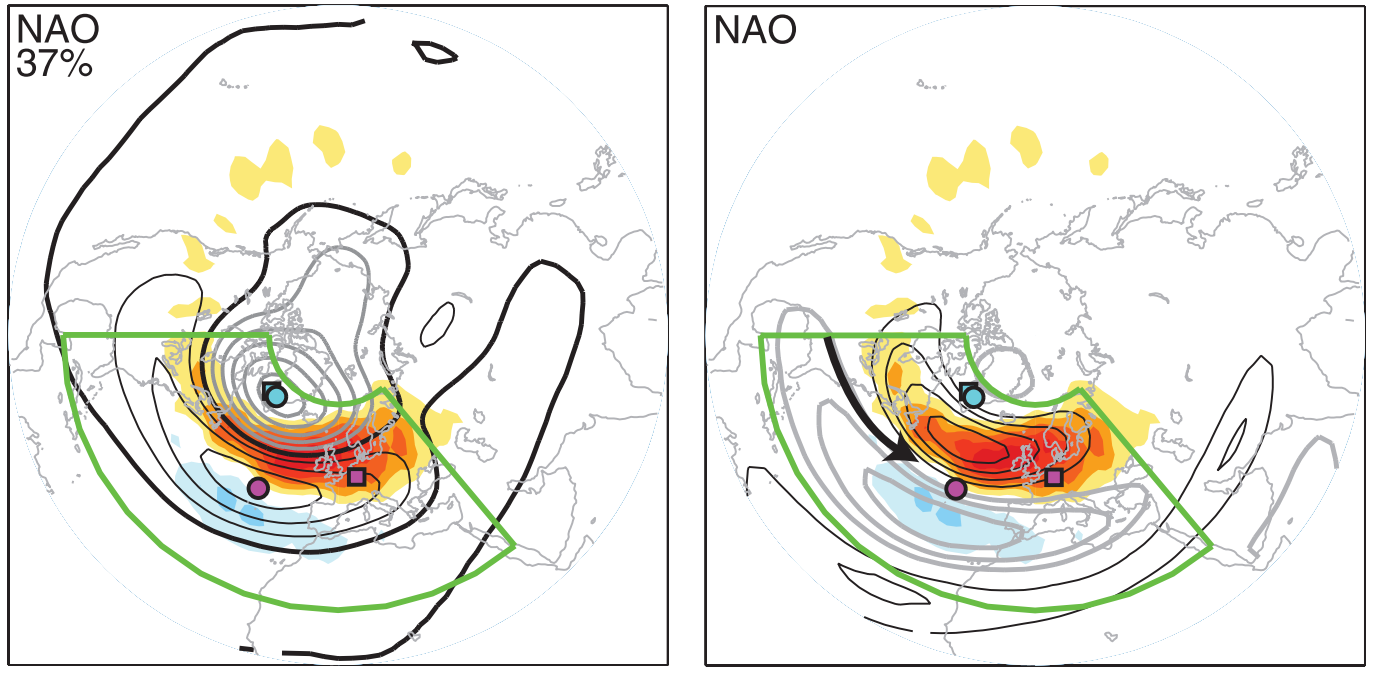}
    \caption{Left panel: Regression of the variance of the meridional eddy velocity (as diagnostic for the stormtrack; v$^\prime$v$^\prime$\textsubscript{300}; m$^{2}$ s$^{-2}$; color shadings) and geopotential height (z\textsubscript{300}; m; contours) at 300\,hPa onto standardized NAO index during winter season (DJFM). The NAO index has been defined as the leading empirical orthogonal function of SLP data over the North Atlantic sector (20$^{\circ}$--70$^{\circ}$N, 90$^{\circ}$W--40$^{\circ}$E; green box). Contour intervals for the v$^\prime$v$^\prime$\textsubscript{300} and z\textsubscript{300} regressions are 10\,m$^{2}$ s$^{-2}$ and 20\,m per NAO standard deviation, respectively. Positive (negative) z\textsubscript{300} regression contours are black (gray) and the zero contour is bold. The magenta and cyan markers represent the position of the maximum and minimum in the z\textsubscript{300} regressions, respectively. The magenta and cyan squares represent the position of the maximum and minimum in the z\textsubscript{300} regressed onto the NAM index. The NAM index has been defined as the leading empirical orthogonal function of SLP data over the Northern Hemisphere (poleward of 20$^{\circ}$N). Right panel: As in the left panel, but for v$^\prime$v$^\prime$\textsubscript{300} (m$^{2}$ s$^{-2}$; color shadings) and zonal winds at 300\,hPa (u\textsubscript{300}; m s$^{-1}$; color shadings). Contour intervals for the v$^\prime$v$^\prime$\textsubscript{300} and u\textsubscript{300} regressions are 10\,m$^{2}$ s$^{-2}$ and 2\,m s$^{-1}$ per NAO standard deviation, respectively. Black (gray) contours in the u\textsubscript{300} regressions represent westerly (easterly) anomalies. The black arrow indicates the latitude of maximum climatological u\textsubscript{300} over a longitude range where the winter upper-tropospheric jet is well defined. The figure is a partial reproduction of Figure 7 and Figure 8 in \citet{wettstein2010}.}
    \label{fig:wettstein2010.png}
\end{figure}

Furthermore, the NAO influences the position of stromtrack and eddy-driven jet in the North Atlantic. Very briefly, it is here indicated that the stormtrack is the region of high temporal variability on timescales of a few days, mainly linked with the synoptic systems that develop in the North Atlantic. The eddy-driven jet is a narrow band of strong westerly winds located around 30$^{\circ}$--60$^{\circ}$N, mainly driven by intense eddy activity along the stormtrack. The positive NAO is associated with enhanced and reduced eddy activity in the extratropical and subtropical North Atlantic, respectively (Figure \ref{fig:wettstein2010.png}, left panel). Consistently, the westerly winds are intensified at extratropics and reduced at tropics, thanks to the eddy momentum flux convergence/divergence due to transient eddies (Figure \ref{fig:wettstein2010.png}, right panel). The reverse is true for the negative NAO. Such anomalies represent a northward (southward) shift of both stormtrack and eddy-driven jet during the positive (negative) NAO respect their climatological position.

\begin{figure}[ht!]
    \centering
    \includegraphics[max size={\textwidth}{\textheight}]{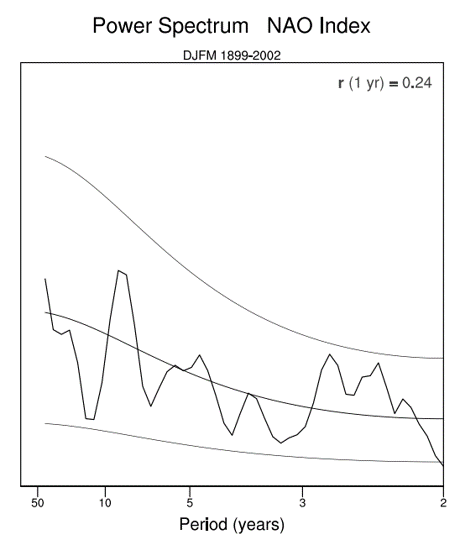}
    \caption{Power spectrum of the winter NAO index over 1899--2022. The NAO index has been defined as the principal component of the leading empirical orthogonal function of SLP data over the North Atlantic sector (20$^{\circ}$--70$^{\circ}$N, 90$^{\circ}$W--40$^{\circ}$E), as in \citet{hurrell2003}. The red noise spectrum and the 5\% and 95\% are also shown. The figure is reproduced from \citet{hurrell2010}.}
    \label{fig:hurrell2010-2.png}
\end{figure}

The NAO has been shown to vary on a broad spectrum of timescales, showing enhanced variability on (sub-)monthly, interannual and decadal timescales \citep{feldstein2000,bellucci2008,reintges2017}. As an example, I show here the power spectrum of the winter-mean NAO index defined as the principal component of the leading empirical orthogonal function of SLP data over the North Atlantic sector \citep[20$^{\circ}$--70$^{\circ}$N, 90$^{\circ}$W--40$^{\circ}$E;][Figure \ref{fig:hurrell2010-2.png}]{hurrell2003}. The winter-mean NAO index shows a slightly red power spectrum, with power increasing with period. Specifically, there is an enhanced NAO variability at [2--3]-year timescale, followed by a reduced power at [3--6]-year timescale and a peak at the sub-decadal timescale. Some studies have also suggested that the NAO can vary at multidecadal timescales \citep{omrani2014,peings2014,woollings2015}. However, the limited length of observational records makes it difficult to verify this peak.

For completeness, it is specified that the NAO index can also be defined with other approaches rather than the one presented above \citep{walker1932,wallace1981,rogers1984,hurrell1995,jianping2003}. For example, another widely adopted approach in literature has been to define the NAO index as the normalized difference of SLP in two weather stations located at low- and high-latitudes (Azores--Iceland, \citet{rogers1984}; Lisbon--Iceland, \citet{hurrell1995}; Gibraltar--Iceland, \citet{jones1997}). The NAO indices defined with these two approaches result highly comparable \citep{hurrell2003}.

The NAO variability described above has been explained as the result of several processes, both internal and external to the atmosphere.

The high-frequency NAO variability has been suggested to rise from internal, non-linear atmospheric processes. \citet{feldstein2000} has shown that the interaction between mean flow and transient eddies in North Atlantic region give rise to a peak in the NAO variability at sub-monthly timescale, with oscillations of about 10 days. Some authors have shown that the atmospheric random processes could also explain the low-frequency NAO variability \citep{stephenson2000,thompson2003}. In this context, the enhanced NAO variance at interannual and longer timescales should be understood as the remnant of an infinite number of high-frequency stochastic processes. This interpretation of the atmospheric variability has been referred to as the “climate noise paradigm” \citep{leith1973,madden1976}.

Another driver of NAO variability on short timescales has been suggested to be the stratosphere \citep{baldwin1999,baldwin2001}. Indeed, it has been shown that the anomalies in the lower stratospheric circulation propagate downward and manage to affect the troposphere up to three weeks after. The stratospheric “downward control” on tropospheric variability is linked to the effect of stratospheric flow on the upward propagation of tropospheric planetary waves \citep{shindell2001,ambaum2002}. In this context, it is important to point it out that the lower stratospheric flow is affected by ozone, greenhouse gases and volcanic aerosol concentrations. These substances affect the meridional temperature gradient in the lower stratosphere and, ultimately, the polar vortex via radiative cooling/warming. Thus, changes in their concentration can play a role on the NAO variability by inducing stratospheric circulation anomalies that propagate downward in the troposphere \citep{graf1998,shindell2001}.

The spectral peak in NAO variability at interannual, decadal and longer timescales has been mainly interpreted as the result of air--sea interaction in the extratropics \citep{barsugli1998,rodwell1999,czaja2003,reintges2017}. The huge thermal inertia of the ocean mixed layer provides a marked persistence to the SST anomalies induced by atmospheric forcing \citep{frankignoul1977}. Then, these anomalies can feed back on the atmospheric circulation via SHF \citep{hoskins1981}. In this context, it has been shown that the thermal damping of oceanic and atmospheric temperature anomalies is reduced when the atmosphere and ocean are coupled \citep{barsugli1998}. Thus, both the atmospheric and oceanic thermal variance are enhanced on timescales longer than the heat capacity of ocean mixed layer (several months). This paradigm has been referred to as the “reduced thermal damping” paradigm.

The SST anomalies forced by large-scale atmospheric circulation can also feed back on the atmospheric flow by changing the surface baroclinicity \citep{kushnir2002}. Indeed, the SST patterns affect the surface temperature gradient in the atmosphere which is directly linked with baroclinicity. Thus, the changes in transient eddy activity force the atmospheric circulation via eddy—mean flow interaction \citep{hoskins1983,novak2015}.

The ocean plays an important role in enhancing the NAO variance on long timescales also through its dynamics. Several studies have shown that the NAO can force SST anomalies in the STG which are then transported northeastward along the NAC in a decade \citep{sutton1997,krahmann2001}. Furthermore, other studies have shown that the NAO influences the deep convection in the LS via anomalous SHF, with a consistent delayed response in the AMOC on the decadal timescale \citep{joyce2000,eden2001,bellucci2008,reintges2017,wills2019}. Such dynamical modes of oceanic variability redden the spectrum of the SST in the North Atlantic and, ultimately, enhance the low-frequency NAO variability.

The NAO variability can be also driven by non-local SST anomalies \citep[outside the extra-tropical North Atlantic;][]{trenberth1998,robertson2000,sutton2000,czaja2002}. The SST patterns in tropical Pacific and Atlantic can affect the atmospheric convection and then the rate and the location of diabatic heating within the troposphere. Such diabatic heating force Rossby wave trains and/or changes in Hadley circulation which can condition the North Atlantic climate inducing atmospheric circulation anomalies partly projecting on the NAO \citep{hoerling2001}.

Another external driver of the interannual NAO variability is the Northern Hemisphere cryosphere \citep{cohen2016,wegmann2020}. As an example, the sea-ice loss in the European Arctic during autumn season has been shown to be associated with upward Rossby wave propagation and a consistent weakening of stratospheric polar vortex \citep{cohen2014}. These anomalies in the low-stratosphere propagate downward into the troposphere in following winter, determining negative NAO-like atmospheric circulation anomalies over the Euro-Atlantic sector \citep{kretschmer2018}. A similar atmospheric response has been found for increased snow cover over eastern Eurasia during winter season \citep{ghatak2010}. The changes in sea-ice and snow cover described above have been shown to be linked each other \citep{wegmann2015,gastineau2017}.  Being the Barents—Kara sea-ice extension deeply affected by oceanic circulation in the North Atlantic, such anomalies in sea-ice extension over Euro-Atlantic sector and snow cover over Eurasia can also play a role on decadal NAO variability \citep{wegmann2020}. Other studies have shown that anomalies in sea-ice over the LS and snow cover in North America also play a role on decadal NAO variability \citep{watanabe1999,kvamsto2004}.

Finally, the decadal NAO variability has been shown to be also driven by the 11-year solar cycle, consistently with the stratospheric “downward control” on the tropospheric circulation discussed above \citep{thieblemont2015}.

\section{The Gulf Stream SST front impact on the atmospheric circulation}
\label{GSF_impact_on_atm_circulation}

The GS is a fundamental component of northern hemisphere climate because it strongly contributes to the meridional heat transport, partly balancing the differential solar input between tropics and high latitudes. Furthermore, the GS is fundamental in determining the horizontal distribution of surface diabatic heating, thus setting the midlatitude stationary and quasi-stationary atmospheric eddies. Finally, the GS plays an important role in northern hemisphere climate because it is characterized by a strong SST gradient (oceanic front). Indeed, there are pieces of evidence showing that the oceanic fronts, such as the one associated with the GS, affect the atmospheric circulation both on local- and large-scale and on different timescales \citep{minobe2010,sampe2010,brayshaw2011,joyce2019}. This fact is quite surprising because in contrast with the dominant paradigm of large-scale air--sea interaction at extratropics. Specifically, the low-frequency variability of extratropical large-scale SST has been mainly interpreted as passive oceanic response to stochastic high-frequency atmospheric forcing \citep{frankignoul1977}. At the same time extratropical large-scale SST anomalies have been shown to exert a weak impact on the atmosphere, with the ocean-induced atmospheric response essentially projecting onto intrinsic atmospheric variability modes \citep{kushnir2002}. Several studies have shown that this interpretation of air--sea interaction in extratropics is not valid over meso-scale oceanic features such as the WBC (and their oceanic fronts) and the oceanic eddies \citep{chelton2001,chelton2004,xie2004,bellucci2021}. Thus, taking into account the role of the atmospheric circulation in determining the existence of WBCs, a two-way interaction is expected between the atmosphere and the WBCs (oceanic fronts). Section \ref{mechanisms} describes the main mechanisms through which the oceanic fronts force the atmospheric circulation. Furthermore, it shows the oceanic front impacts on the time-mean atmospheric circulation, with particular focus on the GSF. Then, section \ref{variability} provides some pieces of evidence of the GSF impact on the atmospheric variability.

\subsection{Oceanic-forcing mechanisms and impact on the time-mean atmospheric circulation}
\label{mechanisms}

\subsubsection{Vertical mixing mechanism}

\begin{figure}[ht!]
    \centering
    \includegraphics[max size={\textwidth}{\textheight}]{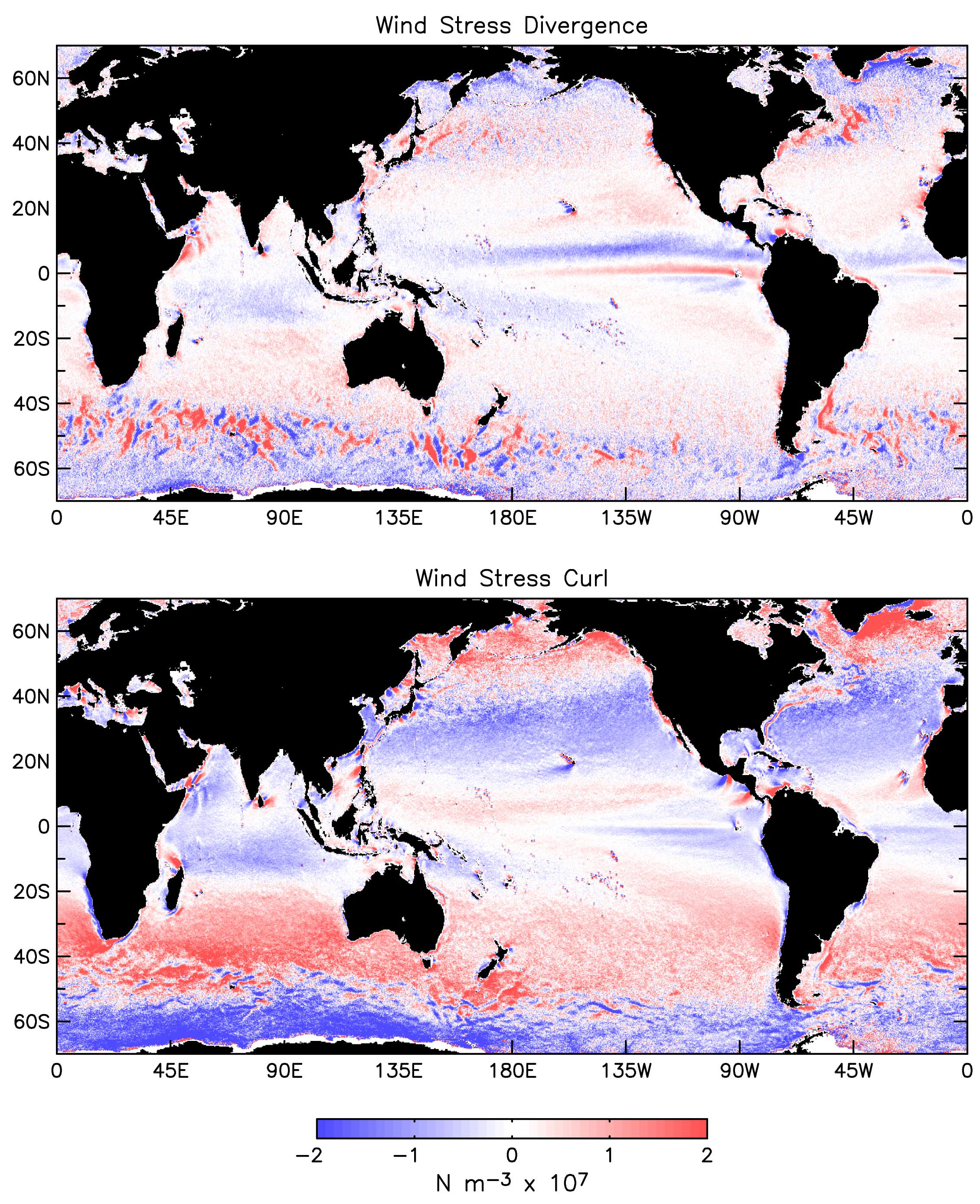}
    \caption{4-year average of the near-surface wind-stress divergence (top) and curl (bottom), computed from 25-km-resolution wind measurements by the QuikSCAT scatterometer. The figure is reproduced from \citet{chelton2004}}
    \label{fig:chelton2004.png}
    \doublespacing
\end{figure}

High-resolution satellite observations have highlighted small-scale patterns in the near-surface wind-stress curl and divergence co-located over meso-scale oceanic features such as the oceanic fronts \citep[Figure \ref{fig:chelton2004.png};][]{chelton2001,nonaka2003,chelton2004,oneilly2005,small2008}. The direct link between the near-surface wind and SST patterns over the oceanic fronts is better shown by the correspondence between the laplacian of the SST (or the SLP) and the wind divergence. Indeed, the laplacian operator acts as a spatial high-pass filter, unveiling the oceanic fronts effect which is generally masked by large-scale atmospheric circulations \citep{xie2004,minobe2008a}. In contrast to what is expected, the correlation between the near-surface wind speed and SST is positive over such oceanic features \citep{battisti1995,hashizume2001}. This means that in the vicinity of oceanic fronts the SST patterns are driven by intrinsic oceanic processes rather than by heat fluxes from the atmosphere. Hence, over such oceanic features it is more the oceanic variability that affects the atmospheric one than the vice versa.

\begin{figure}[ht!]
    \centering
    \includegraphics{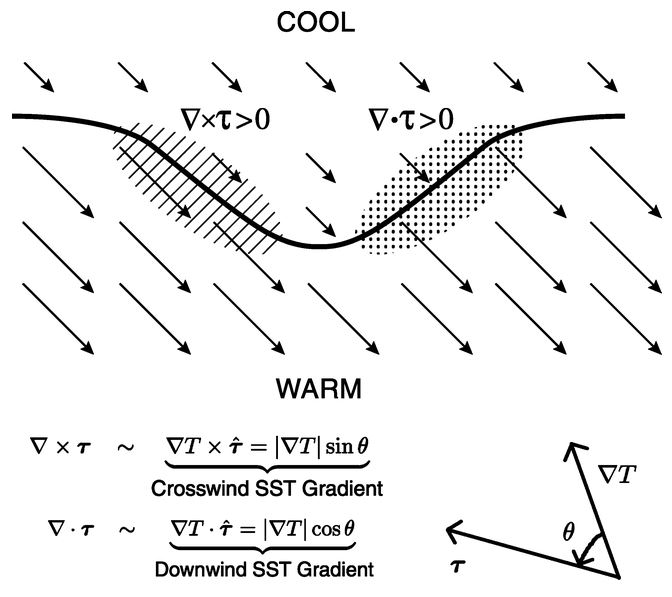}
    \caption{Schematic of the impact of an oceanic front on the near-surface wind-stress due to the vertical mixing mechanism and how the near-surface wind-stress variations across the front generate near-surface wind-stress divergence and curl. The schematic is reproduced from \citet{maloney2006}.}
    \label{fig:maloney2006.JPG}
\end{figure}

The mechanism through which oceanic fronts are able to induce zones of near-surface wind-stress curl and divergence has been referred to as the vertical mixing mechanism \citep{wallace1989}. The SST differences across the oceanic fronts induce distinct thermodynamic conditions in the near-surface atmosphere. Specifically, the near-surface atmosphere is less (more) stable over the warm (cold) side of the SST front and then the vertical mixing in the MABL is intensified (reduced). Consistently, the vertical gradient in the profiles of air temperature, humidity and velocity in unstable (stable) conditions is lower (greater) than the typical neutral logarithmic profile due to the turbulent fluxes \citep{stull1988}. One effect of the vertical mixing within the boundary layer is to transfer horizontal momentum from the upper layers to the near-surface layers during the unstable conditions. Thus, because of the differential vertical mixing across the oceanic fronts, the near-surface winds over the warm side of an oceanic front tend to be stronger than the ones over its cold side \citep{wallace1989}, creating areas of near-surface wind divergence and curl \citep{chelton2004}. The wind divergence is linearly dependent on the component of the SST gradient parallel to the wind (downwind component), whereas the wind-stress curl is linearly dependent on the component of the SST gradient perpendicular to the wind \citep[crosswind component;][]{chelton2001,oneill2003,maloney2006}. The more intense the downwind and crosswind SST gradient, the more intense the wind-stress divergence/convergence and curl. Refer to Figure \ref{fig:maloney2006.JPG} for a schematic of the impact of an oceanic front on the near-surface wind-stress due to the vertical mixing mechanism and how the differential vertical mixing across the front generates wind divergence and curl. The resulting areas of near-surface divergence and convergence due to vertical mixing induce anomalies in the vertical motion that can impact the atmospheric dynamics and thermodynamics also above the boundary layer \citep{xie2004,small2008}.

\subsubsection{Atmospheric pressure adjustment mechanism}

\begin{figure}[ht!]
    \centering
    \includegraphics{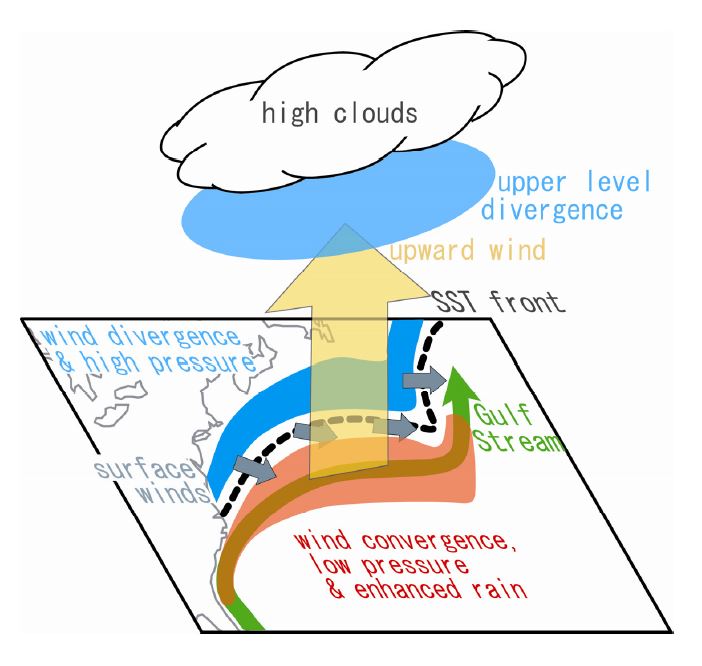}
    \caption{Schematic of the climatic responses to the GSF. The green arrow indicates the warm core of the GS. The black dashed line indicates the GSF. The grey arrows indicate the cross-front near-surface winds. The blue region north of the GSF identifies the area of near-surface wind divergence and high pressure. The orange region south of the GSF identifies the region of near-surface wind convergence, low pressure and enhanced rain. The vertical yellow arrow represents the intense upward motion anchored to the GSF. The blue oval aloft identifies the area of upper-level divergence. The schematic is reproduced from \citet{minobe2008b}.}
    \label{fig:minobe2008b.JPG}
\end{figure}

The GSF has been shown to be associated with time-mean upward motion along its own path extending all the way up to the upper-troposphere \citep{minobe2008a}. The upward motion in the atmospheric boundary layer is induced by the atmospheric pressure adjustment to SST differences across the oceanic front. As shown in Figure \ref{fig:minobe2008b.JPG}, the GSF system is characterized by a low and high pressure system on the warm and cold side of the oceanic front, respectively. Thus, a cross-front secondary circulation is established \citep{small2008}, anchoring a zone of convection on the southern side of the GSF because of wind-convergence over this side of the front. The extension of the upward motion in the free atmosphere is induced by the atmospheric instability due to the latent heat release during the ascent motion. Consistently, a narrow band of precipitation is found along the GSF \citep{pan2002}. Being the GS region an area of intense cyclogenesis, the time-mean ascent motion should be understood as the cumulative effect of frontal systems passing over the GS region \citep{parfitt2016a,parfitt2018} and not as the average daily circulation in that region. In the GSF system, high-cloud formation can occur, with highest frequencies in the winter season \citep{minobe2008a,minobe2010}. The upper-level divergence induced by the deep convection over the GS and the latent heat release in the troposphere represents ways through which the GS can affect the tropospheric circulation not only locally but also in remote areas by forcing stationary planetary waves \citep{hoskins1981, hoskins1990, held2002}.

Before closing this section, we want to specify that the different positions of wind-stress divergence respect an oceanic front shown in Figure \ref{fig:maloney2006.JPG} and Figure \ref{fig:minobe2008b.JPG} are not in contrast each other. Specifically, the area of wind-stress divergence occurs over the SST front in Figure \ref{fig:maloney2006.JPG}, whereas it occurs north of the GSF in Figure \ref{fig:minobe2008b.JPG}. In this context, it should be taken into account that Figure \ref{fig:maloney2006.JPG} shows the impact of a time-varying oceanic front on the near-surface wind-stress, whereas Figure \ref{fig:minobe2008b.JPG} shows the climatic responses to the GSF. The absence of wind-stress divergence over the climatological GSF position in Figure \ref{fig:minobe2008b.JPG}, as expected looking at Figure \ref{fig:maloney2006.JPG}, should be understood as the effect of time averaging over an area of strong eddy activity.

\subsubsection{Oceanic baroclinic adjustment mechanism}

\begin{figure}[ht!]
    \centering
    \includegraphics[max
    size={\textwidth}{\textheight}]{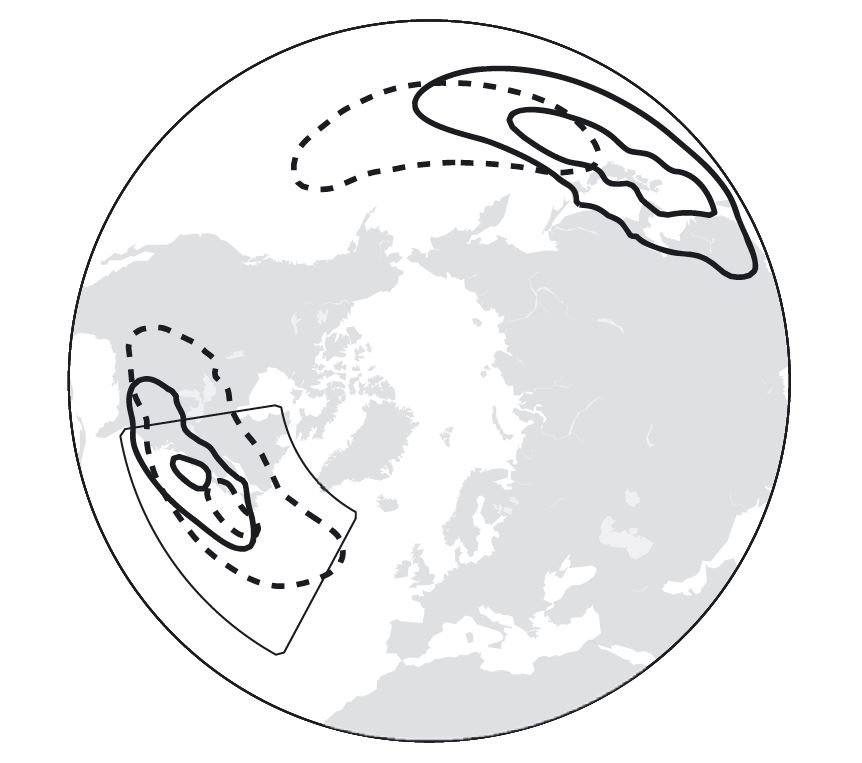}
    \caption{Baroclinicity (0.5 and 0.6 day$^{-1}$; solid contours) at 775\,hPa and meridional eddy heat flux (10 and 20 K m s$^{-1}$; dashed contours) between 700 and 925\,hPa averaged over the 1957--2002 DJF winters. The figure is reproduced from \citet{ambaum2014}. The black sector over the GS represents the area of interest for the analysis developed in that study.}
    \label{fig:ambaum2014.JPG}
\end{figure}

The presence of the GSF has been shown to be important also in shaping the North Atlantic stormtrack and eddy-driven jet \citep{hoskins1990, nakamura2008,brayshaw2011}. Indeed, the GSF intensifies the near-surface atmospheric baroclinicity in the western portion of the North Atlantic, anchoring a zone of intense baroclinic eddy activity \citep{sampe2010, brayshaw2011, novak2015} (Figure \ref{fig:ambaum2014.JPG}). Because of the latter, the southwest--northeast tilt of the stormtrack over the North Atlantic induced by the presence of the Rocky Mountains and the North American continent is further increased \citep{brayshaw2009}. Consistently, the jet stream is tilted thanks to the eddy--mean flow interaction \citep{hoskins1990, brayshaw2011}. The near-surface baroclinicity necessary for baroclinic instability growth is maintained by the GSF through the differential diabatic heating supplied across the front against the erosive effect due to the baroclinic eddies \citep{nakamura2004,sampe2010}. \citet{hotta2011} have shown that the sensible heating plays a more important role than the other diabatic heating sources in maintaining the near-surface baroclinicity.

\begin{figure}[ht!]
    \centering
    \includegraphics{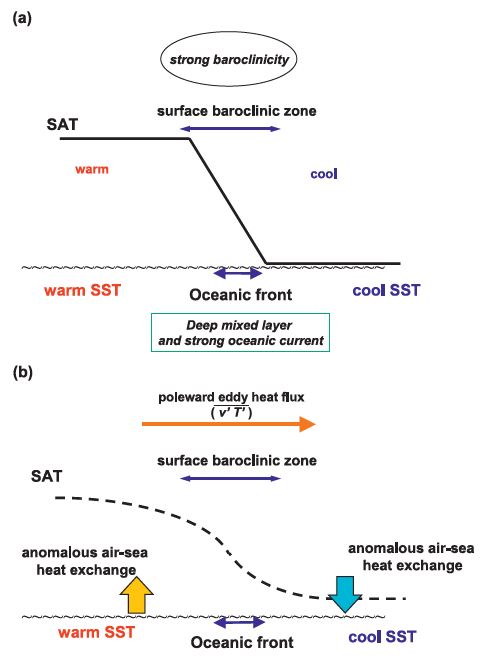}
    \caption{Schematic diagrams showing the restoration of near-surface baroclinicity by an oceanic front through SHF from the ocean against the relaxation by poleward eddy heat transport, reproduced from \citet{sampe2010}.}
    \label{fig:sampe2010.JPG}
\end{figure}

The mechanism through which the sensible heat fluxes maintain the near-surface baroclinicity can be described as follows. The sensible heat fluxes establish an intense surface atmospheric temperature gradient (i.e. baroclinicity) across the oceanic front. As expected, the sharp gradient leads to the development of baroclinic eddies which tend to erode the atmospheric temperature gradient through the associated poleward eddy heat transport. This increases the differences between the SST and the temperature of the overlying air, leading to enhanced upward (downward) sensible heat fluxes south (north) of the GSF. In turn, this leads to a stronger across-front gradient of SHF, the near-surface baroclinicity is restored and the transient eddies can develop again. This mechanism has been referred to as ”oceanic baroclinic adjustment” by \citet{nakamura2008} and it differs from the ”baroclinic adjustment” proposed by \citet{stone1978}, in which it is the differential radiative heating between high and low latitudes to restore the near-surface baroclinicity. The restoration of the near-surface baroclinicity by an oceanic front through the sensible heat fluxes is schematized in Figure \ref{fig:sampe2010.JPG}. More in detail, \citet{ambaum2014} and \citet{novak2015} have shown that the relationship between the near-surface baroclinicity (also induced by the oceanic forcing) and the poleward eddy heat transport can be described by a nonlinear oscillator model.

\subsubsection{Other aspects}

Furthermore, the GSF has been shown to affect the time-mean stratospheric circulation. Results from \citet{omrani2019} indicate that the presence of the GSF warms and weakens the stratospheric polar vortex. This is possible because the GSF enhances the upward wave propagation, leading to wave-activity convergence in the low stratosphere and then to a deceleration of the westerlies \citep{andrews1987}. Recent studies have also suggested a possible effect of the GSF on the troposphere--stratosphere coupling. In particular, in aqua-planet experiments with an atmospheric GCM (AGCM)\nomenclature{AGCM}{atmospheric general circulation model}, \citet{ogawa2015} have shown that the stratospheric changes associated with the ozone depletion in the southern hemisphere can propagate into the troposphere only if the SST front in the South Indian ocean is included. The signal can propagate downward because of the excitation of synoptic and planetary waves by the oceanic front. This means that a realistic representation of the eddy-driven tropospheric jet, only possible in the presence of the oceanic fronts, is important for the troposphere--stratosphere coupling. Consistently with \citet{ogawa2015}, \citet{lubis2018} have demonstrated that the stratospheric changes associated with the polar vortex weakening requires the internal tropospheric activity wave forcing associated with eddy-driven jet dynamics to propagate into the troposphere. Since the GSF directly impacts the tropospheric eddy-driven jet and its dynamics, it is expected that the GSF can impact the troposphere--stratosphere coupling.

The influence of the GSF on the North Atlantic time-mean atmospheric circulation has been mainly analysed using atmosphere-only model simulations forced with realistic and smoothed SSTs in the GS region \citep[e.g.][]{oreilly2016, oreilly2017,omrani2019}. The results show that the presence of a realistic GSF generates stronger baroclinicity over the North Atlantic than simulations with a smoothed or absent front. Consistently, the meridional eddy heat flux is stronger and the eddy-driven jet moves more polewards. Furthermore, the trimodal character of the observed eddy-driven jet latitude distribution is better represented when the GSF is not smoothed, increasing the occurrence of the northern jet position. This has been shown to also improve the representation of blocking frequency and position over Europe \citep{oreilly2016}. Looking at single atmospheric instabilities, \citet{parfitt2016b} have shown not only that the presence of the GSF induces higher occurrence of the atmospheric fronts (i.e. transient eddies) in the western North Atlantic, but also that these changes are mainly due to an higher occurrence of the cold atmospheric fronts. In a similar analysis, \citet{sheldon2017} have shown that the upward motion in the warm conveyor belt of a cyclone is stronger and reaches a higher level of the troposphere when the GSF is realistically represented in GCMs. Consistently, \citet{kuwano2010} have shown that the winter-mean upward motion above the GSF is much weaker in an atmosphere-only model forced by smoothed SST front compared to the same model forced by realistic front condition. Taking into account the role of the atmospheric synoptic eddies on the North Atlantic climate, the latter studies provide further evidence of the GSF role in determining the stormtrack and eddy-driven jet characteristics.

\subsection{Impact on the atmospheric variability}
\label{variability}

As a result of intrinsic oceanic variability as well as atmospheric forcing, the GSF undergoes meridional shifts and meandering, with distinct impacts on the atmospheric variability \citep{lee1995, ducet2000, kwon2010, quattrocchi2012, kwon2013}. Specifically, the SST anomalies corresponding to the GSF shifts are able to meridionally displace the area of intense synoptic variability in the western North Atlantic \citep{joyce2009}. \citet{nakamura2009} have shown that small-scale SST anomalies along the GSF may also affect the atmospheric circulation in larger portions of the North Atlantic. Indeed, positive (negative) SST anomalies along the GSF enhance (reduce) the local and downstream low-level baroclinicity. Then, the baroclinic activity along the North Atlantic stormtrack increases and the eddy-driven jet stream shifts poleward. Consistently, \citet{joyce2009} and \citet{joyce2019}, using observational data, have presented evidence for significant meridional shifts in the North Atlantic eddy-driven jet and stormtrack as a homo-directional response to the GSF shifts, including dynamically consistent changes in the distribution of blocking frequency. Again, \citet{sato2014} have shown that poleward GSF shifts are associated with abnormal sea-ice retreat over the Barent Sea via anomalous southerly warm advection and cold anomalies over the Eurasian continent via planetary waves triggered over the GS. In the context of the GSF influence on remote areas, we should also mention the work by \citet{honda2001a}. Their results suggest that the low-boundary conditions over the GS region can affect the propagation of wave structures coming from the North Pacific basin and induced by oceanic forcing in that region. This means that the GSF variability can affect the atmospheric circulation in downstream areas not only by triggering wave-train as suggested by \citet{sato2014}, but also by modulating the North Pacific--North Atlantic interaction. 

Differently from the atmospheric response to the GSF shifts, there are no studies directly dealing with the atmospheric response to meandering of the GSF. However, some authors have analyzed the atmospheric response to the meandering of the Kuroshio Extension SST front. For example, \citet{oreilly2015} have observed that periods of strong and elongated SST fronts are associated with an intensified low-level baroclinicity and eddy heat transport in the western North Pacific. These changes in the western oceanic basin force anomalous barotropic flow in the eastern North Pacific via eddy--mean flow interaction, with greater occurrence of blocked days. \citet{ma2017} have analyzed the impact of SST front meandering on the atmospheric circulation forcing a Weather Research and Forecasting (WRF)\nomenclature{WRF}{Weather Research and Forecasting} model with solved and smoothed oceanic eddies along the Kuroshio Extension region. Their results show that the realistic representation of mesoscale SST anomalies leads to the local cyclogenesis enhancement and the northward shift of the stormtrack and jet stream in the eastern North Pacific. Even if the GS and the Kuroshio WBCs are very different systems, these results shed light on the possible impact of the GSF meandering on the atmospheric circulation.
\chapter{\textit{\textbf{The atmospheric response to meridional shifts of the Gulf Stream SST front and its dependence on model resolution}}}
\label{atmospheric_response}

\fancyhead[R]{\textit{Chapter \ref{atmospheric_response} - The atmospheric response to GSF shifts}}

\section{Introduction}

Recent observational and modelling studies have shown that the GSF variability affects both the local and large-scale atmospheric circulation (refer to section \ref{GSF_impact_on_atm_circulation}), acting as a source of atmospheric predictability especially on interannual and decadal time scales \citep{joyce2019, athanasiadis2020}. 

Despite these advancements, the real character of the atmospheric response to oceanic front variability is still not well established. Some observational studies have shown that the GSF shifts increase the baroclinic activity in North Atlantic, inducing homo-directional shifts of the eddy-driven jet and the stormtrack and dynamically consistent changes in the distribution of blocking frequency \citep{nakamura2009,joyce2009,sato2014,kwon2013,joyce2019}. However, such anomalous atmospheric circulation has been elusive to detect in other observational studies \citep{joyce2000, frankignoul2001}. For example, \citet{wills2016} have shown that positive SST anomalies over the GS area determine negative NAO-like circulation anomalies, i.e. an opposite response compared to the one described above. Following the same methodological approach of this study, \citet{yook2022} have extracted similar results for SST variability over the Kuroshio current. 

Using atmosphere-only simulations could partly help to disentangle this issue as they allow to isolate the impact of the oceanic variability on the atmosphere. However, even if state-of-the-art AGCMs with horizontal resolutions close the oceanic deformation radius ($\sim$50\,km) are more reliable than previous models and better represent the observed climate, important inter-model discrepancies exist \citep{czaja2019}. \citet{smirnov2015} have shown that, in an AGCM with horizontal resolution of 1$^{\circ}$, the atmospheric response to shifts in the Oyashio Extension SST front results to be weak and exhibits features generally consistent with a steady linear response to a near-surface heat source \citep{hoskins1981}. On the other hand, in the AGCM with horizontal resolution of 0.25$^{\circ}$\, the near-surface circulation is substantially weaker, while the vertical motion is stronger and deeper (affecting the upper-troposphere) and the surface meridional eddy heat transport largely balances the SST-induced diabatic heating anomalies. The differences in vertical motion between these two AGCMs recall results from \citet{feliks2004}. Using a very idealized framework, they showed that a narrow oceanic front is able to force the atmospheric circulation above the MABL through thermal pumping of vertical velocity. However, since oceanic fronts have a width of 100\,km or less, only sufficiently high-resolution models can represent this mechanism. In line with this, \citet{nakamura2009} argued that only models with a grid spacing no larger than 50\,km are able to adequately resolve slight meridional shift of the GS and its impact on the large-scale atmospheric circulation. \citet{ma2017} showed that a WRF model with a 27-km resolution is much more sensitive to mesoscale SST anomalies along the Kuroshio extension compared to the model with a 162-km resolution. The high-resolution model version exhibits a southward shift of Eastern North Pacific stormtrack and jet stream when the mesoscale SST anomalies are smoothed; in contrast, the low-resolution model is not able to resolve the small-scale diabatic processes associated with the mesoscale SST forcing and therefore shows a weak response in eddy activity and large-scale circulation for the same smoothing in oceanic boundary conditions. Looking at single atmospheric instabilities, \citet{willison2013} have shown that a WRF model with a 20-km horizontal resolution exhibits enhanced frontal dynamics in North Atlantic mid-latitude cyclones compared to the same WRF model with a 120-km resolution. The authors have highlighted a more intense positive feedback between cyclone intensity and latent heat release in the high-resolution model, resulting in the intensification of the stormtrack as well as in the strengthening of the jet stream. Following these results, \citet{sheldon2017} have argued that the diabatic heating in the warm conveyor belt of cyclones travelling close to the GSF is directly proportional to the number of air-parcel trajectories feeding the upward motion. The higher the model resolution, the higher the number of air-parcel trajectories reaching the upper-troposphere.

Despite these efforts, the number of studies dealing with the impact of horizontal resolution on the atmospheric response to oceanic forcing is still limited. In addition, previous analyses are based on idealised experimental frameworks forcing the atmosphere with fixed and unrealistic SST anomalies. Finally, such past studies have been limited to single-model assessments. A multi-model analysis to systematically investigate differences between the low-resolution and high-resolution atmospheric response to realistic SST variability linked to the meridional shifts of the GSF is still lacking.

The objective of this part of the PhD thesis is twofold: on the one hand it aims at assessing the dependence of the atmospheric response to GSF shifts on model resolution, in particular the atmospheric horizontal resolution, and on the other it aims at providing further evidence on the character of the large-scale atmospheric response by assessing the latter in isolation from the coupled variability in which is embedded, using atmosphere-only simulations. The atmospheric response has been investigated in the context of the High Resolution Model Intercomparison Project (HighResMIP)\nomenclature{HighResMIP}{High Resolution Model Intercomparison Project}, by analysing historical simulations performed with three AGCMs, each run at two different horizontal resolutions. The AGCMs have been forced with the same observed SSTs. Understanding the impact of horizontal resolution on air--sea interaction can shed light on the influence of extratropical oceanic variability on the atmosphere, with important implications for climate predictions and climate change studies.

The chapter \ref{atmospheric_response} of this PhD thesis is structured as follows. In section \ref{data} the HighresMIP dataset, the SST anomalies associated with the GSF shifts and the methodological approach are described. In section \ref{response} the atmospheric response to the GSF shifts is presented. In section \ref{heatbudget} we present a heat budget analysis along, across and above the GSF as well as in the North Atlantic basin, in order to investigate ways through which diabatic heating is balanced locally and at a wider scale. In section \ref{discussion} we discuss some large-scale features of the atmospheric response to the GSF shifts. Finally, in section \ref{conclusion} we summarize the results, highlighting the most salient outcomes of this part of the PhD thesis.

\section{Data and methodological approach}
\label{data}

\subsection{Data}
The HighResMIP is part of the wider Coupled Model Intercomparison Project 6, and it was designed with the specific objective of investigating the impact of increasing model horizontal resolution on the representation of the observed climate and of an array of important physical processes \citep{haarsma2016}. As such, it provides an ideal framework for a multi-model analysis of the impact of increasing model resolution on the atmospheric response to oceanic forcing. In this study, an ensemble of six atmosphere-only historical simulations have been analysed. Three different models have been used, each run with two configurations differing only in their horizontal resolution. Hereafter we will refer to model configurations with a nominal resolution coarser than 50\,km as R100 models, and those with a nominal resolution finer than or equal to 50\,km as R50+ models. Each model has been forced with the HadISST2 sea ice concentration and SST dataset, provided at daily frequency in the period 1950--2014 on a 0.25$^{\circ}$ grid \citep{kennedy2017}. For each model a multi-member ensemble of simulations has been used, and the results in the following sections refer to the respective ensemble means. This specific experimental design and the use of multi-member ensembles allows a more robust  identification of the atmospheric response forced by the observed oceanic variability as the ensemble averaging aids the forced response to emerge from the chaotic atmospheric variability, which is particularly strong at midlatitudes. The models considered in this study are documented as follows: EC-Earth3P \citep{haarsma2020}, ECMWF-IFS \citep{roberts2018}, HadGEM3-GC31 \citep{roberts2019}. Table \ref{table:models} shows the respective model configurations, nominal resolutions (km), number of vertical levels, number of members and reference (model documentation). Additional details about the experimental set-up can be found in \citet[][see \enquote{Tier 1 - highresSST-present} experiment]{haarsma2016}. Finally, model results have been compared to ERA5 reanalysis \citep{hersbach2020}, here used as a surrogate of observations in the period 1950--2014.

\begin{table}[t]
\caption{\label{table:models} HighResMIP models. Columns detail the institution name, the model name, the nominal resolution, the number of vertical levels, the number of members used for analysis and the model reference.}
\begin{center}
\resizebox{\textwidth}{!}{\begin{tabular}{l l c c c c}
\hline\hline\\
\textbf{Institution} & \textbf{Model} & \textbf{Nominal Resolution (km)} & \textbf{Vertical Levels} & \textbf{Members} & \textbf{Reference} \\
[2ex]
\hline\\
EC-Earth-Consortium & EC-Earth3P & 100 & 91 & 3 & \citet{haarsma2020} \\
& EC-Earth3P-HR & 50 & 91 & 3 & \\
[2ex]
Met Office Hadley Centre & HadGEM3-GC31-MM & 100 & 85 & 3 & \citet{roberts2018} \\ 
(MOHC) & HadGEM3-GC31-HM & 50 & 85 & 3 & \\
[2ex]
ECMWF & ECMWF-IFS-LR & 50 & 91 & 8 & \citet{roberts2019}\\
& ECMWF-IFS-HR & 25 & 91 & 6 & \\
[2ex]
\hline
\end{tabular}
}
\end{center}
\end{table}

\subsection{Gulf Stream sea surface temperature front shift}
\label{GSF_index}

\begin{figure}[ht!]
    \centering
    \includegraphics[max
    size={\textwidth}{\textheight}]{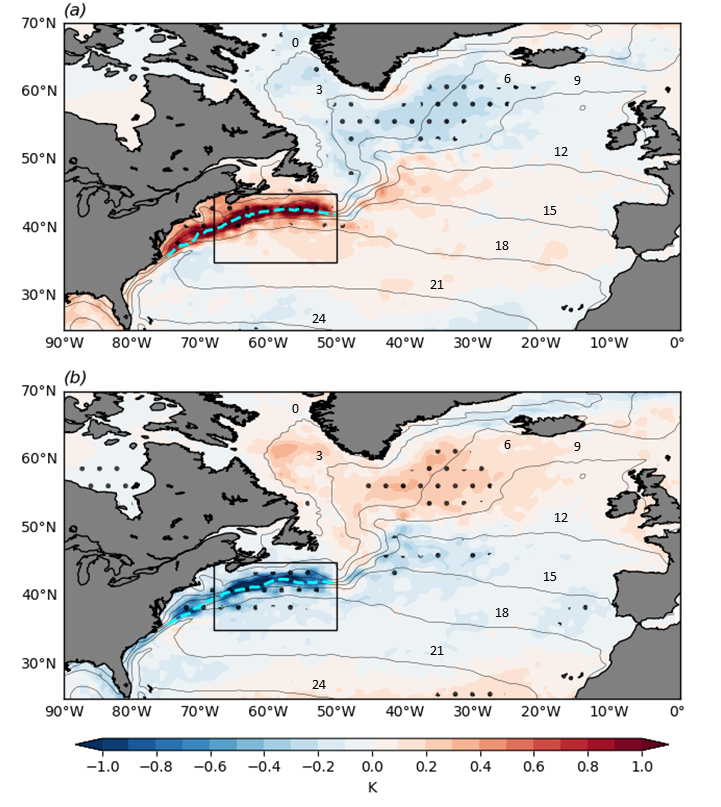}
\caption{SST (K; color shaded) anomalies associated to “North” (a) and “South” (b) phases of the GSF in winter (DJF) in HadISST2 dataset. The “North” (“South”) phase has been defined as the upper (lower) tercile of GSF mean latitude. The respective climatological position of the SST front is indicated by the cyan dashed line. The black contours indicate winter SST climatology. The black rectangular frames indicate the longitudinal range over which the GSF mean latitude and subsequent diagnostics (Figure \ref{fig:diabatic_heating}, Figure \ref{fig:budget}, Figure \ref{fig:omega}) have been calculated. Black dots denote anomalies that were found to be statistically significant at the 90\% confidence level (details in section \ref{data}).
}
\label{fig:sst}
\end{figure}

In this study, the atmospheric response to interannual GSF meridional shifts in the winter season (December--February) has been investigated. Previous studies have shown that the north--south shift of GSF represents the leading variability mode of SST variability in the GSF area on interannual and longer timescales \citep{joyce2009, kwon2013}. The winter season has been selected because this is the time of the year characterised by the most intense heat exchanges between ocean and atmosphere, resulting in a stronger impact of the ocean variability on the atmosphere \citep{kallberg2005}. The GSF has been defined as the line of maximum SST gradient. The SST gradient magnitude has been calculated for winter mean SST fields, smoothed with a 2D spatial Gaussian--Kernel filter applied to a 7x7 gridpoint box, with standard deviation equal to 2. The smoothing has been applied in order to remove isolated points of strong SST gradient not representative of the GSF position. The latitude of the GSF has been averaged in the 50$^{\circ}$--68$^{\circ}$W longitudinal range, where the GS is more zonally oriented. Then, this zonally averaged latitude of the GSF has been used to define the “North” and “South” phases of the front shift via the respective upper and lower tercile categories. 

\begin{figure}[ht!]
    \centering
    \includegraphics[max
    size={\textwidth}{\textheight}]{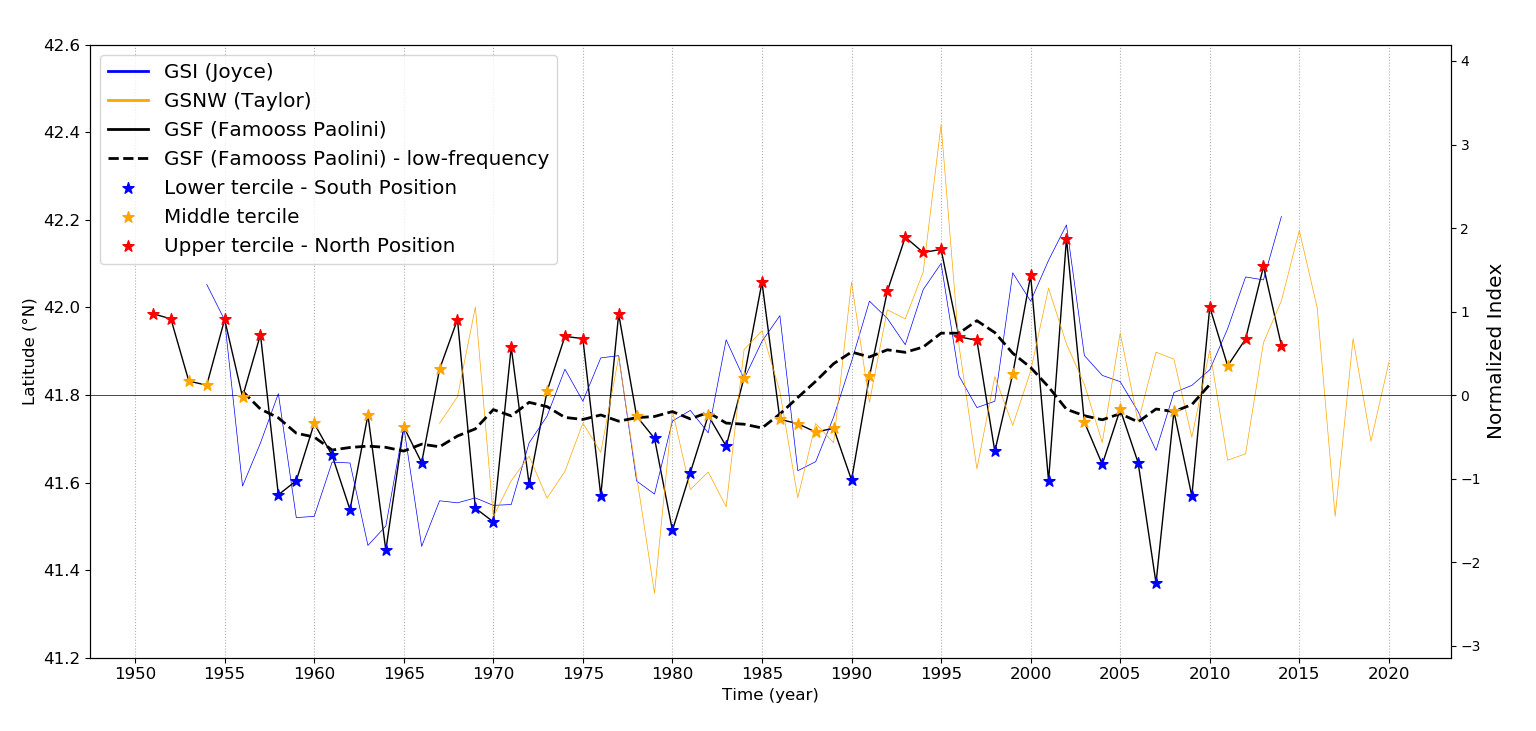}
\caption{Winter-mean latitude of the GSF averaged in the range 50$^{\circ}$--68$^{\circ}$W (solid black line). Red, yellow and blue stars represent years in which the GSF latitude falls, respectively, in the upper, middle and lower tercile categories (i.e. North, Middle and South position). The dashed black line is the 10-year running mean applied to GSF latitude timeseries. The solid and dashed black lines also represent the normalized interannual and smoothed GSF indices for the comparison with the indices suggested in other studies. The blue line represents the winter-mean normalized GSNW index as defined by \citet{taylor1980}. The orange line represents the normalized GSI as defined by \citet{joyce2019}. The GSI is averaged over January and March as representative of the winter season. Refer to the y-axis on the right for the normalized indices.}
\label{fig:GSF}
\end{figure}

Figure \ref{fig:sst} shows the SST composites obtained by averaging all years over the upper and below the lower terciles, i.e. the “North” and “South” phases of the GSF, consisting of 22 and 21 years, respectively. The resulting SST pattern shows a tripolar structure extending to the entire North Atlantic in both GSF phases but of opposite sign (Figure \ref{fig:sst}). The “North” (“South”) phase is associated dominantly with positive (negative) anomalies along the climatological position of the GSF in winter, as well as in the midlatitude North Atlantic to the south of the GSF and the NAC, and negative (positive) anomalies further north in the subpolar gyre region and south of 30$^{\circ}$N. The anomalies are strongest close to the GSF with values exceeding 1\,K (absolute departures). In the remaining part of the basin, the SST anomalies are much weaker and lower than 0.4\,K. The corresponding SST anomalies in other parts of the global ocean have been assessed, and their significance in remotely forcing an atmospheric circulation response over the extratropical North Atlantic is discussed in the following.

In Figure \ref{fig:GSF} the GSF latitude time series (black line) in the period 1950--2014 is presented. The years corresponding to the “North” (“South”) phase are highlighted with red (blue) stars. The index features an interannual variability component associated with SST front latitudinal shift of about 0.2$^{\circ}$--0.5$^{\circ}$. A lower frequency, decadal-scale component is also evident associated with weaker amplitude meridional displacements. The North Atlantic SST variability described above has been previously interpreted as the oceanic response to NAO forcing via SHF and Ekman currents, explaining an important portion of the extratropical SST variability on seasonal and interannual timescale \citep{cayan1992, deser2010}. However, it has been shown that close to the GSF the SST variability is primarily driven by oceanic processes, such as oceanic heat transport and diffusion, with the former admittedly remaining subject to the atmospheric forcing \citep{kelly1995, dong2004, kelly2010, kwon2010, minobe2010, patrizio2021}. Therefore, the SST anomalies described here are the surface oceanic signature of both historical atmospheric forcing and intrinsic oceanic variability. This aspect must be taken into account to avoid erroneous considerations during the analysis of data, especially observations. A number of studies \citep[e.g.][]{czaja2002, ciasto2004, wills2016} present evidence that SST anomalies in the vicinity of the GS may affect the large-scale atmospheric circulation up to several months ahead. Some authors have argued that meridional shifts of the GSF may be even key in explaining decadal NAO variability and predictability \citep{feliks2011, joyce2019, athanasiadis2020}.

Before closing this section, it is specified that other approaches to define the latitudinal position of the GS and its front have been proposed in literature. Amongst them, the GS North Wall (GSNW) index defined by \citet{taylor1980} and the GS Index (GSI) defined by \citet{joyce2000} have been widely used. The former defines the GS position through the principal component analysis of the surface temperature front position in the 79$^{\circ}$--65$^{\circ}$W longitudinal range. The latter defines the GS position through the principal component analysis of the 15$^{\circ}$C isotherm position at 200\,m depth in 9 fixed locations along the GS path and between 75$^{\circ}$W and 55$^{\circ}$W. Due to their widespread use, the GSF index is here compared with the GSNW index and the GSI. To do so, we have used the monthly GSNW data provided in \citet{mccarthy2018}, extracting the index for the winter season (DJF). Furthermore, we have used the GSI data provided in \citet{joyce2019}. The approach to define the GS position adopted by \citet{joyce2019} is slightly different compare to the one defined in \citet{joyce2000}. However, the two indices are very comparable being both based on the principal component analysis of the 15$^{\circ}$C isotherm position at 200\,m depth in 9 fixed positions along the GS. Being the GSI data in \citet{joyce2019} provided with a 3-month time step, we have adopted the GSI averaged over January and March as representative of the winter season. As shown in Figure \ref{fig:GSF}, the GSF index (black line) captures the general trend of both GSNW index (orange line) and GSI (blue line), especially from 1975 onward. This is not always the case on the interannual timescale, where the indices can also be in anti-phase. Consistently, the zero-lag cross-correlation between the GSF index and the GSI (GSNW index) is not particularly high, with a value of 0.52 (0.39) \citep[significance greater than 99\%; the statistical significance of the correlations has been assessed through bootstrap test in][]{ebisuzaki1997}. The different trends of the indices on short (interannual) timescale may be linked to differences in the adopted longitudinal ranges, which capture sectors of the GS with distinct behaviours and characteristics. \citet{gangopadhyay2016} have shown that the signature of the interannual variability in the GS path changes along the stream’s path from 75$^{\circ}$W to 55$^{\circ}$W. Furthermore, it should be taken into account that the GSI is based on sub-surface temperature data, which are largely unaffected by surface diabatic processes being the wintertime mixed layer west of 60$^{\circ}$W less than 200\,m \citep{qiu1995}. This is not the case for the GSNW and the GSF indices which are based on surface temperature data. The added value of the GSF index here defined stays on the fact that it specifically captures the surface oceanic variability and that it does it in the GS region with the most intense oceanic-eddy activity \citep[east of 70$^{\circ}$W;][]{ducet2000}. Both aspects are really important if we take into account that the air--sea interaction associated with small- and meso-scale oceanic features (such as oceanic fronts and eddies) has been shown to strongly influence the atmospheric circulation on the large spatial scale \citep{ma2017}.

\subsection{Methods}
The atmospheric circulation anomalies associated with the GSF shifts have been analysed using composites of “North” minus “South” phase of the GSF. For assessing the statistical significance of the differences between the two GSF phases, a two-sided Student's t-test against the null hypothesis of no-difference has been applied at the 90\% significance level. Specifically, the impact of the GSF meridional displacement on the atmospheric circulation has been characterised through the analysis of winter-mean near-surface winds, SLP, SHF (i.e. the sum of turbulent sensible and latent heat fluxes), zonal winds at 850\,hPa (representing the eddy-driven jet, U850) and meridional temperature gradient (related to baroclinicity) at 925\,hPa and, starting from daily data, [2--6]-day high-pass meridional eddy heat flux (MEHF)\nomenclature{MEHF}{meridional eddy heat flux} at 850\,hPa, variance of [2--6]-day high-pass eddy meridional wind at 250\,hPa and blocking frequency. The eddy-driven jet response has been analysed both in terms of the associated anomalous field and jet-latitude variability. As in \citet{woollings2010}, the daily zonal winds at 850\,hPa have been zonally averaged in the 0$^{\circ}$--60$^{\circ}$W longitudinal range, masking out Greenland and the Atlas mountains that intersect the 850\,hPa isobaric surface. Then a 10-day low-pass Lanczos filter with a window of 61 days \citep{duchon1979} has been applied to the resulting fields. The jet stream latitude has been defined as the latitude of maximum westerly wind speed in the 15$^{\circ}$--75$^{\circ}$N latitudinal range. To detect atmospheric blocking, the 2D large-scale blocking index defined in \citet{davini2012} has been adopted including the condition to avoid the detection of false blocking events at low-latitudes \citep{athanasiadis2014}. In addition to the above mentioned diagnostics, the monthly zonal-mean circulation and the atmospheric heat budget in the vertical--meridional cross-section in the 50$^{\circ}$--68$^{\circ}$W longitude range have been computed. 

\subsection{SST front-following coordinate system}
\label{coordinate_system}
For analyses in the vertical--meridional cross-section an SST front-following coordinate system has been devised. In order to define the SST-front coordinate system, the position of the GSF computed on the original HadISST2 0.25$^{\circ}$ grid has been assigned to the nearest grid point of the native grid of each AGCM. Since the oceanic grid is finer than those of AGCMs, the  GSF position in each model is slightly different from the original one. However, this choice has no significant effect on atmospheric diagnostics calculated along the vertical--meridional plane because the AGCMs would not have been affected by subgrid SST front features anyway. Starting from the interpolated GSF position, a 3D space has been considered with the abscissa representing the displacement along the GSF, the second horizontal axis representing the meridional direction and the vertical axis representing pressure levels. Then, the vertical--meridional plane has been derived averaging the diagnostics in the along-front direction. The zonal average has been taken avoiding grid points over land north of the GSF. The SST front-following coordinate system has allowed for taking into account the poleward tilt of the SST front and carrying out the analysis purely along the cross-front direction. Such a system represents a novelty in the scientific literature dealing with the oceanic fronts variability and their impact on the atmospheric circulation.

\section{Atmospheric response to Gulf Stream SST front shift}
\label{response}

\begin{figure}[ht!]
    \centering
    \includegraphics[width=28pc]{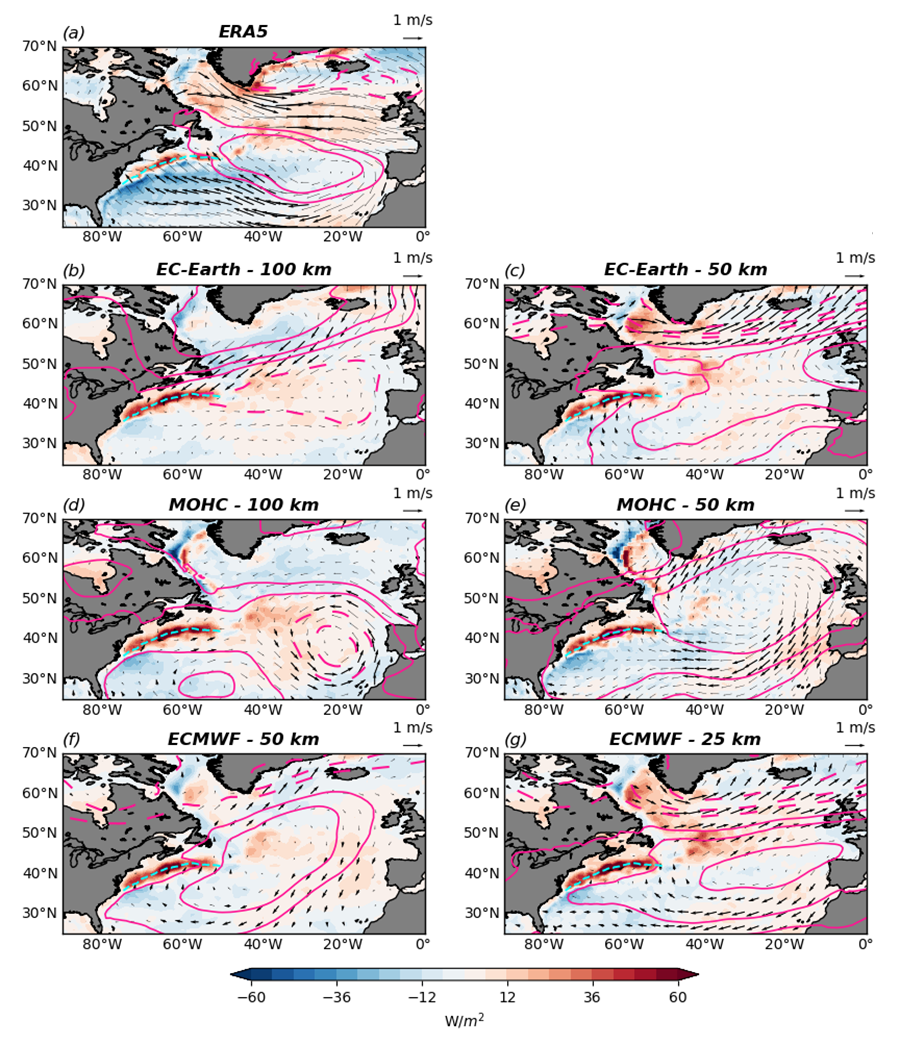}
\caption{SHF (W m$^{-2}$; color shading), SLP (Pa; contours, solid for positive) and near-surface wind (m s$^{-1}$; vectors) response to the GSF shifts in winter (DJF). (a) ERA5. (b,\,c) EC-Earth. (d,\,e) MOHC. (f,\,g) ECMWF. Beside institution name, the model nominal resolution in km is reported. SHF are considered to be positive upwards, namely from the ocean to the atmosphere. For models the magenta contours represent SLP anomalies at -120, -60, -30, 30, 60 and 120\,Pa; for ERA5 the magenta contours represent SLP anomalies at -180, -150, 150 and 180\,Pa. Thick vectors indicate wind anomalies that were found to be significant at the 90\% confidence level (details in section \ref{data}). The winter climatological position of the GSF is indicated by the cyan dashed line. The arrow in the upper-right corner of each panel is the unit vector for surface wind (m s$^{-1}$).}
\label{fig:shf}
\end{figure}

Figure \ref{fig:shf} shows near-surface wind, SLP and SHF composite differences on the GSF meridional displacement in the models and the observations. R100 models exhibit a low-pressure anomaly downstream of the GSF that is largely consistent with meridional temperature advection in the area along the GSF tending to balance the anomalous diabatic heating. The general features of the atmospheric response in R100 models recall what one would expect at the surface for an extratropical shallow heat source in theoretical linear models \citep{hoskins1981}. On the other hand, R50+ models show a high-pressure anomaly downstream of SST anomalies, with winds blowing from the southern portion of North Atlantic poleward in the vicinity of the SST front. This is in contrast to what is expected from the linear-theory. The R50+ response resembles results from other studies,  showing  that the atmospheric response is strongly mediated by transient eddy fluxes balancing the anomalous diabatic heating \citep{peng1997, peng1999, watanabe2000}. It is specified that in R50+ models the SLP anomalies are statistically significant over a large part of the North Atlantic at the 90\% confidence level. This is not the case for R100 models, for which the statistically significant anomalies over the same region are more limited.

SHF anomalies are particularly intense close to the GSF climatological position both in R100 and R50+ models, reaching values that correspond to an important portion of winter climatology (about 15--20\%). These largely coincide with the area of strongest along-front SST anomalies. For R100 models meridional temperature advection tends to balance this anomalous diabatic heating associated with the GSF shifts (Figure \ref{fig:budget_R100}d). Indeed, the anomalous atmospheric circulation transports cold air from higher latitudes toward the GSF area. In contrast, in R50+ models meridional temperature advection does not tend to balance the anomalous diabatic heating but to exacerbate the induced temperature tendencies (Figure \ref{fig:budget_R50+}d). This is consistent with the northward advection of warm air from lower latitudes. This fact indicates that other processes, different from  horizontal temperature advection, have a key role in balancing heating anomalies above the GSF. This aspect will be discussed in more detail in section \ref{heatbudget}. Unlike in the frontal area, SHF anomalies differ in other portions of the ocean basin. R100 models show negative SHF anomalies south of Greenland and on the southern flank of the GSF and positive ones in the central North Atlantic. In contrast, R50+ models develop positive flux anomalies south of Greenland and on the eastern North Atlantic, with negative values mainly confined at the southern flank of the GSF. For the ECMWF model with a nominal resolution of 50\,km (Figure \ref{fig:shf}f) and the Met Office Hadley Centre (MOHC) R50+ model (Figure \ref{fig:shf}e) the negative SHF anomalies extend more northward compared to the other two R50+ models, reaching the southern Greenland coast. These discrepancies are consistent with near-surface wind differences among R50+ models. In Figure \ref{fig:shf}c and Figure \ref{fig:shf}g, positive SHF anomalies close to the Greenland coast are associated with surface wind blowing from inland of North America and then transporting cold and dry continental air that is warmed by the ocean. In Figure \ref{fig:shf}e and Figure \ref{fig:shf}f, negative SHF anomalies close to Greenland are associated with near-surface wind blowing from the southern North Atlantic and transporting warm and wet air that reduces the thermal air--sea contrast. Apart from these discrepancies, R50+ models are more comparable to each other than R100 ones. Furthermore, R50+ models reproduce general features of surface atmospheric anomalies found in the ERA5 dataset. As depicted in Figure \ref{fig:shf}a, observations show a zonally elongated anticyclonic circulation anomaly in surface winds, consistent with the positive SLP anomaly downstream the heating source (statistically significant at 90\% confidence level). SHF anomalies are positive far north of the GSF and negative far south as the respective wind anomalies suggest in an area of climatological westerly surface flow. Consequently, the SHF anomalies are spatially anti-correlated with the SST anomalies as expected for atmospheric forcing to the ocean. This implies that the large-scale SST anomalies seen in Figure \ref{fig:sst} away from the GSF are the fingerprint of local atmospheric forcing that occurred in the real system during and prior to the GSF shifts.

The SHF, SLP and SST patterns in observations recall positive NAO-like forcing on the ocean. This is in line with previous studies showing that the GS meridional shifts are correlated with the low frequency NAO variability at a positive lag of about 1--2 years with the ocean following the atmospheric forcing \citep{taylor1998b, frankignoul2001, sanchez2016}. However, the comparison between AGCMs and observations suggests that, as soon as the GSF shift is established, the ocean provides a positive feedback on the atmospheric circulation, and that the realistic representation of this feedback in AGCMs requires a sufficiently high horizontal resolution (as R100 models fail  to reproduce it).

\begin{figure}[ht!]
    \centering
    \includegraphics[width=28pc]{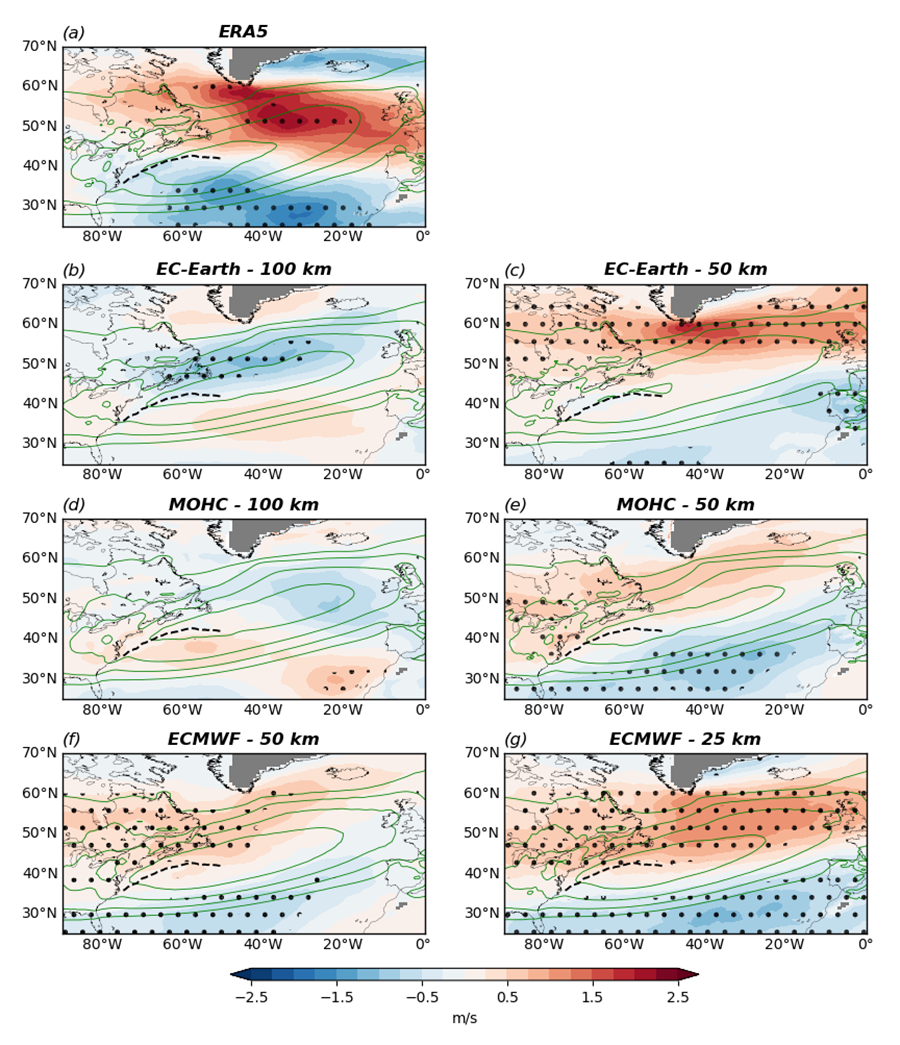}
\caption{850\,hPa zonal wind (m s$^{-1}$; color shaded) response to the GSF shifts in winter (DJF). (a) ERA5. (b,\,c) EC-Earth. (d,\,e) MOHC. (f,\,g) ECMWF. Beside institution name, the model nominal resolution in km is reported. Black dots denote anomalies that were found to be statistically significant at the 90\% confidence level (details in section \ref{data}). Green contours indicate the winter climatology of zonal wind at 850\,hPa every 2 m s$^{-1}$ from 4 m s$^{-1}$. The winter climatological position of the GSF is indicated by the black dashed line. The grey masking corresponds to areas where the orography intersects the climatological 850\,hPa isobaric surface.}
\label{fig:u850}
\end{figure}

In Figure \ref{fig:u850} zonal wind anomalies at 850\,hPa are presented. R100 models generate a southward shift of the eddy-driven jet, with negative anomalies to the north of the climatological jet position and positive anomalies to the south, even though significant differences can also be seen between the two models. In contrast, R50+ models exhibit anomalies of the opposite sign, indicating a northward shift of the jet. These anomalies recall the pattern seen in the observations (ERA5), yet with a lower amplitude. Then, for both the R100 and the R50+ models, similar but stronger zonal wind anomalies were found aloft (not shown) indicating an equivalent barotropic structure extending throughout the troposphere.

\begin{figure}[ht!]
    \centering
    \includegraphics[width=28pc]{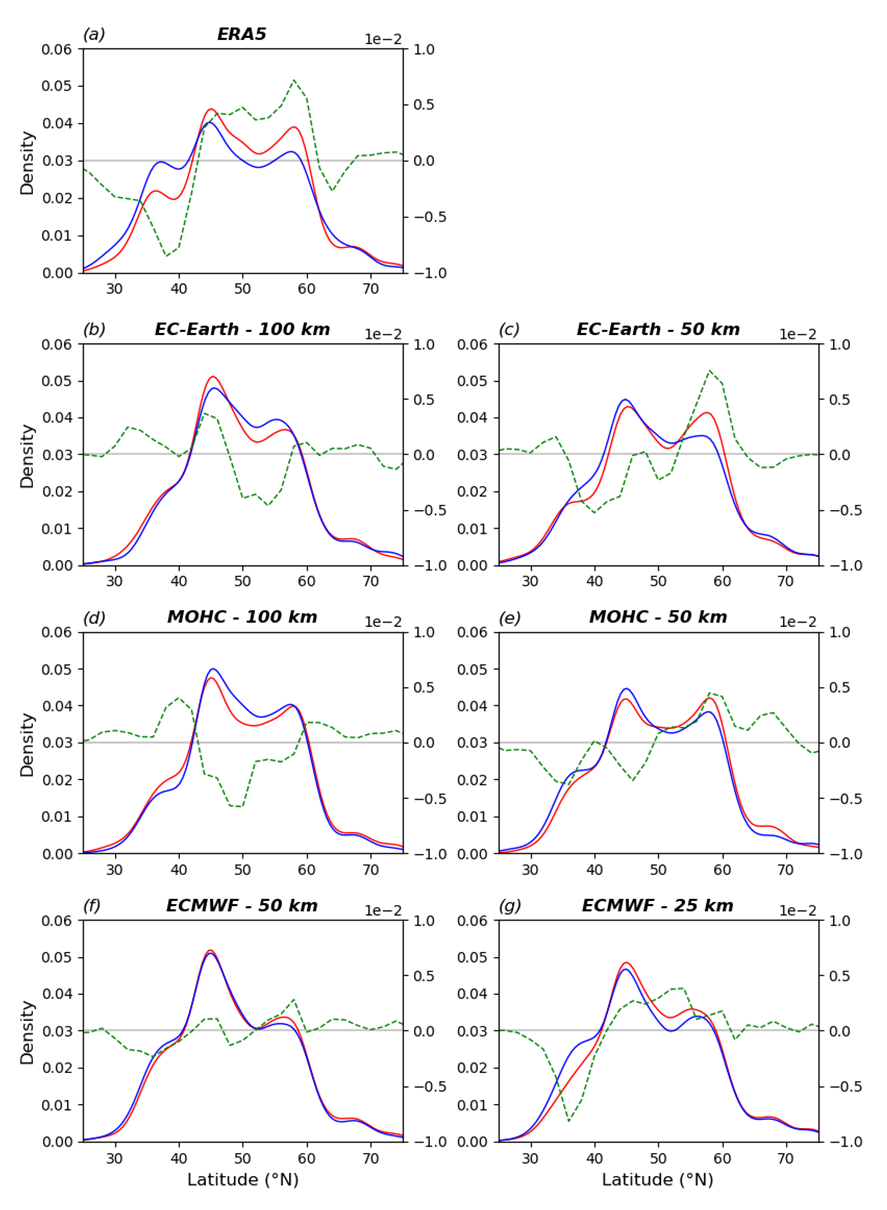}
\caption{Jet latitude distributions during winter (DJF). As specified in section \ref{data} and following the method in \citet{woollings2010}, the jet latitude has been defined through the zonal winds at 850\,hPa, zonally averaged in the 0$^{\circ}$--60$^{\circ}$W longitudinal range and maximized in the 15$^{\circ}$--75$^{\circ}$N latitudinal range. (a) ERA5. (b,\,c) EC-Earth. (d,\,e) MOHC. (f,\,g) ECMWF. Beside institution name, the model nominal resolution in km is reported. Red (blue) lines represent the distribution during the “North” (“South”) phase of the GSF position in the respective model. Green lines represent the difference between the two distributions (“North” minus “South”) and correspond to the right y-axis.}
\label{fig:jet_stream}
\end{figure}

To assess in more detail the response of the North Atlantic eddy-driven jet to the GSF shifts, the jet latitude distributions (probability density functions, PDFs)\nomenclature{PDF}{probability density function} for each phase of the GSF position are shown for each model in Figure \ref{fig:jet_stream}. As specified in section \ref{data}, it is recalled that the jet stream latitude has been defined through the zonal winds at 850\,hPa, zonally averaged in the 0$^{\circ}$--60$^{\circ}$W longitudinal range and maximized in the 15$^{\circ}$--75$^{\circ}$N latitudinal range. Even though there are some differences between the two R100 models and between the four R50+ models, the common characteristics in each group are consistent with the results shown for the zonal wind anomalies at 850\,hPa, indicating a northward shift in R50+ models and a southward shift in R100 models. Again, only the anomalies in the jet latitude distribution of the R50+ models resemble the respective observed (ERA5) anomalies, being consistently weaker than the latter. Indeed, in ERA5 the GSF displacements are associated with anomalies of opposite sign in the jet latitude distribution north and south of the GSF position (approximately at 42$^{\circ}$N) indicating shifts in the eddy-driven jet that are homo-directional to the GSF shifts. It should be noted that none of the six models can reproduce the trimodal character of the observed jet latitude distribution in wintertime, originally shown by \citet{woollings2010} and seen here for ERA5. Particularly, the southern-jet regime is too weakly represented by the models, a problem that is arguably linked to the absence of coupled feedbacks in atmosphere-only simulations. Models unable to represent the observed circulation regimes \citep{madonna2017} are also expected to feature a “distorted” response to the GSF variability as it has been noted that the atmosphere tends to respond to oceanic and other moderate forcings by changes in the frequency of occurrence of its dominant circulation regimes \citep{palmer1993}.

\begin{figure}[ht!]
    \centering
    \includegraphics[width=28pc]{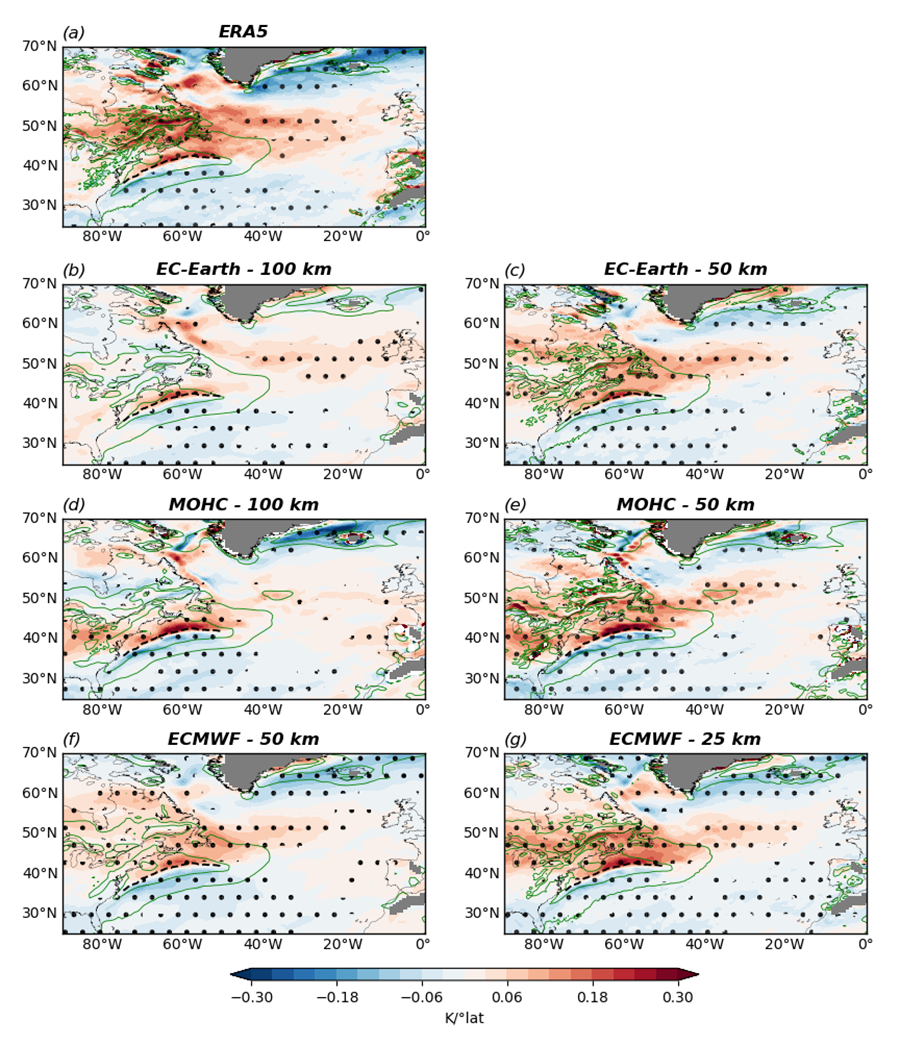}
\caption{925\,hPa meridional air temperature gradient (K $^{\circ}$lat$^{-1}$; color shaded) response to the GSF shifts in winter (DJF). The temperature gradient is reversed, i.e. equatorward. (a) ERA5. (b,\,c) EC-Earth. (d,\,e) MOHC. (f,\,g) ECMWF. Beside institution name, the model nominal resolution in km is reported. Black dots denote anomalies that were found to be statistically significant at the 90\% confidence level (details in section \ref{data}). Green contours indicate the winter climatology every 0.8\,K $^{\circ}$lat$^{-1}$ from 0.8\,K $^{\circ}$lat$^{-1}$. The winter climatological position of the GSF is indicated by the black dashed line. The grey masking corresponds to areas where the orography intersects the climatological 925\,hPa isobaric surface.}
\label{fig:t925_grad}
\end{figure}

\begin{figure}[ht!]
    \centering
    \includegraphics[width=28pc]{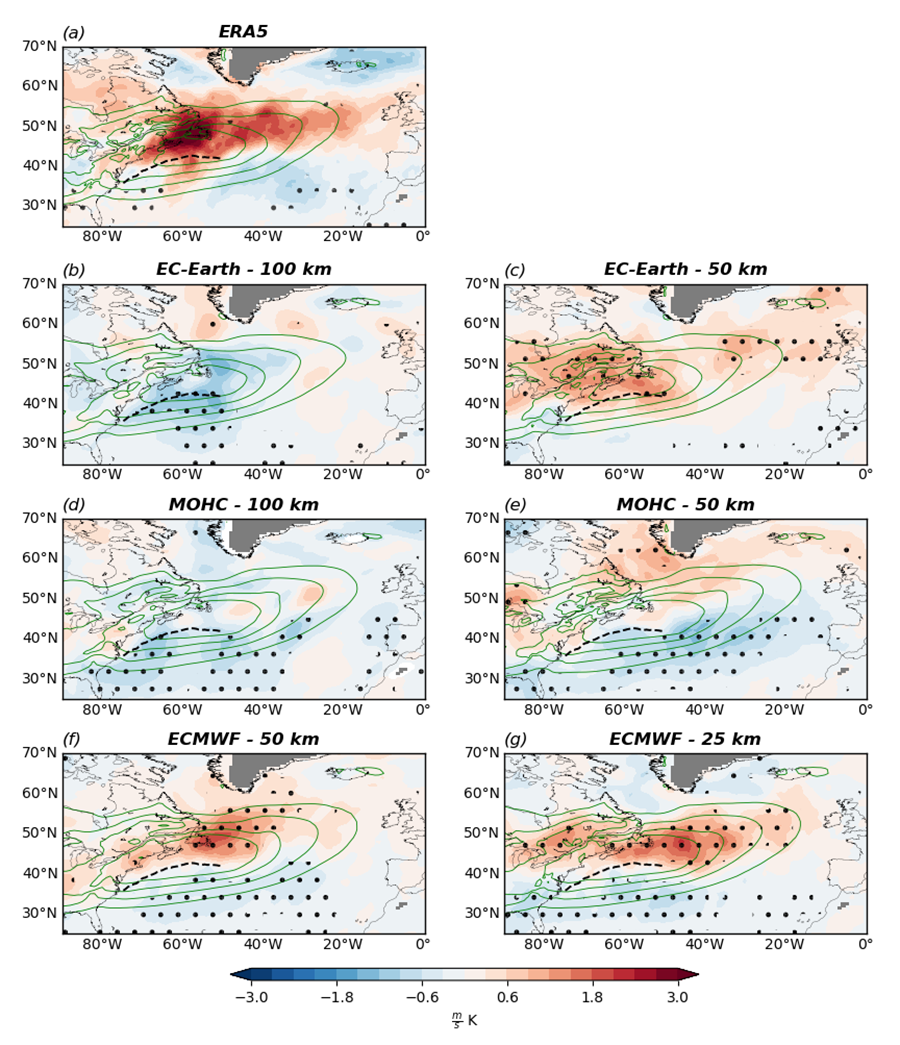}
\caption{MEHF anomalies (v$^\prime$T$^\prime$; m s$^{-1}$ K; color shaded) at 850\,hPa induced by the GSF shifts in winter (DJF). (a) ERA5. (b,\,c) EC-Earth. (d,\,e) MOHC. (f,\,g) ECMWF. Beside institution name, the model nominal resolution in km is reported. Black dots denote anomalies that were found to be statistically significant at the 90\% confidence level (details in section \ref{data}). Green contours indicate the winter climatology of eddy heat flux at 850\,hPa every 3 m s$^{-1}$\,K from 6 m s$^{-1}$\,K. The winter climatological position of the GSF is indicated by the black dashed line. The grey masking corresponds to areas where the orography intersects the climatological 850\,hPa isobaric surface.}
\label{fig:vt850}
\end{figure}

The jet latitude variability has been previously interpreted as a response to variations in the stormtrack activity, which influences the jet stream through an upstream baroclinic and a downstream barotropic effect \citep{novak2015, oreilly2017}. The former is based on the non-linear oscillator relationship between low-level baroclinicity and MEHF \citep{ambaum2014}; the latter is the downstream representation of the upstream low-level baroclinicity changes, which induce variations in eddy anisotropy and wave breaking, with direct impact on jet latitude \citep{orlanski2003}. To assess whether the previously discussed homo-directional jet stream response to the GSF shifts is the result of changes in low-level baroclinicity and then in stormtrack activity, in Figure \ref{fig:t925_grad} and Figure \ref{fig:vt850} we show the meridional temperature gradient (here used as a proxy for baroclinicity) and MEHF anomalies at 925\,hPa and 850\,hPa respectively. As expected, the strong positive SHF anomaly (corresponding to the “North” minus “South” composites difference) along the GSF induces local changes in low-level baroclinicity both in R100 and R50+ models. Specifically, baroclinicity is found to increase at the northern flank of the GSF and to decrease at the southern flank. In the immediate vicinity of the GSF, these changes are understood as forced by the displacement of the SST front itself and are consistent with the meridional gradient of the above-mentioned SHF anomaly. Beyond these local changes, in R50+ models low-level baroclinicity undergoes significant changes also in a meridionally broader zone to the north of the GSF. As discussed later, these changes may be explained considering the meridional gradient in zonal temperature advection (dipole of opposite temperature tendencies) associated with the meridional shift in the low-level jet. In fact, stronger westerlies to north and weaker westerlies to the south, in an area where westerlies in winter are associated with cold temperature advection from the North American continent over the relatively much warmer ocean, tend to increase atmospheric baroclinicity in between. Similar large-scale anomalies in atmospheric baroclinicity have been detected also by \citet{nakamura2009}, who have argued that SST anomalies associated with the GS variability, despite being limited to the GS area, may be responsible for such broader-scale anomalies in atmospheric baroclinicity.

Given the changes in low-level baroclinicity in the broader area to the north of the GSF (around and to the east of Newfoundland, between 45$^{\circ}$--55$^{\circ}$N), R50+ models are found to respond with increased MEHF there as expected for baroclinic adjustment \citep{stone1978}. This can be seen clearly in Figure \ref{fig:vt850}, while it is not the case for R100 models. This discrepancy relates to how R100 models respond (differently, as shown in the next section) to the anomalous diabatic heating along the GSF.

\begin{figure}[ht!]
    \centering
    \includegraphics[width=28pc]{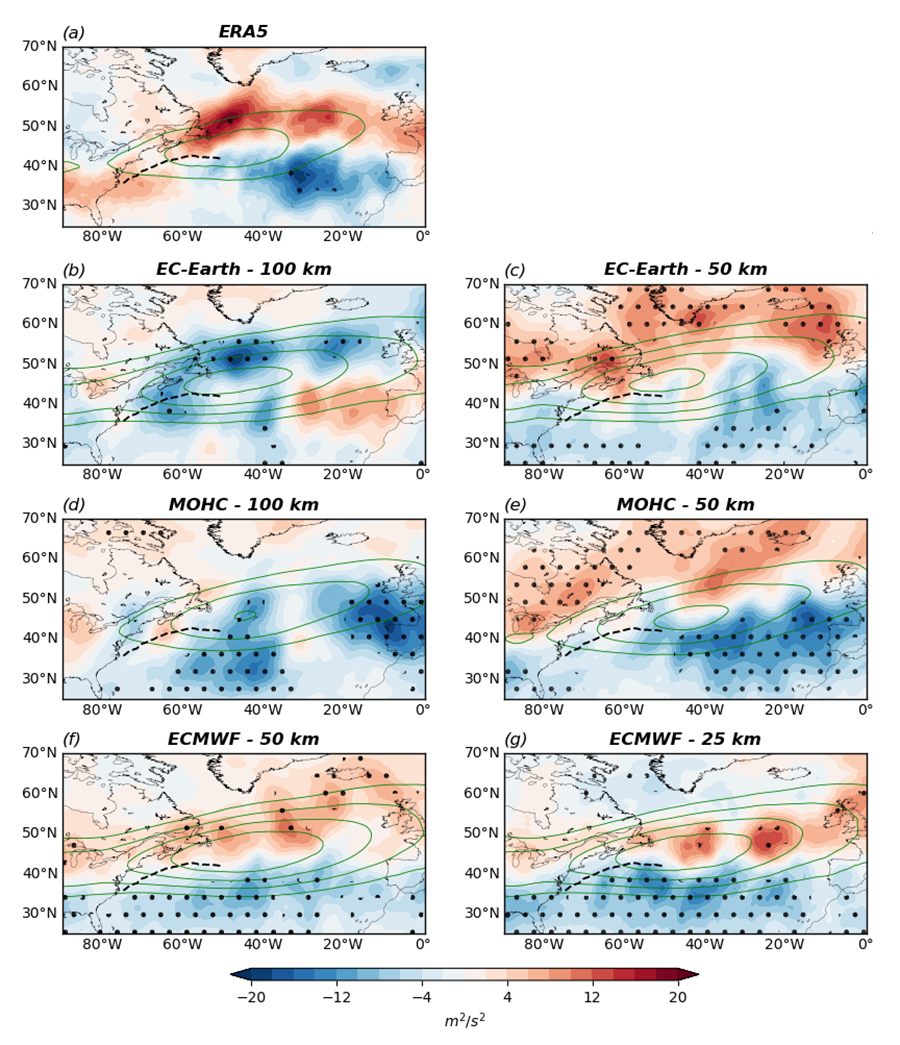}
\caption{Stormtrack anomalies (diagnosed via the variance of the meridional eddy velocity aloft, v$^\prime$v$^\prime$ at 250\,hPa; m$^{2}$ s$^{-2}$; color shaded) induced by the GSF shifts in winter (DJF). (a) ERA5. (b,\,c) EC-Earth. (d,\,e) MOHC. (f,\,g) ECMWF. Beside institution name, the model nominal resolution in km is reported. Black dots denote anomalies that were found to be statistically significant at the 90\% confidence level (details in section \ref{data}). Green contours indicate the winter climatology of stormtrack at 250\,hPa every 30 m$^{2}$ s$^{-2}$ from 120 m$^{2}$ s$^{-2}$. The winter climatological position of the GSF is indicated by the black dashed line.}
\label{fig:vv250}
\end{figure}

In turn, the increased MEHF found for R50+ models in the broader area to the north of the GSF is considered to be part of the causal chain leading to the shift of the eddy-driven jet as it indicates westward acceleration in the lower troposphere following the E-vector formulation of eddy--mean flow interaction \citep{hoskins1983}. In fact, the stormtrack (diagnosed via the variance of the meridional eddy velocity aloft, v$^\prime$v$^\prime$ at 250\,hPa; Figure \ref{fig:vv250}) is found to move with the jet as it exhibits a meridional dipole pattern of anomalies similar to the one seen for the zonal wind in R50+ models. However the horizontal E-vector (v$^\prime$\textsuperscript{2} - u$^\prime$\textsuperscript{2}; -u$^\prime$v$^\prime$) divergence at 250\,hPa (not shown) did not reveal a significant forcing on the zonal mean flow. 

The detected changes in low-level baroclinicity, MEHF and storm activity for R50+ models are found to be similar and more pronounced in the observations (ERA5, in Figure \ref{fig:t925_grad} and Figure \ref{fig:vt850}), as it was the case for the zonal wind anomalies (Figure \ref{fig:u850}). In this context it must be specified that, while in the atmosphere-only simulations the atmospheric anomalies are only a response to the oceanic forcing, the respective anomalies detected in the observations represent also the atmospheric forcing that causes the GSF shifts in the first place (at negative lags but also at lag-0, i.e. concurrent with and preceding the changes in the GSF position). The study by \citet{frankignoul2001} is quite insightful in this regard, showing that the GS axis moves to the north (south) following positive (negative) NAO-like forcing.

Finally, it is specified that the low-level baroclinicity anomalies have been computed taking into account also the changes in the static stability component, yet they show patterns quite comparable to what described above both for AGCMs and observations (not shown). This means that the low-level baroclinicity anomalies are mostly due to changes in the meridional temperature gradient rather than changes in static stability.

\section{Heat budget}
\label{heatbudget}

In the previous section it has been shown that R100 and R50+ simulations respond differently to the meridional displacements of the GSF, in terms of circulation and transient eddy activity. The resolution-dependent response of the atmospheric circulation to similar SHF anomalies close to the GSF raises the question of what might be different between R100 and R50+ models in terms of the primary, local response to the anomalous diabatic heating associated with the GSF shifts. Moreover, it was found that in R100 models cold meridional advection tends to balance the anomalous diabatic heating, whilst this was not found to be the case for R50+ models. Therefore, our study was naturally led to the analysis of the local thermodynamic balance.

Relevantly to this point, \citet{smirnov2015}, analyzing the response of low- and high-resolution atmosphere-only models to prescribed idealized SST forcing along an oceanic front, have found the local heat budget to depend on the horizontal resolution. Thus, to address the above-posed questions, a similar analysis was conducted for each of the examined models. Specifically, the terms of the time-mean thermodynamic equation have been computed as in \citet{smirnov2015}:

\begin{equation}
\label{heat_budget_equation}
\overline{\dot{Q}} - \underset{\lower.9em\hbox{I}}{\overline{u}\frac{\partial \overline{T}}{\partial x}} - \underset{\lower.9em\hbox{II}}{\frac{\partial \overline{u'T'}}{\partial x}} - 
\underset{\lower.7em\hbox{III}}{\overline{v}\frac{\partial \overline{T}}{\partial y}} - \underset{\lower.7em\hbox{IV}}{\frac{\partial \overline{v'T'}}{\partial y}} - \underset{\lower.7em\hbox{V}}{(\overline{\omega}\frac{\partial \overline{T}}{\partial p} - \frac{k}{p}\overline{\omega}\overline{T})} - \underset{\lower.7em\hbox{VI}}{(\overline{\omega' \frac{\partial T'}{\partial p}} - \frac{k}{p}\overline{\omega'T'})} = 0
\end{equation}

Here u and v represent the zonal and meridional wind component respectively; $\omega$ is the pressure tendency, proportional to the vertical wind; p and T represent the pressure and the temperature, while \.{Q} is the diabatic heating rate. It is specified that \.{Q} is not provided as model output and then it has been calculated as a residual from the other heat budget terms. The overbars indicate climatological monthly means during winter, while the primes indicate  departures thereof, k is R/C\textsubscript{p}, with R equal to 287 J kg\textsuperscript{-1} K\textsuperscript{-1} and C\textsubscript{p} equal to 1004 J kg\textsuperscript{-1} K\textsuperscript{-1}. Terms I, III and V represent the zonal, meridional and vertical mean thermal advection, respectively. Terms II, IV and VI represent the respective zonal, meridional and vertical eddy heat flux convergences. Terms V and VI also include the mean and eddy component of the adiabatic heating rate, respectively. All terms have been calculated for each winter calendar month and both positions of the GSF (i.e., “North and “South”). Then, monthly differences between the “North” and “South” positions have been derived and averaged to form the respective winter means. 

\begin{figure}[ht!]
    \centering
    \includegraphics[width=36pc]{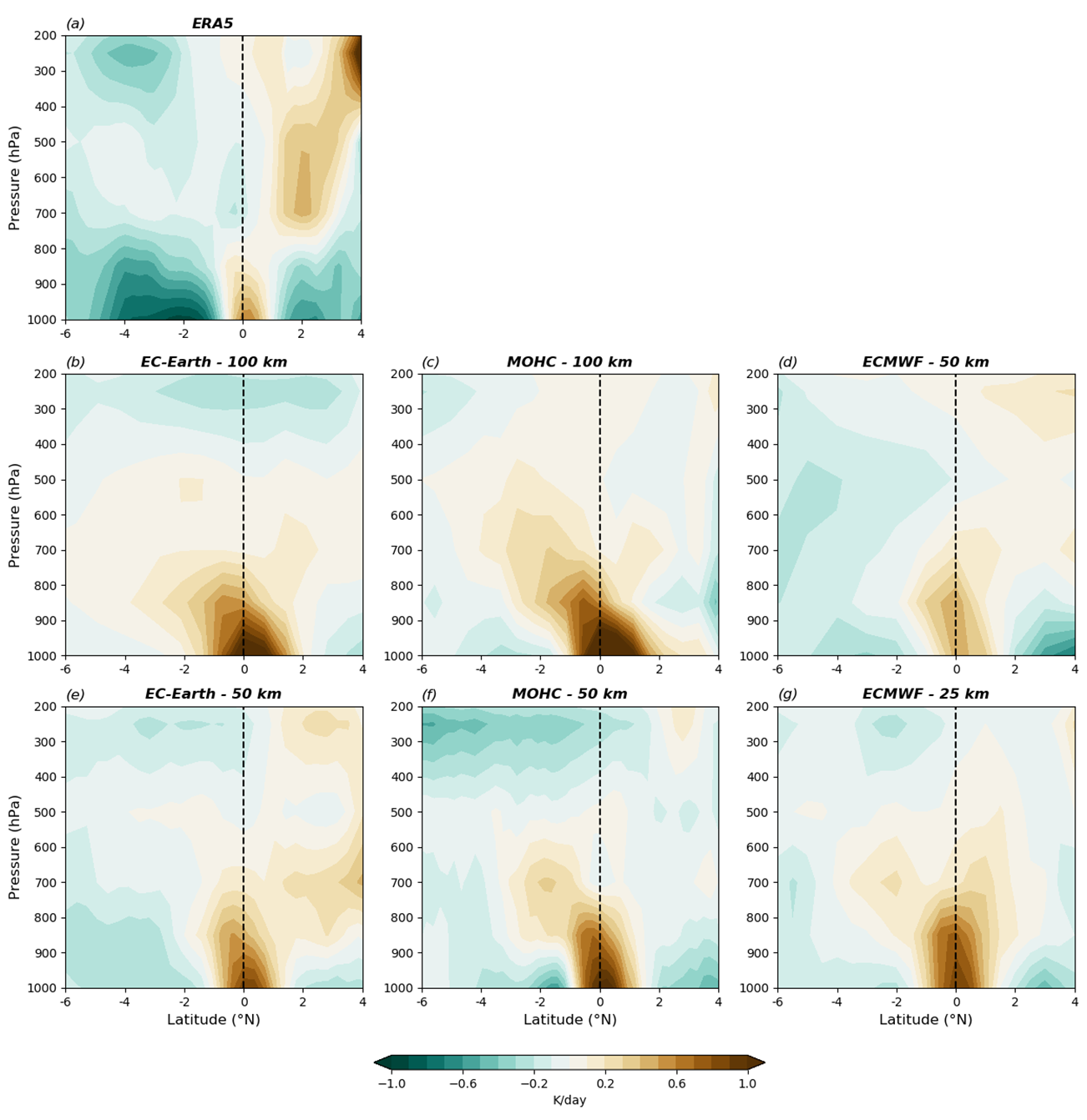}
\caption{Cross-front vertical section of diabatic heating term (K day$^{-1}$; color shading) in the thermodynamic budget (equation \ref{heat_budget_equation}), calculated as a residual. (a) ERA5. (b,\,e) EC-Earth. (c,\,f) MOHC. (d,\,g) ECMWF. Beside institution name, the model nominal resolution in km is reported. The vertical dashed line represents the GSF position in relation to which the cross-front section has been constructed.}
\label{fig:diabatic_heating}
\end{figure}

Figure \ref{fig:diabatic_heating} shows the cross-front vertical section of the diabatic heating term calculated in the thermodynamic budget (equation \ref{heat_budget_equation}). At the vicinity of the GSF both R100 and R50+ models show strong diabatic heating anomalies in the lower troposphere, below 600--700 hPa, with maximum values near the surface. This is expected considering the collocated positive SHF anomalies shown in Figure \ref{fig:shf}. It is noted, also, that R100 models exhibit significantly larger diabatic heating near the surface (Figure \ref{fig:diabatic_heating}b, Figure \ref{fig:diabatic_heating}c). Furthermore, it should be mentioned that the diabatic heating rate has been calculated as a residual from the thermodynamic equation and thus it includes other heating terms. In general, the diabatic heating in the boundary layer of a limited geographical area (as that straddling the GSF) depends on the underlying SHF (sensible and, if condensation occurs, also latent) distributed in the vertical by the boundary layer parameterization scheme, as well as to the locally absorbed radiation and the net heat fluxes entering through the lateral boundaries of the area. Considering these aspects, the diabatic heating rates shown in Figure \ref{fig:diabatic_heating} are not directly comparable with the SHF shown in Figure \ref{fig:shf}. Finally, it should be mentioned that the diabatic heating rate calculated as a residual is affected by any errors in computing all the terms in the thermodynamic budget. In ERA5 the diabatic heating in the lower troposphere is weaker than in the models (Figure \ref{fig:diabatic_heating}a), something that is understood considering that ERA5 data come from a coupled world/model. 

\begin{figure}[ht!]
    \centering
    \includegraphics[width=36pc]{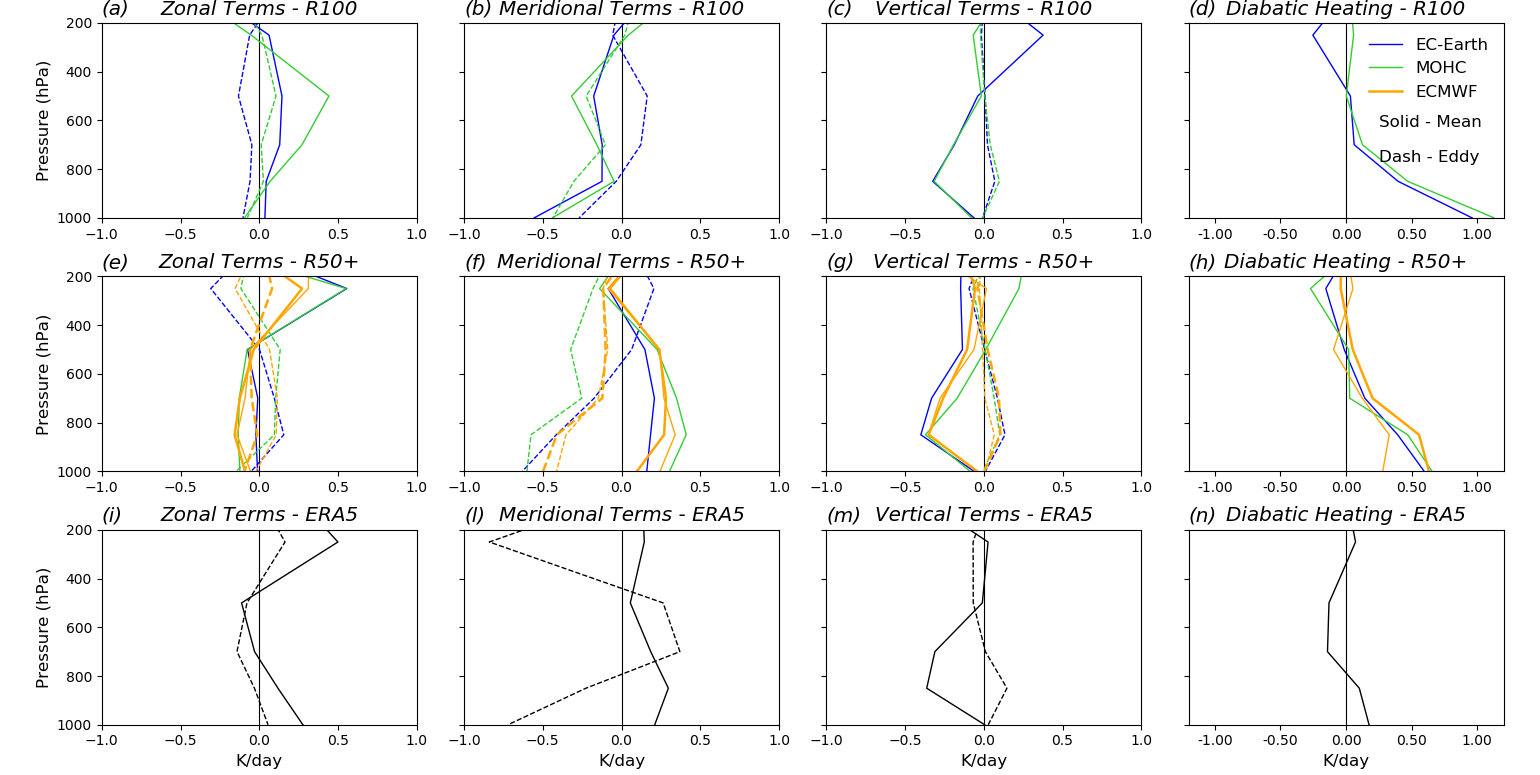}
\caption{Vertical profiles of composited differences for zonal (terms I and II), meridional (terms III and IV) and vertical (terms V and VI) terms in the thermodynamic budget (equation \ref{heat_budget_equation}) (K day$^{-1}$). Last column shows the vertical profile of the diabatic heating term (K day$^{-1}$), calculated as residual. (a)–(d) R100 models. (e)–(h) R50+ models. (i)–(n) ERA5. Each term has been calculated as a difference between the “North” and “South” phase of the GSF, averaged zonally along the front in the 50$^{\circ}$--68$^{\circ}$W longitudinal range and meridionally in the range -1$^{\circ}$ and +1$^{\circ}$ north of the SST front position (using the SST front-following coordinate system described in section \ref{data}). Blue lines refer to the EC-Earth model, green lines to the MOHC model, orange lines to the ECMWF model, and black lines refer to ERA5. Orange bold lines in (d)–(f) refer to the ECMWF model with a nominal resolution of 25\,km. Solid lines refer to terms involving heat transport by time-mean fields (terms I, III and V), whilst dashed lines refer to the respective eddy components of heat transport (terms II, IV and VI).}
\label{fig:budget}
\end{figure}

Figure \ref{fig:budget} shows the heat budget as indicated above, zonally averaged along the SST front and meridionally in the range -1$^{\circ}$ and +1$^{\circ}$ north of the GSF position (using the previously described SST front-following coordinates). This latitude range has been chosen so as to match the meridional span of the intense SST anomalies induced by the meridional shifts of the GSF (Figure \ref{fig:sst}) and the induced SHF anomalies (Figure \ref{fig:shf}) in its vicinity.

Near the surface the diabatic heating anomalies described above are largely balanced by horizontal heat terms, with a dominance of the meridional components (terms III and IV) over the zonal ones (terms I and II) in all cases. However, while in ERA5 and R50+ models the anomalous diabatic heating at and near the surface is balanced exclusively by the meridional eddy heat flux divergence (term IV) with the meridional mean advection (term III) opposing this action, in R100 models the meridional mean advection co-operates with the meridional eddy heat flux divergence in balancing diabatic heating (Figure \ref{fig:budget}b, Figure \ref{fig:budget}f, Figure \ref{fig:budget}l). The different role played near the surface by the meridional mean advection in R50+ and R100 models is consistent with the near-surface wind anomalies presented in Figure \ref{fig:shf}, while similar differences in mechanisms balancing the heating at the surface emerged also in \citet{smirnov2015}. Vertical terms (terms V and VI) are negligible at the surface as expected for vanishing vertical motions at the lower boundary (Figure \ref{fig:budget}c, Figure \ref{fig:budget}g, Figure \ref{fig:budget}m). Instead, considering that the vertical profile of diabatic heating exhibits a gradual reduction from the surface up to about 600\,hPa (Figure \ref{fig:diabatic_heating}, Figure \ref{fig:budget}), it is noted that for all models and the observations (ERA5) the anomalous diabatic heating in the lower troposphere is partly balanced by mean vertical motion. In the vicinity of the GSF, which is an area of intense cyclogenesis and low-level convergence associated with the passage of frontal systems \citep{parfitt2018}, the “mean vertical motion” should be understood as the aggregated effect of pulses in vertical motion. Zonal terms (terms I and II) seem to play a less important role, especially in the lower troposphere where the diabatic heating takes place (Figure \ref{fig:budget}a, Figure \ref{fig:budget}e, Figure \ref{fig:budget}i).

\begin{figure}[ht!]
    \centering
    \includegraphics[width=34pc]{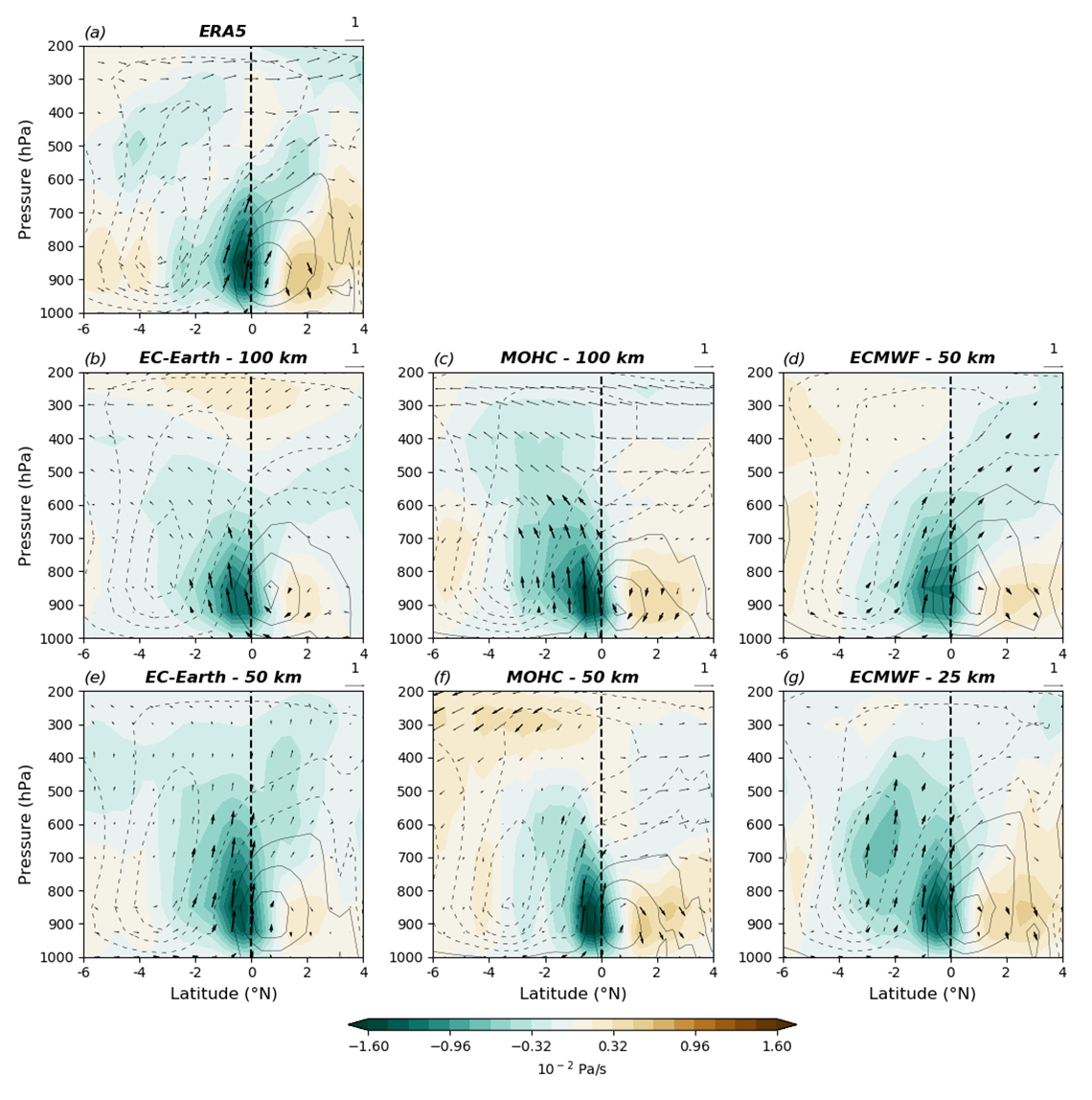}
\caption{Zonally-averaged omega composited differences (Lagrangian pressure tendency in 10$^{-2}$ Pa s$^{-1}$; color shading) and vertical--meridional winds (vectors) in response to the GSF shifts in wintertime (with the meridional wind component in m s$^{-1}$). Both the terms have been zonally averaged along the SST front in the 50$^{\circ}$--68$^{\circ}$W longitudinal range, using the SST front-following coordinate system described in section \ref{data}. A vector scale for vertical--meridional winds is shown in the top-right corner of each panel. (a) ERA5. (b,\,e) EC-Earth. (c,\,f) MOHC. (d,\,g) ECMWF. Beside institution name, the model nominal resolution in km is reported. Thick vectors indicate wind anomalies that were found to be significant at the 90\% confidence level (details in section \ref{data}). Negative (positive) omega values correspond to upward (downward) motion. Grey contours indicate the winter climatology of vertical motion in the vicinity of the GSF, contour interval: 10$^{-2}$ Pa s$^{-1}$ with dashed contours for negative. The vertical dashed line represents the GSF position in relation to which the cross-front section has been constructed.}
\label{fig:omega}
\end{figure}

In order to further examine the atmospheric response in the vicinity of the GSF, in Figure \ref{fig:omega} we present the atmospheric circulation anomalies (composite differences: “North” minus “South”) along the vertical--meridional cross-section in the 50$^{\circ}$--68$^{\circ}$W longitudinal range. All models show upward motion anomalies directly to the south of the GSF (warm sector) and downward motion anomalies to the north, thus indicating an anomalous cell-like circulation similar in character to the one seen for the time-mean circulation (contours). The upward motion anomalies are strongest in the vicinity of the GSF and in the lower troposphere (below 600\,hPa) reaching their maximum at about 850\,hPa. This is in agreement with the role the mean vertical motion (term V) was found to play for the local heat budget in all models and the observations. In contrast, R100 and R50+ models differ in the meridional component of the atmospheric circulation response in proximity to the GSF. R100 models exhibit equatorward motion, whilst R50+ models rather show a poleward anomaly. This finding is also in agreement with the different role that the meridional mean advection (term III) was found to play for the local heat budget in R100 and R50+ models.

\begin{figure}[ht!]
    \centering
    \includegraphics[width=28pc]{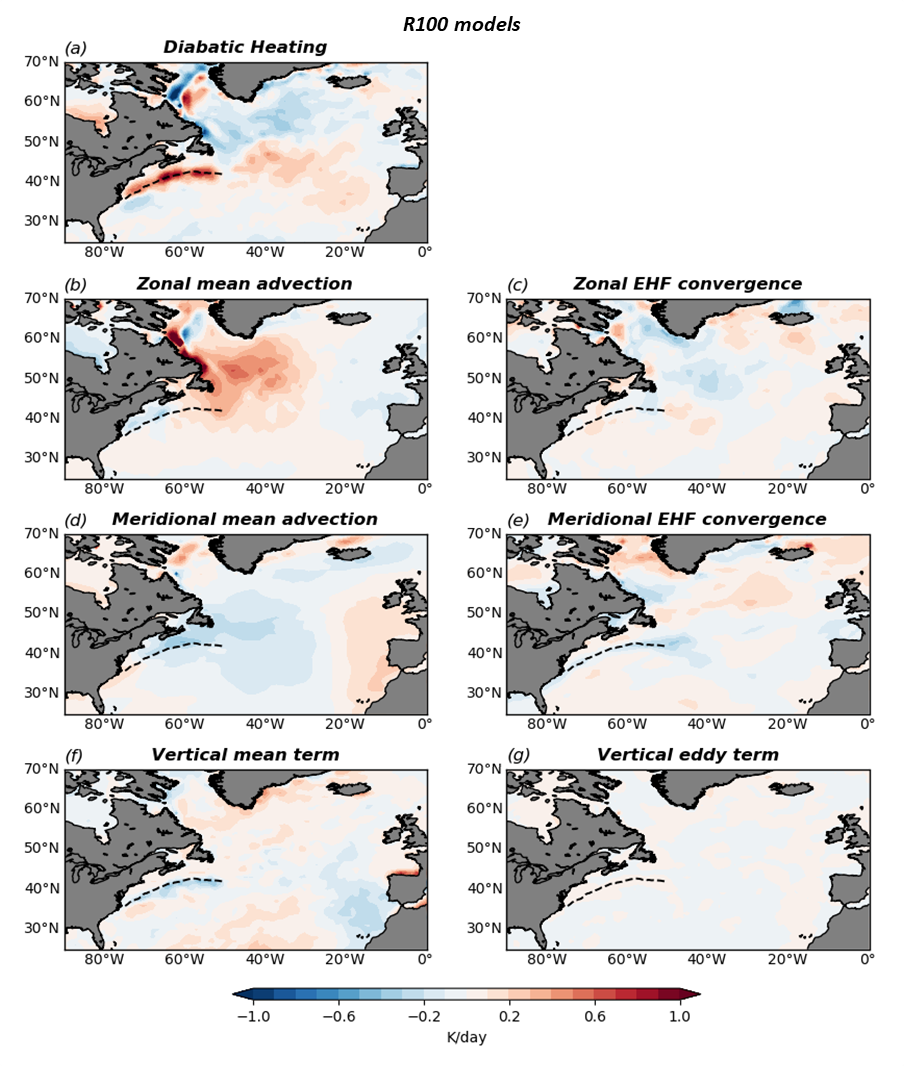}
\caption{Horizontal distribution of composited differences for zonal (terms I and II), meridional (terms III and IV) and vertical (terms V and VI) terms in the thermodynamic budget (equation \ref{heat_budget_equation}) (K day$^{-1}$), averaged in 700--1000\,hPa, as ensemble mean of models with a nominal resolution of 100\,km (R100 models). (a) Diabatic heating. (b) Zonal mean advection (term I). (c) Zonal eddy heat flux (EHF)\nomenclature{EHF}{eddy heat flux} convergence (term II). (d) Meridional mean advection (term III). (e) Meridional EHF convergence (term IV). (f) Vertical mean term (term V). (g) Vertical eddy term (term VI). The winter climatological position of the GSF is indicated by the black dashed line.}
\label{fig:budget_R100}
\end{figure}

\begin{figure}[ht!]
    \centering
    \includegraphics[width=28pc]{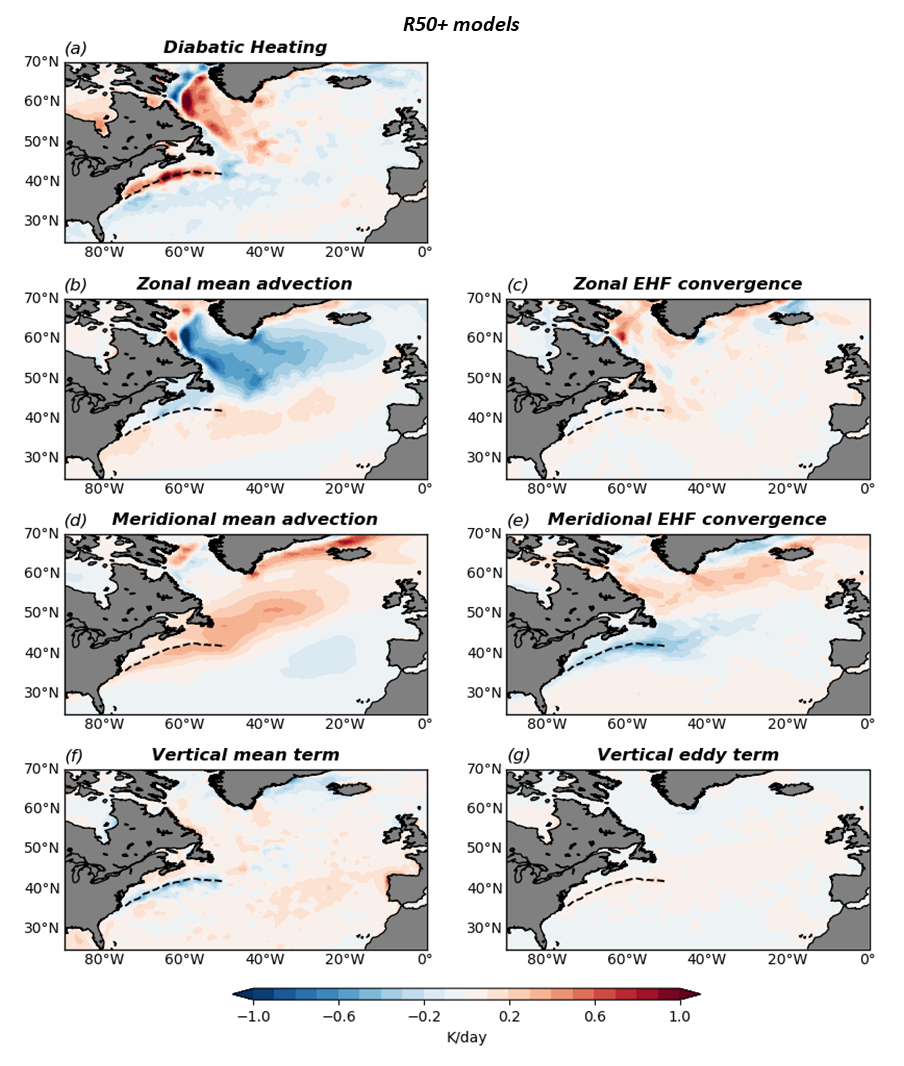}
\caption{Horizontal distribution of composited differences for zonal (terms I and II), meridional (terms III and IV) and vertical (terms V and VI) terms in the thermodynamic budget (equation \ref{heat_budget_equation}) (K day$^{-1}$), averaged in 700--1000\,hPa, as ensemble mean of models with a nominal resolution greater than 50\,km (R50+ models). (a) Diabatic heating. (b) Zonal mean advection (term I). (c) Zonal eddy heat flux (EHF) convergence (term II). (d) Meridional mean advection (term III). (e) Meridional EHF convergence (term IV). (f) Vertical mean term (Term V). (g) Vertical eddy term (term VI). The winter climatological position of the GSF is indicated by the black dashed line.}
\label{fig:budget_R50+}
\end{figure}

\begin{figure}[ht!]
    \centering
    \includegraphics[width=28pc]{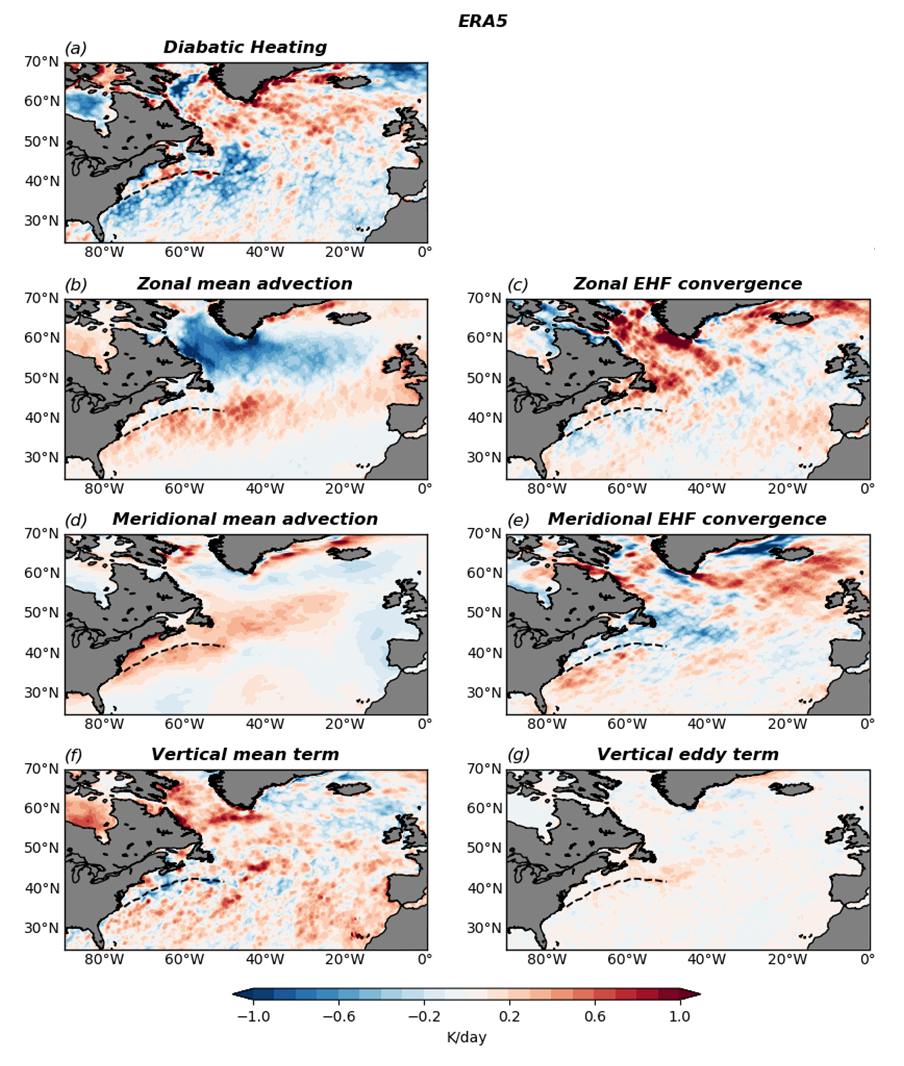}
\caption{Horizontal distribution of composited differences for zonal (terms I and II), meridional (terms III and IV) and vertical (terms V and VI) terms in the thermodynamic budget (equation \ref{heat_budget_equation}) (K day$^{-1}$), averaged in 700--1000\,hPa, for ERA5. (a) Diabatic heating. (b) Zonal mean advection (term I). (c) Zonal eddy heat flux (EHF) convergence (term II). (d) Meridional mean advection (term III). (e) Meridional EHF convergence (term IV). (f) Vertical mean term (term V). (g) Vertical eddy term (term VI). The winter climatological position of the GSF is indicated by the black dashed line.}
\label{fig:budget_ERA5}
\end{figure}

Based on the results of the local thermodynamic budget and atmospheric circulation, we have extended the heat budget to the entire North Atlantic region, in order to understand more the atmospheric response on the basin-scale. Here we show the results for the multi-model ensemble means of R100 and R50+ models in Figure \ref{fig:budget_R100} and Figure \ref{fig:budget_R50+} respectively. The results for each model confirm the general features of what is described in the following (not shown). Results for ERA5 are provided in Figure \ref{fig:budget_ERA5}. It is specified that each heat term in the thermodynamic budget extended to the entire North Atlantic region has been averaged in 700--1000 hPa considering that the diabatic heating anomalies are maximum in the lower troposphere \citep[Figure \ref{fig:diabatic_heating};][]{kallberg2005}. Figure \ref{fig:budget_R100}, Figure \ref{fig:budget_R50+} and Figure \ref{fig:budget_ERA5} show that the effect induced by meridional heat terms is not limited to areas above the GSF. Indeed, the cooling (warming) effect induced by meridional mean advection (term III) in the R100 models (R50+ models and ERA5) extends farther north and downstream the GS (Figure \ref{fig:budget_R100}d, Figure \ref{fig:budget_R50+}d and Figure \ref{fig:budget_ERA5}d), consistent with the surface circulation anomaly shown in Figure \ref{fig:shf}. In R100 models, such large-scale temperature tendencies are largely balanced by zonal mean advection (term I), i.e. reduced cold advection from the continent associated with reduced westerlies (Figure \ref{fig:budget_R100}b). On the other hand, in R50+ models and ERA5 the warming induced by meridional mean advection north and downstream the GSF is partly balanced by zonal advection of cold air coming from the inland of North America (term I; Figure \ref{fig:budget_R50+}b and Figure \ref{fig:budget_ERA5}b) and partly by meridional eddy heat flux divergence (term IV; Figure \ref{fig:budget_R50+}e and Figure \ref{fig:budget_ERA5}e). The former is consistent with the zonal wind intensification north of the GSF shown in Figure \ref{fig:u850}; the latter is consistent with the northward shifts of the stormtrack shown in Figure \ref{fig:vt850}. In the northern portion of the North Atlantic, the effect induced by zonal mean advection is particularly intense in both models and observations. However, in R100 models the zonal mean advection is largely counterbalanced by negative diabatic heating anomalies (Figure \ref{fig:budget_R100}a), whereas in R50+ models and observations it is balanced partly by positive diabatic heating anomalies (Figure \ref{fig:budget_R50+}a and Figure \ref{fig:budget_ERA5}a) and meridional eddy heat flux convergence (Figure \ref{fig:budget_R50+}e and Figure \ref{fig:budget_ERA5}e). Also Hotta and Nakamura (2011) have shown that the eddy heat flux convergence is an important heating source at high latitudes. For both models and observations, the diabatic heating is quite consistent with SHF shown in Figure \ref{fig:shf}. Finally, both in models and observations, it is still possible to see the cooling effect induced by vertical motion above the GSF, even if weaker because of the averaging on different vertical levels (Figure \ref{fig:budget_R100}f, Figure \ref{fig:budget_R50+}f and Figure \ref{fig:budget_ERA5}f).

Overall, the large-scale thermodynamic budget is consistent with the local heat budget. Its added value lies in highlighting which are the mechanisms maintaining the large-scale baroclinicity anomalies shown in Figure \ref{fig:t925_grad}, shedding light on the large-scale atmospheric circulation as discussed in more detail in section \ref{discussion}.

\section{Discussion}
\label{discussion}

The results discussed in the previous sections show that R100 and R50+ models exhibit a substantially different atmospheric response to the GSF shifts. The primary forcing for the atmosphere is the anomalous diabatic heating arising from the GSF shifts (effectively “replacing” cold waters with significantly warmer waters). The north--south displacement of the maximum SST gradient is an additional forcing directly affecting low-level baroclinicity, yet only locally.

The different large-scale responses featured by R50+ (realistic response) and R100 models (unrealistic response) are linked to how these models locally react to the primary oceanic forcing (SHF anomalies) to maintain their thermodynamic balance above the GSF. In R100 models, the local diabatic heating anomalies are partly balanced by meridional mean advection, while in R50+ models the meridional mean advection plays the opposite role, thus requiring a stronger sub-monthly meridional eddy heat flux divergence to maintain local thermodynamic balance.

Then, in R50+ models the local intensification of baroclinic eddy activity (which resembles the observed one) extends farther north and downstream of the GSF leading to a poleward shift in the stormtrack and the eddy-driven jet that is homo-directional to the GSF shifts, as in the observations (ERA5). Instead, R100 models fail to reproduce this large-scale response because, in the first place, they do not have the correct local circulation response to the diabatic heating anomalies.

Our results show that the mechanisms maintaining the local- and the large-scale baroclinicity are different. As previous studies show \citep[e.g.][]{nakamura2004, nakamura2010}, differential surface heating across oceanic fronts is key in maintaining low-level baroclinicity close to the GSF against the erosive effect of baroclinic eddy fluxes. Given that the GSF shifts are quite limited in meridional distance (approximately 50--100\,km as indicated by Figure \ref{fig:GSF} for the two tercile categories), the zone of maximum atmospheric baroclinicity is expected to shift with the GSF by a similar distance. Therefore, this effect alone cannot explain the detected large-scale baroclinicity anomalies. The detected large-scale changes in low-level baroclinicity in R50+ and in ERA5 can be explained by the differential zonal temperature advection induced by the detected changes in zonal wind near the east coast of the North American continent. Given that in that area (around and east of Newfoundland) in winter there is strong cold advection by the westerly flow, stronger westerlies to the north and weaker westerlies to the south (as in Figure \ref{fig:u850}) imply a differential temperature tendency (Figure \ref{fig:budget_R50+}, Figure \ref{fig:budget_ERA5}) that tends to increase baroclinicity in between (Figure \ref{fig:t925_grad}). We actually suggest that there is a positive feedback between the intensification of the MEHF and the large-scale zonal wind anomalies, triggered by the local atmospheric response in the vicinity of the GSF.

\begin{figure}[ht!]
    \centering
    \includegraphics[width=28pc]{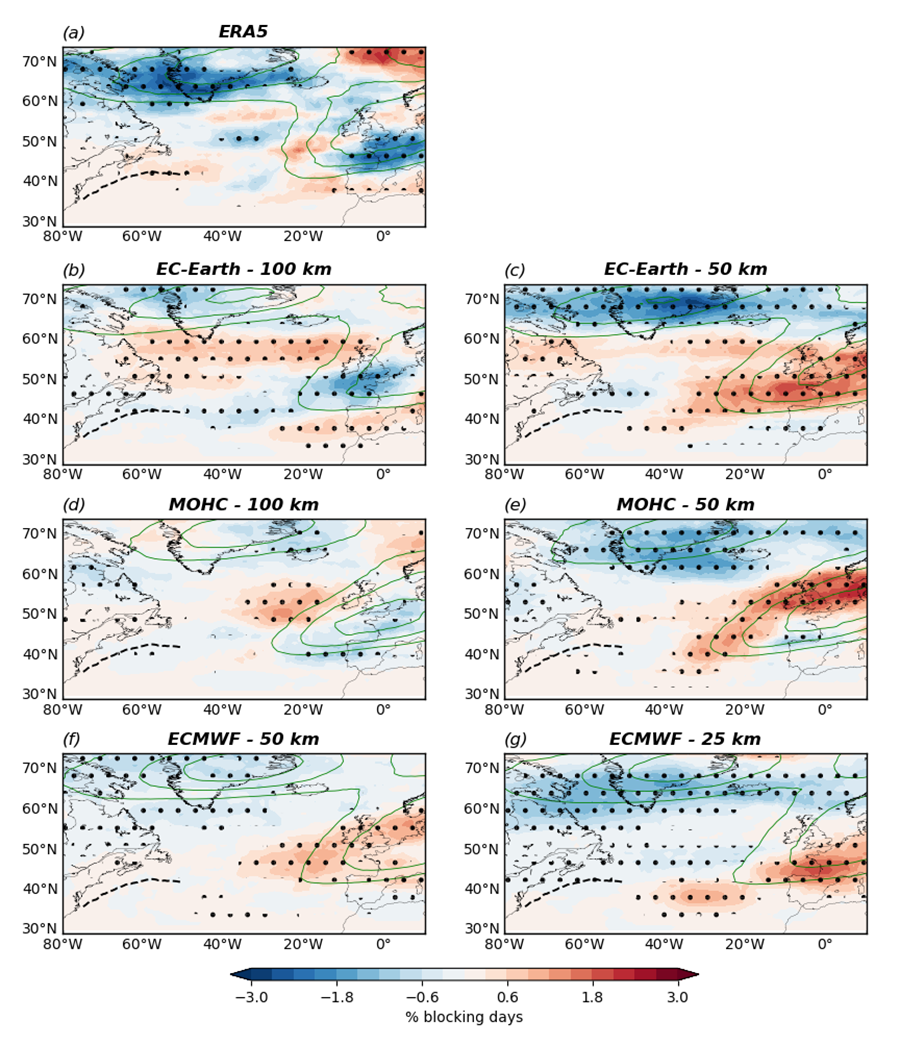}
\caption{Blocking frequency anomalies (\% of blocked days on total days; color shaded) induced by the GSF shifts in winter (DJF). (a) ERA5. (b,\,e) EC-Earth. (c,\,f) MOHC. (d,\,g) ECMWF. Beside institution name, the model nominal resolution in km is reported. Black dots denote anomalies that were found to be statistically significant at the 90\% confidence level (details in section \ref{data}). Green contours indicate the winter climatology of blocking frequency every 2\% from 4.5\%. The winter climatological position of the GSF is indicated by black dashed line.}
\label{fig:blocking}
\end{figure}

In this context, another point that requires discussion is what causes the detected shift in the stormtrack and the jet (referring to R50+ models and ERA5). The respective anomalies in MEHF shown in Figure \ref{fig:vt850} are, indeed, consistent with collocated changes in the low-level jet following eddy--mean flow interaction arguments \citep{hoskins1983, novak2015}. The former, however, cannot fully explain the pattern of the zonal wind anomalies (Figure \ref{fig:u850}), while the divergence of the horizontal E-vector components aloft (not shown) was not found to play the expected role aiding the understanding of the changes in the jet. In contrast, the analysis of atmospheric blocking was found to be insightful as the detected anomalies are dynamically consistent with the jet and stormtrack changes. Figure \ref{fig:blocking} shows blocking frequency anomalies associated with the GSF shifts (composite differences “North” minus “South”, as for the other diagnostics). The reduced blocking frequency over Greenland seen for R50+ models and ERA5 is in agreement with a northerly displaced jet and stormtrack, as Greenland blocking tends to displace the jet to the south leading to higher occurrences in the southern-jet regime \citep{woollings2008, woollings2010, madonna2017}. Our findings are very much in agreement with \citet{joyce2019} who showed that periods of northerly GS path are associated with a reduction in Greenland blocking frequency and increased excursions of the stormtrack to the northeast over the Labrador Sea (namely a poleward shift). Furthermore, our results are consistent with previous studies showing that increased horizontal resolution in climate models leads to reduced biases in blocking frequency \citep{matsueda2009,jung2012,anstey2013}. Thus, a better representation of Greenland blocking in R50+ models could explain why those models realistically represent the jet stream response to GSF shifts, in contrast to R100 models. However, as in the typical chicken-egg problem, further analysis are necessary to finally disentangle if they are the westerly wind anomalies (induced by eddy-mean flow interaction) to determine the anomalies in the blocking frequency or vice versa. Apart from the reduction in high-latitude blocking frequency, the observations show opposite anomalies over the European continent than models. Such anomalies are dynamically consistent with the enhanced zonal winds over the central-western European coast (Figure \ref{fig:u850}a) and the higher occurrence of both the central- and the northern-jet regimes (Figure \ref{fig:jet_stream}a) in ERA5. 
 
\begin{figure}[ht!]
    \centering
    \includegraphics[width=28pc]{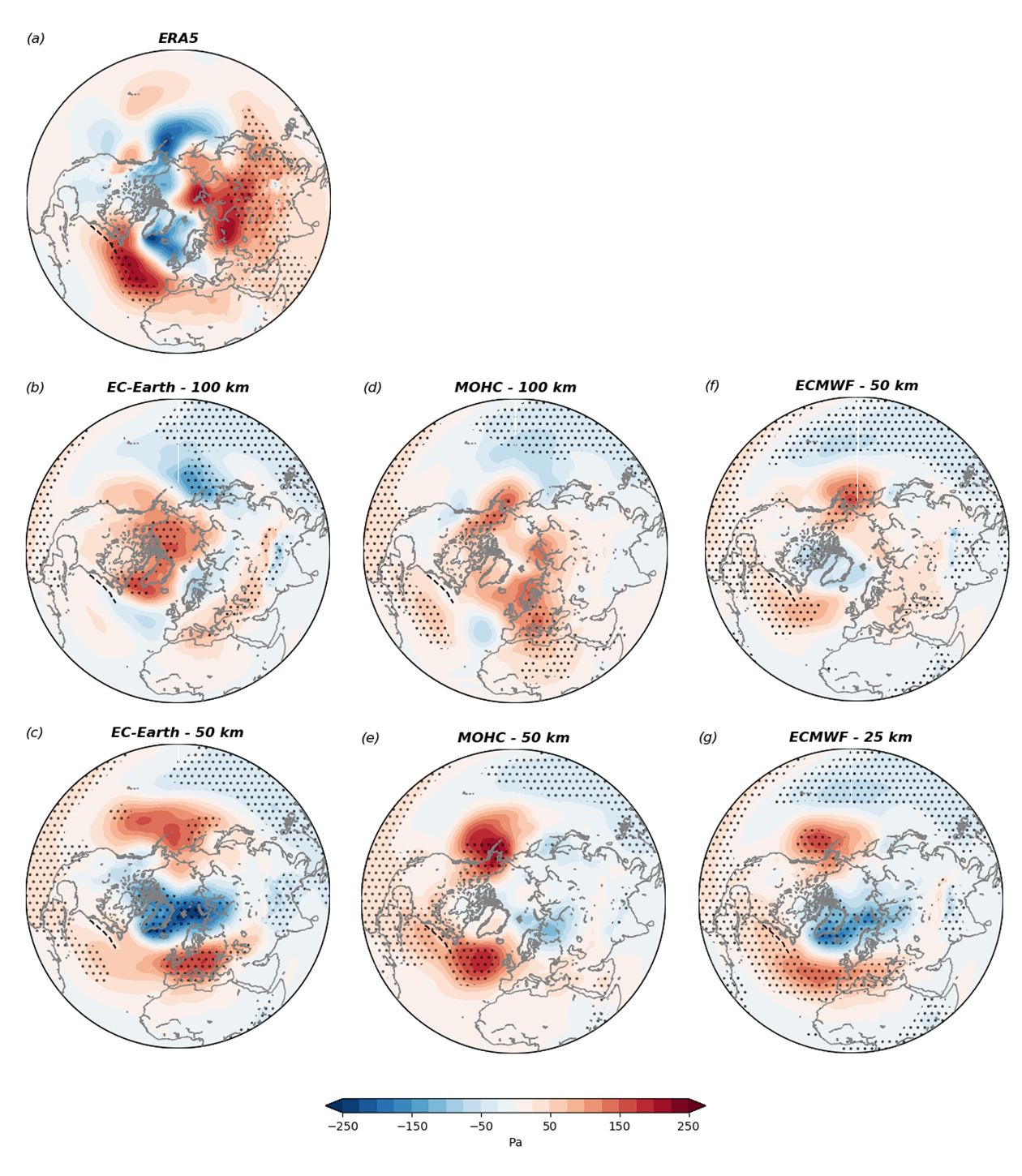}
\caption{Sea level pressure anomalies (Pa) induced by the GSF shifts in winter (DJF). (a) ERA5. (b,\,e) EC-Earth. (c,\,f) MOHC. (d,\,g) ECMWF. Beside institution name, the model nominal resolution in km is reported. Black dots denote anomalies that were found to be statistically significant at the 90\% confidence level (details in section \ref{data}). The winter climatological position of the GSF is indicated by the black dashed line.}
\label{fig:slp}
\end{figure}

Before closing this discussion it should be mentioned that although the methodology adopted in this study aimed at isolating the oceanic forcing related to the GSF meridional shifts, the atmospheric response to this variability may be contaminated by concurrent oceanic influences from other parts of the global ocean. To assess this possibility the authors first examined the global SST composites on the “North” and “South” tercile categories defined via the GSF position and found, in fact, anomalies recalling the typical patterns of the El Ni\~no--Southern Oscillation (ENSO)\nomenclature{ENSO}{El Ni\~no--Southern Oscillation}, with amplitudes reaching 0.4\,K in the eastern equatorial Pacific (not shown). Consistent with these SST anomalies, Figure \ref{fig:slp} shows negative (positive) SLP anomalies over the western (eastern) tropical Pacific in all model simulations. On the other hand, no significant SST and SLP anomalies are found over the tropical Pacific in the ERA5 dataset. To assess the possibility of a dominant ENSO influence, the zonal wind has been linearly regressed onto ENSO (NINO3.4 index) and then the composite differences for the zonal wind at 850\,hPa (Figure \ref{fig:u850}) have been repeated after having removed the linearly regressed zonal wind from the original data. For the models it was found that the general character and the amplitude of the detected atmospheric response in the North Atlantic is largely insensitive to this test (not shown). In ERA5 the results indicate a northward shift of the eddy-driven jet, with an amplitude and statistical significance greater than the one in the original analysis. Since no ENSO-like anomalies are present in the tropical Pacific in ERA5, such differences are understood as the effect of other phenomena concurrent with ENSO.

However, excluding completely all possible influences from remote parts of the global ocean is practically impossible as the historical atmosphere-only simulations are forced with global observed SST fields. There exist elaborate statistical methods for defining a GSF signal with minimal contamination by other specific processes (such as the ENSO variability), yet it is not possible to exclude every potential influence foreign to the GSF variability. As an example, Figure \ref{fig:slp} shows SLP anomalies associated with the GSF shifts that are statistically significant over various areas of the northern hemisphere. For this, one would need to repeat all simulations with observed SSTs in the area of interest and climatological SSTs elsewhere.

\section{Chapter conclusions}
\label{conclusion}

The objective of this part of the PhD thesis was twofold. First, the study aimed at assessing the atmospheric response to meridional shifts in the GSF via the use of multi-model atmosphere-only simulations forced with observed SSTs. Using multiple realizations for each model was key in letting the forced signal emerge from the chaotic atmospheric variability, while comparing results between different models and the observations enhanced our confidence in the interpretation of the findings. In turn, the use of atmosphere-only simulations allowed us to study the atmospheric response to the oceanic forcing in isolation, namely, focusing on a single direction of the two-way interaction between the two realms. These choices allowed the present study to provide robust evidence on how the observed GSF variability may be impacting the North Atlantic midlatitude atmospheric circulation. Secondly, the present study aimed at assessing the role of atmospheric horizontal resolution for the realistic simulation of the above-discussed response. In fact, past studies \citep[e.g.][]{smirnov2015} have examined the dependence on model resolution of the atmospheric response to idealized SST forcings mimicking the meridional shift of an oceanic front, yet our study does so for realistic SST anomalies and in the more robust context of a multi-model, protocol-driven, coordinated framework. Here we refer to model configurations with a nominal resolution coarser than 50\,km as R100 models, and those with a nominal resolution finer than or equal to 50\,km as R50+ models.

The main findings can be summarized as follows. The interannual variability in the meridional position of the GSF was found to relate to intense localized SST anomalies, positive for the “North” position and negative for the “South” position (Figure \ref{fig:sst}). These SST anomalies later induce collocated anomalies in SHF (sensible and latent) representing an important portion (15--20\%) of the winter SHF climatology over a significant area (Figure \ref{fig:shf}). This anomalous diabatic heating forces a local atmospheric circulation response which is fundamentally different between R100 and R50+ models. The response in the latter was found to be similar to the respective observed anomalies, which is not the case for R100 models.

Discussing first the local response of the R50+ models, it was found that in the presence of the anomalous diabatic heating along the GSF, thermodynamic balance near the surface is maintained mainly by anomalous meridional eddy heat flux divergence (Figure \ref{fig:budget}f), while the anomalous meridional mean advection (term III in equation \ref{heat_budget_equation}) tends to exacerbate the temperature tendency induced by the heating. Anomalous vertical motion is generated (Figure \ref{fig:omega}) and this contributes significantly to counter-balancing the anomalous heating in the lower troposphere (Figure \ref{fig:budget}g). Effectively, the baroclinic eddy activity is modified to equilibrate the anomalous baroclinicity close to the GSF, possibly intensifying the zonal winds north of the GS. This is consistent with the detected large-scale changes in low-level baroclinicity (Figure \ref{fig:t925_grad}), which is indeed mostly maintained by advection of cold air from inland of North America (Figure \ref{fig:budget_R50+}, Figure \ref{fig:budget_ERA5}). Given the latter, in response through baroclinic adjustment, the baroclinic eddy activity is further intensified. As a result, the eddy-driven jet is also modified (Figure \ref{fig:u850}) through the action of the synoptic eddy fluxes (vertical divergence of E-vectors). Finally, the detected changes in blocking frequency (Figure \ref{fig:blocking}) are dynamically consistent with the changes in the stormtrack and the jet and in agreement with previous findings \citep{joyce2019}.

\begin{figure}[ht!]
    \centering
    \includegraphics[width=22pc]{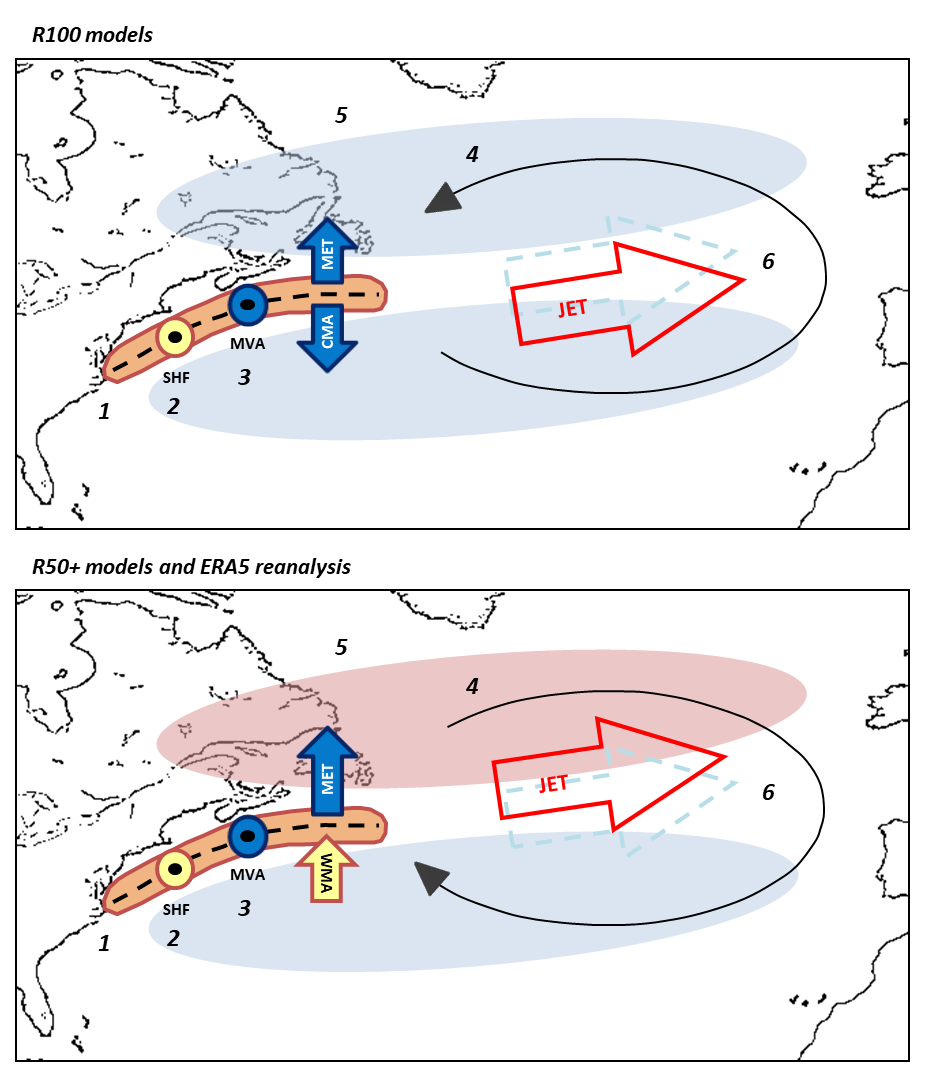}
\caption{Schematic of the atmospheric response to the GSF shifts in winter for R100 models (top) and R50+ models and the ERA5 reanalysis (bottom). The winter climatological position of the GSF is represented by the black dashed line. (1) The brown zone straddling the GSF represents the area of intense SST anomalies induced by a shift of the oceanic front from its “South” to its “North” position. (2) The strong along-front SST anomalies induce intense anomalies in surface heat fluxes (SHF), represented by yellow circles with a black point in the center, meaning that the ocean is warming the atmosphere. (3) In R100 models the diabatic heating anomaly in the vicinity of the GSF is largely balanced by mean vertical advection (MVA)\nomenclature{MVA}{mean vertical advection}, meridional eddy heat transport (MET, with “eddy” meaning monthly departures from climatology)\nomenclature{MET}{meridional eddy heat transport} and cold meridional advection (CMA)\nomenclature{CMA}{cold meridional advection}. In R50+ models MVA has a similar role, but the warm meridional advection (WMA)\nomenclature{WMA}{warm meridional advection} induces a positive temperature tendency. Thus, to maintain balance, MET is significantly stronger than in R100 models. The direction of the arrows in respect to the GSF is supposed to indicate heat transport convergence (warming for WMA, yellow) and divergence (cooling for MET and CMA, blue). (4) Downstream of the GSF, the R100 (R50+) models exhibit surface cyclonic (anticyclonic) circulation anomalies consistent with the CMA and WMA, respectively. (5) In R100 models the GSF shifts are associated with a reduction in synoptic eddy heat fluxes (v$^\prime$T$^\prime$) in most of the North Atlantic, whereas R50+ models exhibit positive (negative) anomalies north (south) of the GSF, represented by the red (blue) shadows. (6) The eddy-driven jet shifts poleward (equatorward) in R50+ (R100) models.}
\label{fig:sketch}
\end{figure}

To conclude, R50+ models produce a local as well as a large-scale atmospheric circulation response with patterns similar to the respective observed anomalies. The latter consist of a shift in the North Atlantic eddy-driven jet and stormtrack that is homo-directional to the GSF shifts (Figure \ref{fig:jet_stream}).
Low-resolution (R100) models produce a strongly different local circulation response to the anomalous diabatic heating associated with the GSF shifts. It is argued that these differences in the local circulation response are key in understanding also the differences in the large-scale response detected in these models. Arguably, the most important of the above-mentioned aspects is that R100 models were found to maintain thermodynamic balance near the surface not through an intense anomaly in meridional eddy heat transport but through the action of anomalous meridional mean temperature advection (Figure \ref{fig:sketch}). This does not agree with the circulation changes found in the observations and in high-resolution (R50+) models.

It is plausible that the improvement of the response in high-resolution models is conveyed by a better representation of mesoscale ocean-to-atmosphere forcing that is not resolved in their low-resolution counterparts and/or a more realistic representation of small-scale key atmospheric features, such as fronts and conveyor belts. After all, it should be remembered that the atmosphere does not interact with a smooth and quasi-linear time mean SST front but with a much more convoluted, time-evolving SST front with gradients that are even more pronounced locally. 

In the comparison between AGCMs and observations, it should be pointed out that in ERA5 the atmospheric anomalies associated with concurrent GSF shifts are, at least in part, the representation of the atmospheric forcing that caused the GSF shifts. This is in line with previous studies \citep{taylor1998b, frankignoul2001, sanchez2016} showing that the GSF latitude is positively correlated with the NAO at zero and at negative lags (NAO leading). Therefore, in contrast to the AGCMs, the concurrent atmospheric anomalies in ERA5 cannot be interpreted purely as an atmospheric response to the GSF shifts. Considering this important disparity between AGCMs and ERA5, the larger amplitude of the atmospheric anomalies detected in ERA5 compared to the anomalies in the R50+ models (e.g. see Figure \ref{fig:u850}) is indicative of a positive feedback between ocean and atmosphere: as soon as the GSF shift is established the associated SST anomalies tend to force a NAO-like response (as the R50+ models indicate) that strengthens / prolongs the original atmospheric forcing (present only in ERA5). Notably, the existence of a positive feedback of this kind, i.e. between the NAO-induced tripole SST anomalies (recalling the SST anomalies associated with GSF shifts) and the NAO, has been proposed also by other authors \citep{czaja2002, joyce2019}.  However, such a positive feedback is not confirmed by \cite{wills2016} who found that the atmospheric variability pattern forcing SST changes in the Gulf Stream Extension area and the atmospheric anomalies subsequently forced by the same SST changes are spatially anticorrelated and temporally distant. Such discrepancies show that the character of the atmospheric response to SST anomalies near the GSF should be further investigated in the observations. This is especially true for monthly and shorter timescales, when the atmospheric variability is expected to be dominated by internal atmospheric processes which could hamper the emergence of the atmospheric response to SST variability.

\chapter{\textit{\textbf{Non-stationarity in the NAO--Gulf Stream SST front interaction}}}
\label{decadal_variability}

\fancyhead[R]{\textit{Chapter \ref{decadal_variability} - Non-stationarity in the NAO--GSF interaction}}

\section{Introduction}
\label{introduction_decadal}
Recent studies have shown that the extratropical atmospheric circulation in the Northern Hemisphere is partly controlled by the GSF \citep{joyce2019}. Specifically, the presence of the GSF has been shown to shape the climatological structure of the North Atlantic storm-track and eddy-driven jet \citep{brayshaw2011,oreilly2016,omrani2019}. Furthermore, the meridional displacements of the GSF have been shown to affect the North Atlantic tropospheric variability, with atmospheric circulation anomalies extending up to the Eurasian continent \citep{joyce2009,nakamura2009,kwon2013,sato2014,seo2017}. Therefore, studying the drivers of the GSF variability is important for understanding the associated coupled (ocean--atmosphere) variability and predictability.

The GSF latitudinal position has been linked to different climate modes, such as the Atlantic meridional mode\nomenclature{AMM}{Atlantic meridional mode} \citep[AMM;][]{hameed2018, wolfe2019}, the Atlantic multidecadal oscillation\nomenclature{AMO}{Atlantic multidecadal oscillation} \citep[AMO;][]{nigam2018}, the AMOC \citep{decoetlogon2006,joyce2010,perez2014,ezer2015,sanchez2015,goddard2015,zeng2016}, the ENSO \citep{taylor1998a,kwon2010,perez2014,sanchez2016} and the NAO \citep{taylor1998b,joyce2000,frankignoul2001,hameed2004,kwon2010,perez2014,sanchez2016,wolfe2019}. In this context, particular attention has been directed onto the NAO, which is the dominant mode of variability of the surface atmospheric circulation in the North Atlantic region \citep{hurrell1995}. For example, \citet{taylor1998a} have shown that the NAO explains a much larger portion of annual variance in the GSF latitudinal position than ENSO.

The NAO can affect the GSF latitudinal position through anomalous wind-stress over the North Atlantic, by directly changing the wind-driven oceanic circulation and by exciting westward propagating Rossby waves \citep{gangopadhyay2016}. For example, the positive phase of the NAO is associated with anomalous northward Ekman transport over the GS region \citep{visbeck2003} and a northward shift of the zero wind-stress curl line. The resulting wind-driven oceanic circulation anomalies are generally largest at the confluence of the SPG and the STG \citep{marshall2001,reintges2017}, and may thus influence the position of the GSF. Indeed, previous studies have shown that a positive NAO results in a northward displacement of the GSF, which can lag the NAO forcing by less than one month up to 2--3 years \citep{taylor1998b,joyce2000,frankignoul2001,taylor2001,joyce2019,watelet2020}. Similarly, a negative NAO results in a southward displacement of the GSF with some time lag. The time lags have been related to the time that fast barotropic and slow baroclinic planetary waves require to cross the ocean, travelling from the eastern to western North Atlantic \citep{veronis1973,gangopadhyay1992,marshall2001,bellucci2006}. As an example, in a numerical study, \citet{sasaki2011} have shown that the NAO can affect the GS position through westward propagating sea-level height anomalies generated by anomalous Ekman convergence in the central North Atlantic, which reach the GS area after about 2 years. Consistently, in another numerical study, \citet{decoetlogon2006} have shown some evidence of NAO-forced baroclinic Rossby wave propagation in the southern part of the SPG, which affects the GS path and transport after about 2 years. However, there is still no consensus on the role of Rossby wave propagation on the GSF path. Indeed, several studies have documented the characteristics of oceanic Rossby waves in North Atlantic propagating with phase velocities in the 0,93--4,17 cm s$^{-1}$ range \citep{cipollini1997,cipollini2001,osychny2004,decoetlogon2006,arthun2020,watelet2020}. These phase velocities imply a lag between the NAO and the GSF shifts greater than what is actually observed, even for waves that are generated in the mid-Atlantic \citep{frankignoul2001,wolfe2019}. 

Furthermore, the NAO can indirectly affect the GSF position through buoyancy fluxes in the LS \citep{joyce2000,pena2008,watelet2020}. The NAO is associated with anomalous SHF over the SPG, with impacts on the deep convection and thus on the export of LS water (LSW)\nomenclature{LSW}{Labrador Sea water} via the DWBC. Once the deep water is formed, it is partly recirculated within the LS and along the NAC and partly travels southward, flowing along the Slope Sea\citep[i.e. the ocean band between the GS and the eastern continental shelf of North America from Cape Hatteras to the Grand Banks;][]{stommel1958,csanady1988,bower2009,bower2011}. 
\citet{zhang2007} have shown that the bottom vortex stretching of the DWBC transport is able to affect the GS latitudinal position by changing the intensity of the NRG, a barotropic cyclonic circulation north of the GS \citep{hogg1992}. The stronger (weaker) the DWBC transport close to the Grand Banks, the stronger (weaker) the NRG and the more southward (northward) the GS. The effect of the DWBC transport on the GS latitudinal position has been suggested to be particularly relevant on the decadal and multi-decadal timescales \citep{zhang2006}. Indeed, the DWBC feeds the deeper branch of the AMOC, which has been shown to be affected by the NAO on that range of timescales \citep{bellucci2006,bellucci2008,reintges2017,omrani2022}.

In the context of the NAO forcing changes in the GS latitudinal position through modifying the DWBC transport, \citet{joyce2000} have speculated on the existence of a self-sustained decadal NAO--GS coupled oscillation using a simple oceanic model. In this oscillation, a positive NAO enhances the southward LSW transport in the DWBC. The anomalies propagate southward along the eastern continental shelf of North America and reach the GS area after some years. This propagation time lag depends on the values of model parameters. The enhanced southward LSW transport in response to the positive NAO acts to move the GS separation point close to Cape Hatteras southward \citep{pickart1993,spall1996a}. Consequently, negative SST anomalies are established along the GS, reducing the atmospheric transient eddy activity and influencing the NAO phase. Finally, the reversal of the NAO phase changes the buoyancy forcing in the LS and the cycle starts again with the opposite sign. In agreement with this mechanism acting on decadal timescale, \citet{pena2011} have shown in an observational study that anomalies in LSW require approximately 4--9 years to propagate from the LS to the north-east of Cape Hatteras. Similarly, \citet{molinari1998} have estimated a transit time from the LS to 26.5°N of 10 years. Using chlorofluorocarbon and hydrographic data, \citet{smethie1993} has inferred an even longer transit time. However, it is unclear whether such a decadal NAO–GS oscillation operates in reality since observational support is lacking.

Furthermore, some studies have shown that a significant part of the LSW is not exported from the LS via the DWBC but it follows other, southward-oriented, interior pathways \citep{rossby1999,rossby2000,rossby2005,bower2009,bower2011}. In agreement with these studies, analyzing observational and reanalysis data, \citet{hameed2004} and \citet{sanchez2016} have shown that the pressure and the longitude of the Icelandic Low (IL; i.e. one NAO center of action)\nomenclature{IL}{Icelandic Low} are good predictors for the GS latitudinal position. In particular, enhanced (reduced) and westward (eastward) shifted IL intensifies (reduces) the transport of the surface LC, thus reducing (enhancing) the cold water sources along the Slope Water (i.e. the upper part of the water column in the Slope Sea – first 500 m depth) and shifting northward (southward) the GS \citep{peterson2017,holliday2020,new2021}. The highest prediction skill has been obtained when enhanced and westward shifted IL leads the GSF latitudinal position by 2--3 years. However, other studies have estimated that the temperature and salinity anomalies in the Slope Water would take about 4--18 months to travel from the Newfoundland to the GS separation point \citep{rossby2005,pena2008,peterson2017,new2021}. This suggests a shorter time lag between the NAO forcing and the GS shifts compared to the ones proposed by \citet{hameed2004} and \citet{sanchez2016}.

\citet{hameed2004} have also shown that the effect of the Azores High (i.e. the other NAO center of action) on the GS latitudinal position is insignificant during 1965--2000. For this reason, they have concluded that the fluctuations in the southward flow of the LSW is dominant compared to the influence of the NAO on the GS latitudinal position through Rossby waves excited in the extratropical North Atlantic. However, they have analyzed the time period 1965--2000 as a whole. On the other hand, \citet{sasaki2011}, looking at the non-stationarity of time series in an oceanic model forced with observed atmospheric fluxes, provide evidence of wave propagation during 1965--1985, affecting the GS path. The different methodological approach of these two studies show that the wave-structure propagation can only be seen by analyzing the non-stationarity of time series.

\begin{figure}[ht!]
    \centering
    \includegraphics[max
    size={\textwidth}{\textheight}]{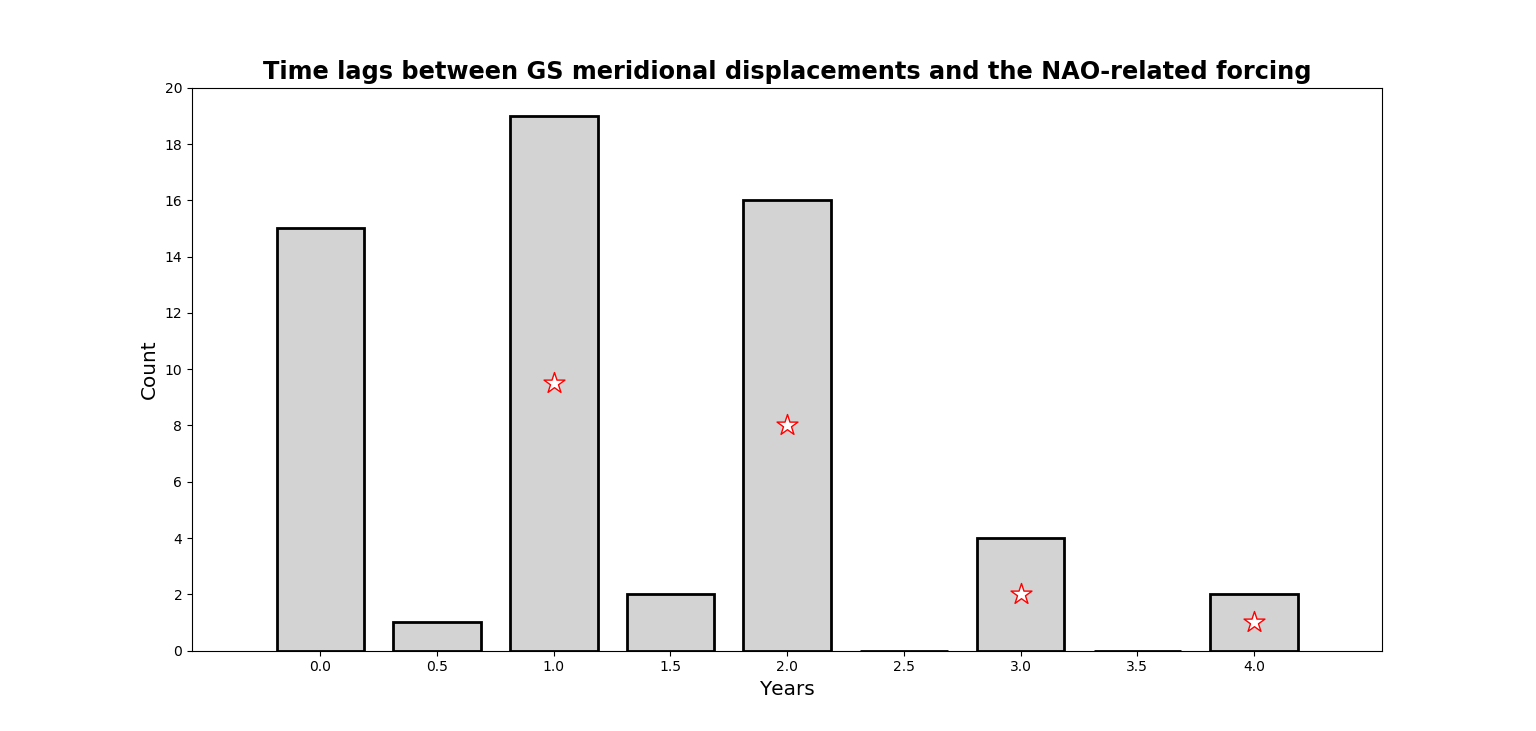}
\caption{Histogram of the time lags between GS meridional displacements and the NAO-related forcing suggested in literature, as reported in Table \ref{table:NAO-GSF}. The stars identify the time lags where the correlation between the GSF and NAO indices is statistically significant at 95\% significance level in the present work.}
\label{fig:histogram_lag}
\end{figure}

Previous studies have shown that a variety of mechanisms can contribute to interactions between the NAO and the GSF, but the primary mechanisms behind the response of the GSF to the NAO in observations remain unclear. The lack of consensus is emphasized by a wide range of time lags between meridional displacements of the GSF and the NAO reported among prior studies, ranging between 0 and 4 years (Table \ref{table:NAO-GSF}, Figure \ref{fig:histogram_lag}). The discrepancies between studies may be linked to differences in the definition of the GS latitudinal position, which has been based both on surface \citep{taylor1998b,watelet2020} and subsurface data \citep{joyce2000,hameed2018} and has adopted different space-domain \citep{hameed2018}. These discrepancies might also be linked to differences in the datasets (models, observations and reanalyses) and methodological approaches. In particular, analyses that do not consider the non-stationarity of the mechanisms involved can miss capturing some of them \citep[e.g. the wave propagation;][]{hameed2004,sasaki2011}. Finally, the discrepancies might be linked to differences in the analyzed time period. In some cases, this could reflect disparities in the quality and coverage of observations, but this could also suggest that the NAO--GSF interaction is non-stationary in time \citep{frankignoul2001,hameed2004,sasaki2011,lillibridge2013,hameed2018}. For example, using a revised index for the latitudinal position of the GS north wall, \citet{hameed2018} showed that the NAO leads GS shifts by a lag of 1 year in 1940--2014, 0 year in 1961--2014 and no correlation in 2005--2014.

Table \ref{table:NAO-GSF} shows some examples of the time lags proposed in literature, with the respective index for the GS latitudinal position, the atmospheric forcing, the typology of the analysed dataset, the seasonality and the time period. As shown in Figure \ref{fig:histogram_lag}, time lags of 0--2 years between the GS meridional displacements and the NAO-related forcing are those most common in literature.

To summarize, there is a lack of consensus on the key mechanisms underlying the response of the GSF to the NAO. There are also indications of non-stationarity in the NAO--GSF interaction. However, it is difficult to draw robust conclusions due to differences among methodologies used in prior studies, including the indices of GSF variability. The goal of this part of the PhD thesis is to assess the possible non-stationarity in the NAO--GSF interaction during the winter season and to help clarify the mechanisms underlying this interaction over the last few decades using ERA5 and ORAS5 reanalysis data.

\newcommand{\splitcell}[2][c]{%
  \renewcommand{\arraystretch}{1}
  \begin{tabular}[c]{@{}c@{}}\strut#2\strut\end{tabular}%
}

\renewcommand{\arraystretch}{2}

\begin{table}[]
\caption{\label{table:NAO-GSF} Time lags between the GS meridional displacements and the NAO-related forcing. Columns detail the index to define the GS latitudinal position, the index to define the NAO-related forcing, the typology of the analysed dataset, the period over which the data have been averaged, the analyzed time period, the time lag and the reference. Only the time lags where the correlation between the GS index and the NAO-related index is statistically significant at 90\% and 95\% significance levels are shown in the table. Refer to the respective reference for more details.}
\begin{center}
\resizebox{\textwidth}{!}{\begin{tabular}{ccccccc}
\hline\hline\

\textbf{GS index} & \textbf{NAO-related index} & \textbf{Data} & \textbf{Time mean} & \textbf{Time period} & \textbf{Time lag} & \textbf{Reference} \\\hline\

GSNW\nomenclature{GSNW}{Gulf Stream North Wall} & NAO & Observations & DJFM & 1966--1996 & 2 years & \citet{taylor1998b} \\\hline\

GSI\nomenclature{GSI}{Gulf Stream index} & NAO & Observations & JFM & 1954--1989 & 0--1 year & \citet{joyce2000} \\\hline\

Surface salinity data & NAO & Observations & DJF & 1978--1998 & 1.5 years & \citet{rossby2000} \\\hline\

\multirow{2}{*}{GSI} &  \multirow{2}{*}{NAO} & \multirow{2}{*}{Observations} & Monthly & \multirow{2}{*}{1955--1998} & 12--18 months & \multirow{2}{*}{\citet{frankignoul2001}} \\
& & & Annual & & 0--2 years & \\\hline\

\multirow{6}{*}{GSNW} &  IL pressure & \multirow{6}{*}{Observations} & \multirow{6}{*}{DJF} & \multirow{6}{*}{1966--2000} & 1--3 years & \multirow{6}{*}{\citet{hameed2004}} \\\cline{2-2}\cline{6-6}\
& IL latitude & & & & 1--2 years & \\\cline{2-2}\cline{6-6}\
& IL longitude & & & & 3--4 years & \\\cline{2-2}\cline{6-6}\
& AH pressure & & & & 0--2 years & \\\cline{2-2}\cline{6-6}\
& AH latitude & & & & No correlation & \\\cline{2-2}\cline{6-6}\
& AH longitude & & & & 2 years & \\\hline\

GSNW & \multirow{2}{*}{NAO} & \splitcell{Observations (ocean);\\reanalysis (atmosphere)} & \multirow{2}{*}{Annual} & 1954--1998 & 0--3 years & \multirow{2}{*}{\citet{decoetlogon2006}}\\\cline{1-1}\cline{3-3}\cline{5-6}\
\splitcell{GST index\nomenclature{GST}{Gulf Stream temperature}\\(similar to GSNW)} & & \splitcell{Model (ocean);\\reanalysis (atmosphere)} & & 1948--2000 & 0--2 years &  \\\hline\

GSNW & NAO & Observations & DJFM & 1980--1999 & 1 year & \citet{chauduri2009}\\\hline\
SSH data & NAO & Model & DJFM & 1960--2003 & 2 years & \citet{sasaki2011}\\\hline\

GSI & \splitcell{Synoptic transient\\eddy heat flux} & \splitcell{Observations (ocean);\\reanalysis (atmosphere)} & JFM & 1979--2009 & 2 years & \citet{kwon2013} \\\hline\

SSH data & NAO & Observations & Monthly & 1992--2011 & 6 months & \citet{lillibridge2013}\\\hline\

\splitcell{16-point GS index\\(based on SSH data)} & NAO & Model & DJFM & 1960--2003 & 2 years & \citet{perez2014}\\\hline\

\multirow{3}{*}{GSNW} &  IL pressure & \multirow{3}{*}{\splitcell{Observations (ocean);\\reanalysis (atmosphere)}} & \multirow{3}{*}{Annual} & \multirow{3}{*}{1966--2014} & 0--2 years & \multirow{3}{*}{\citet{sanchez2016}} \\\cline{2-2}\cline{6-6}\
& IL longitude & & & & 3--4 years & \\\cline{2-2}\cline{6-6}\
& NAO & & & & 0--2 years & \\\hline\

GSNW-revised & NAO & \splitcell{Observations (ocean);\\reanalysis (atmosphere)} & Annual & 1940--2014 & 1 year & \citet{watelet2017}\\\hline\

\multirow{2}{*}{eGSI-0m}\nomenclature{eGSI}{extended Gulf Stream index} & \multirow{10}{*}{NAO} & \multirow{10}{*}{\splitcell{Observations (ocean);\\reanalysis (atmosphere)}} & \multirow{6}{*}{Annual} & 1961--2015 & 0 year & \multirow{10}{*}{\citet{hameed2018}} \\\cline{5-6}\
& & & & 2005--2015 & No correlation & \\\cline{1-1}\cline{5-6}\
\multirow{2}{*}{\splitcell{eGSI-200m\\(comparable to GSI)}} & & & & 1961--2015 & 0--1 year & \\\cline{5-6}\
& & & & 2005--2015 & No correlation & \\\cline{1-1}\cline{5-6}\
\multirow{2}{*}{GSNW} & & & & 1966--2014 & 0--2 years & \\\cline{5-6}\
& & & & 2055--2014 & No correlation & \\\cline{1-1}\cline{4-6}\
\multirow{4}{*}{wGSNW} & & & Annual & 1940--2014 & 1 year & \\\cline{4-6}\
& & & DJF & 1940--2014 & 0 year & \\\cline{4-6}\
& & & \multirow{2}{*}{Annual} & 1961--2014 & 0 year & \\\cline{5-6}\
& & & & 2005--2014 & No correlation & \\\hline\

\multirow{6}{*}{eGSI-0m} & IL pressure & \multirow{6}{*}{\splitcell{Observations (ocean);\\reanalysis (atmosphere)}} & \multirow{6}{*}{DJF} & \multirow{6}{*}{1961--2015} & 0--2 years & \multirow{6}{*}{\citet{wolfe2019}}\\\cline{2-2}\cline{6-6}\
& IL latitude & & & & 0--1 year & \\\cline{2-2}\cline{6-6}\
& IL longitude & & & & 1 year & \\\cline{2-2}\cline{6-6}\
& AH pressure & & & & 0--2 years & \\\cline{2-2}\cline{6-6}\
& AH latitude & & & & No correlation & \\\cline{2-2}\cline{6-6}\
& AH longitude & & & & No correlation & \\\hline\

\end{tabular}
}
\end{center}
\end{table}

\section{Data and methodological approach}
\label{data_decadal}
\subsection{Data}
The interaction between the GSF and the NAO has been assessed using ERA5 \citep{hersbach2020} and ORAS5 \citep{zuo2019} reanalysis data in the period 1950--2020 and 1958--2018, respectively. The ORAS5 reanalysis has been used as a reference for oceanic variables no t available in ERA5 reanalysis, in order to assess the mechanisms through which the NAO drives the GSF latitudinal position (see below). The ORAS5 reanalysis is forced by ERA-40 and ERA-Interim reanalysis during 1958--1978 and 1979--2014, respectively. From 2015 onwards, the ORAS5 reanalysis is forced by ECMWF/IFS operational analysis. These atmospheric products are consistent with ERA5 reanalysis in terms of representation of large-scale modes of atmospheric variability such as the NAO. Thus, using the more recent and accurate ERA5 reanalysis as a reference for the atmospheric state instead of the other ECMWF products does not affect the NAO--GSF interaction and the interpretation of results presented here.

The GSF and its latitude have been defined as described in section \ref{GSF_index}. The GSF latitude (i.e. the GSF index) has been computed using ERA5 data. SST data are also provided in ORAS5 reanalysis, however this reanalysis exhibits SST biases in the GS region \citep{zuo2019} which may affect the representation of the GSF latitudinal position. 

Similarly to the approach followed by \citet{hurrell2003}, the NAO has been defined using a principal component analysis of winter SLP over the North Atlantic sector (0$^{\circ}$--80$^{\circ}$N, 90$^{\circ}$W--40$^{\circ}$E). Prior to the index calculation, SLP data have been scaled by a factor $\sqrt{cos(latitude)}$ so that different grid cells contribute to the variance of the field proportionally to the area they represent. Then, the NAO index has been defined as the leading principal component of SLP.

\subsection{Methods}
\subsubsection{Cross-wavelet analysis}
In order to assess the NAO--GSF interaction and its spectral features, a cross-wavelet analysis has been applied \citep{torrence1998, grinsted2004}. The NAO and GSF time series have been linearly detrended and standardized before performing the cross-wavelet analysis. The cross-wavelet analysis is a mathematical tool developed for the analysis of the relationship between two time series in the time--frequency space \citep{hudgins1993}. This tool allows us to study how the co-spectral features evolve in time, which is useful in assessing also the non-stationarity of the link between two variables. With this tool, the relation between two time series is assessed through the cross-wavelet power spectrum, which shows domains in the time--frequency space where the two time series share high power \citep{torrence1998}. However, peaks in the cross-wavelet power spectrum can appear in a given frequency range when two time series are actually independent and only one of them features significantly high power in this range \citep{maraun2004}. For this reason, another common measure for the relationship between two time series is the squared wavelet coherence, i.e. the cross-wavelet power spectrum normalized by the wavelet power spectrum of both time series. This coefficient provides a good estimate of the correlation between two time series in the time--frequency space \citep{grinsted2004}, although spurious peaks can occur in areas with low wavelet power. Finally, the time lag between two time series is estimated through the phase of the cross-wavelet power spectrum, defined as the arctan of the ratio between its real and imaginary part.

The Morlet function has been adopted as the wavelet basis function, with non-dimensional frequency $\omega_{0}$ equal to 6 to satisfy the admissibility condition \citep{farge1992}. The statistical significance of the NAO--GSF cross-wavelet power spectrum has been assessed with the procedure detailed in \citet[][refer to their section 6c]{torrence1998}, assuming the white- and the red-noise spectra as background spectra for the NAO and the GSF, respectively. The statistical significance of the squared wavelet coherence has been assessed through a Monte Carlo test, performing 300 simulations. In both cases, the 90\% significance level has been used. The phase relationship between the NAO and the GSF has been considered statistically significant when the squared wavelet coherence exceeds  the 90\% significance level \citep{kohyama2021}.

The spectral features of both time series have been further assessed through the wavelet transform, in order to verify whether peaks in the NAO--GSF cross-wavelet power spectrum and squared wavelet coherence are associated with peaks in the NAO and GSF wavelet power spectra. Results from the wavelet analysis applied to the single time series are presented here as global wavelet power spectra, i.e. wavelet power spectra averaged along the time dimension. The global wavelet power spectrum provides a consistent estimate of the power spectrum of a time series \citep{percival1995} and it closely approximates the Fourier spectrum \citep{hudgins1993,torrence1998}.  The NAO and GSF power spectra have been assessed also with the fast Fourier transform, in order to make possible the comparison between the global wavelet and the fast Fourier power spectra. The statistical significance of global wavelet power spectra has been assessed following the procedure described in \citet{torrence1998} (refer to their section 5a), assuming the white- and the red-noise spectra as background spectrum for the NAO and the GSF respectively.

\subsubsection{Lanczos band-pass filter}
As described later in section \ref{results}, the cross-wavelet analysis shows that the NAO--GSF covariability is particularly prominent at [6--11]-year period range. In order to assess the mechanisms through which the NAO may be forcing the GSF latitudinal position on the decadal timescale, data have been temporally filtered using a Lanczos [6--11]-year band-pass filter with a window of 12 coefficients \citep{duchon1979}. The number of coefficients has been subjectively chosen to balance the filtering with the number of data lost at the beginning and end of the time series. However, the general features of the lead--lag relationships between time series discussed in later sections are not highly sensitive to the number of coefficients ranging from 5 to 25.

\subsubsection{Indices}
The NAO can affect the GSF latitudinal position through anomalous wind-stress over the North Atlantic, influencing the wind-driven oceanic circulation and exciting Rossby waves \citep{gangopadhyay1992,taylor1998b,joyce2000,marshall2001,taylor2001,gangopadhyay2016}, as well as through buoyancy forcing over the LS, affecting the LSW formation \citep{hameed2004,zhang2007,pena2008,sanchez2016}. In order to assess the mechanisms through which the NAO drives the GSF latitudinal position, the following approaches have been adopted.

Since the NAO affects the meridional Ekman transport over the GSF area \citep{visbeck2003,deser2010}, the response of the wind-driven oceanic circulation to the NAO has been assessed by averaging the winter meridional Ekman transport over the 50$^{\circ}$--68$^{\circ}$W longitudinal range (where the GSF is defined) and the 35$^{\circ}$--40$^{\circ}$N latitudinal range. A positive (negative) Ekman index indicates an anomalous northward (southward) Ekman transport over the GS region.

The presence of Rossby wave-like patterns in the extratropical North Atlantic has been assessed through the Hovmöller diagram \citep{hovmoller1949} of the band-pass filtered SST and sea surface height (SSH)\nomenclature{SSH}{sea surface height} anomalies averaged in the 35$^{\circ}$--38$^{\circ}$N latitudinal range. This region has been selected in order to be close to the GS and, at the same time, avoid relatively high latitudes. Indeed, the Rossby waves detection is particularly challenging at those latitudes, as shown by the small number of studies dealing with the Rossby waves at latitudes higher than 39$^{\circ}$N \citep[e.g.][]{osychny2004,sasaki2011,watelet2020}. However, it is here specified that results about Rossby wave propagation are quite comparable to the ones discussed in later sections if SST and SSH anomalies are averaged in latitudinal range southward to 35$^{\circ}$--38$^{\circ}$N. This is also the case if SST and SSH data are averaged in smaller latitudinal range (e.g. latitudinal range of 1$^{\circ}$N).

The NAO-related atmospheric forcing of Rossby waves has been assessed through the Hovmöller diagram of the band-pass filtered wind-stress curl anomalies averaged in the 35$^{\circ}$--37$^{\circ}$N latitudinal range. This latitudinal range is smaller than the one used for SST and SSH anomalies in order to avoid artecfact in the wind-stress curl pattern induced by the presence of Azores islands at 25$^{\circ}$W.

\begin{figure}[ht!]
    \centering
    \includegraphics[width=28pc]{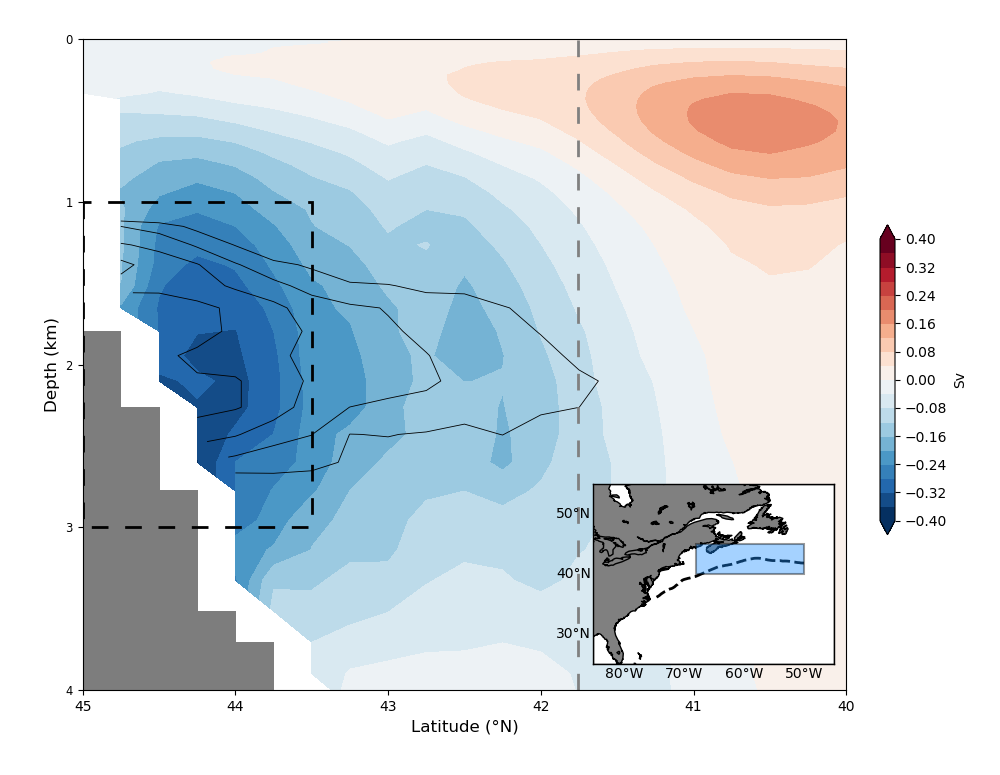}
\caption{Latitude--depth cross section of the winter mean zonal transport (Sv; color shading) in ORAS5 dataset averaged in the 50$^{\circ}$--68$^{\circ}$W longitudinal range. The black contours indicate the winter mean potential vorticity averaged in the 50$^{\circ}$--68$^{\circ}$W longitudinal range every 0.5 10$^{-12}$ m$^{-1}$ s$^{-1}$ from 2.5 10$^{-12}$ m$^{-1}$ s$^{-1}$. It is specified that the potential vorticity is lower close to the coast and it increases towards the open sea. The vertical dashed line represents the GSF latitudinal position averaged in the 50$^{\circ}$--68$^{\circ}$W longitudinal range during the winter season. The black dashed box represents the area where the zonal transport is integrated to define the DWBC index. For the DWBC index, the sign of the zonal transport is reversed in order to have positive (negative) DWBC index for anomalous westward (eastward) flow. The blue box in the bottom--right inset shows where the zonal transport, the potential vorticity and the GSF latitudinal position have been averaged along the longitudinal direction. The black dashed line in the inset indicates the winter climatological position of the GSF.}
\label{fig:dwbc_clim}
\end{figure}

Finally, the NAO influence on the GSF path through buoyancy forcing in the LS has been assessed by analyzing the zonal oceanic transport along the Slope Sea. First of all, the zonal oceanic transport has been averaged along the 50$^{\circ}$--68$^{\circ}$W longitudinal range, where the GSF index has been defined. Then, the zonal transport has been integrated in the 43.5$^{\circ}$--45$^{\circ}$N latitudinal range and over the 1000--3000 m depth interval (refer to the black box in Figure \ref{fig:dwbc_clim}). The sign of the integrated transport has been inverted in order to have positive values for westward transport. The latitude--depth box has been selected in order to include the core of the sub-surface oceanic transport along the Slope Sea during the winter season. This core coincides with the core of the DWBC transport. This is shown by the location of the potential vorticity minima generally used as a tracer to distinguish the DWBC from other water masses \citep[Figure \ref{fig:dwbc_clim};][]{talley1982}. For this reason, the index has been referred to as the DWBC index. It is specified that the meridional component of the DWBC transport has not been taken into account because the oceanic currents along the continental shelf of North America north of the GSF (50$^{\circ}$--68$^{\circ}$W) are mainly zonally oriented (not shown). Furthermore, it is specified that the vertical patterns of the zonal transport and velocity do not perfectly coincide because of the different grid resolution along the vertical direction in ORAS5 reanalysis. Indeed, the vertical resolution in the upper oceanic levels is higher than the one in the deeper levels. Then, in the calculation of the zonal transport, the zonal velocity in the upper levels is multiplied by grid areas that are lower than the ones in the deeper levels. This gives rise to zonal transport values at the upper levels that are lower compared to the ones at greater depths.

\subsubsection{Lead--lag relationship}
The lead--lag relationships between the band-pass filtered NAO, GSF, Ekman and DWBC indices have been assessed performing the lead--lag cross-correlation analysis. The statistical significance of the lead--lag cross-correlations has been assessed through bootstrap test \citep{ebisuzaki1997} at the 95\% significance level. The bootstrap test has been performed by randomly re-sampling one of the two time series 1000 times.

Finally, a lead--lag linear regression analysis based on the band-pass filtered NAO index has been applied to band-pass filtered Ekman transport and SHF data in the North Atlantic and westward transport data along the Slope Sea, in order to extract the respective response to the NAO-forcing. The statistical significance of lead--lag linear regression coefficients has been assessed through a two-tailed Student’s t-test against the null hypothesis of no-correlations at the 90\% significance level.

\section{Results}
\label{results}

\begin{figure}[ht!]
    \centering
    \includegraphics[width=\textwidth]{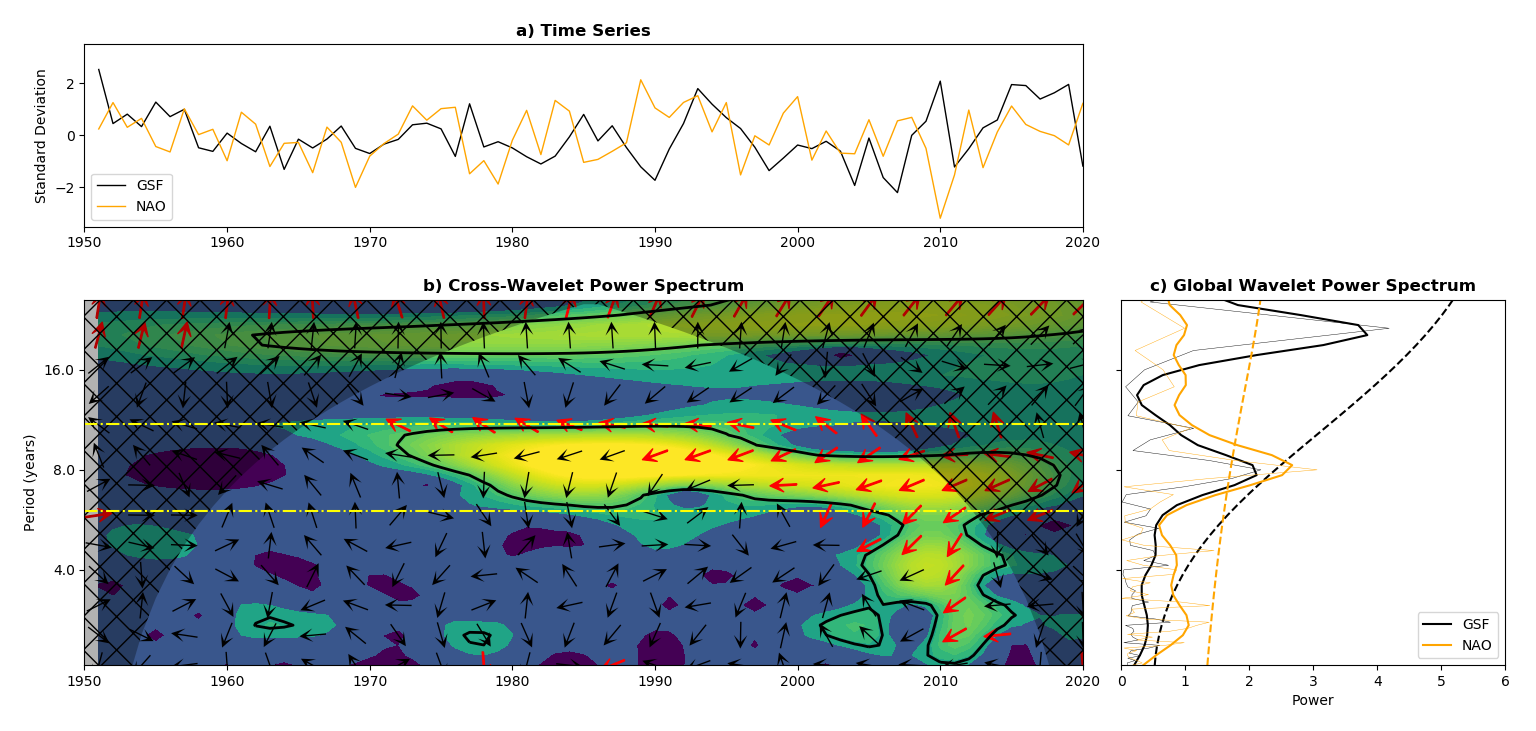}
\caption{a) Detrended and standardized winter GSF (black) and NAO (orange) time series in ERA5 dataset (1950--2020). b) Cross-wavelet transform of the winter GSF and NAO time series. Thick black contours encircle the cross-wavelet power spectrum values that are statistically significant at the 90\% confidence level. The phase relationship between the winter GSF and NAO time series is indicated by the vectors, following the convention in \citet{torrence1999}. In case of in-phase signals, vectors point upwards; in case of anti-phase signals, vectors point downwards. If the GSF leads the NAO, vectors point to the right; if the NAO leads the GSF, vectors point to the left. Thick red vectors indicate phase relationship where squared wavelet coherence is statistically significant at the 90\% confidence level \citep{kohyama2021}. The lower and upper dotted-dashed yellow lines represent the 6-year and 11-year periods, respectively. The Lanczos band-pass filter has been defined to retain periods in the [6--11]-year range. c) Global wavelet (bold) and Fourier (thin) power spectrum of detrended and standardized winter GSF (black) and NAO (orange) time series. The bold dashed lines represent the 90\% confidence level of time-averaged red- (black) and white-noise (orange) spectra.}
\label{fig:xwt_ERA5}
\end{figure}

\begin{figure}[ht!]
    \centering
    \includegraphics[width=\textwidth]{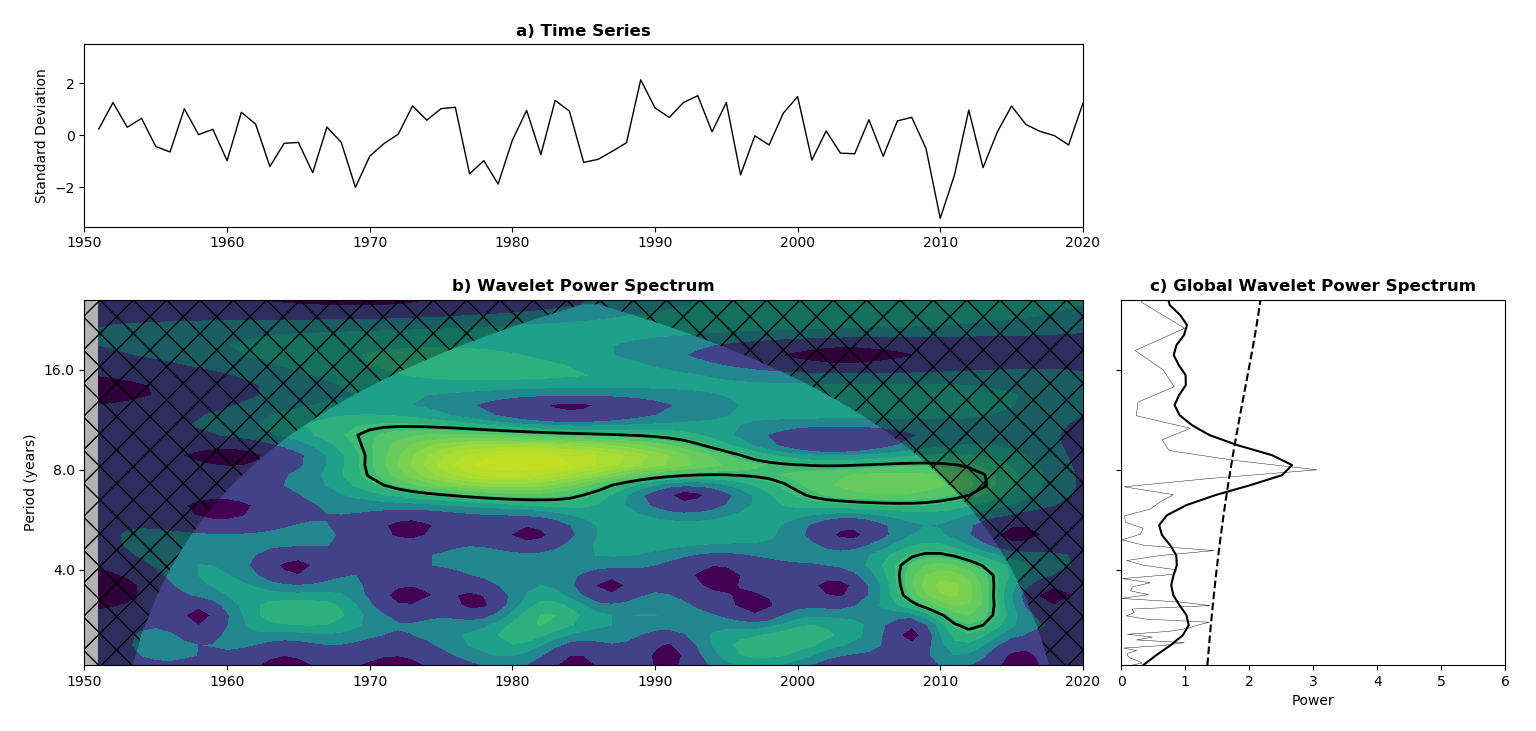}
\caption{a) Detrended and standardized winter NAO time series in ERA5 dataset (1950--2020). b) Wavelet transform of the winter NAO time series. Thick black contours encircle the wavelet power spectrum values that are statistically significant at the 90\% confidence level. c) Global wavelet (bold) and Fourier (thin) power spectrum of detrended and standardized winter NAO time series. The bold dashed lines represent the 90\% confidence level of time-averaged white-noise spectra.
}
\label{fig:cwt_nao_ERA5}
\end{figure}

\begin{figure}[ht!]
    \centering
    \includegraphics[width=\textwidth]{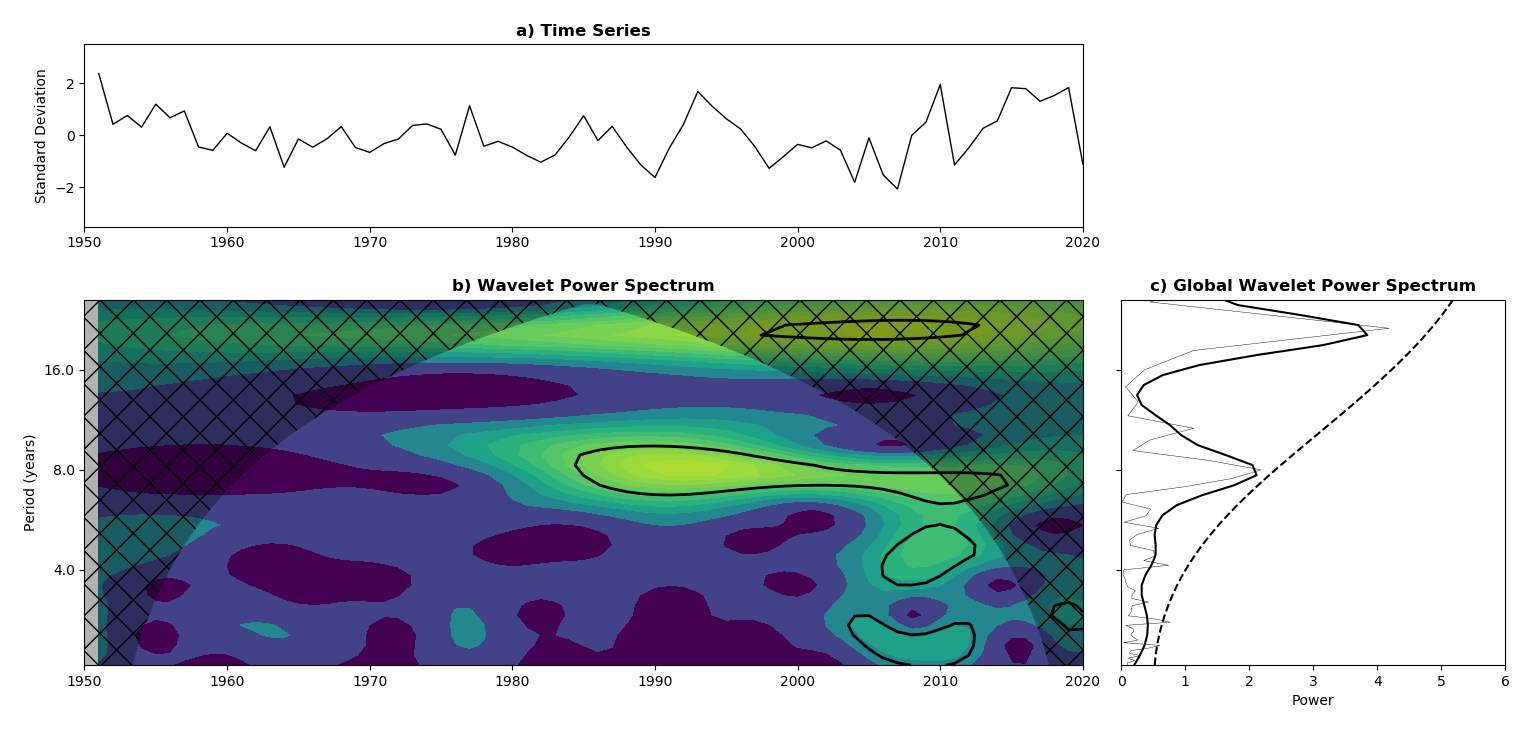}
\caption{a) Detrended and standardized winter GSF time series in ERA5 dataset (1950--2020). b) Wavelet transform of the winter GSF time series. Thick black contours encircle the wavelet power spectrum values that are statistically significant at the 90\% confidence level. c) Global wavelet (bold) and Fourier (thin) power spectrum of detrended and standardized winter GSF time series. The bold dashed lines represent the 90\% confidence level of time-averaged red-noise spectra.}
\label{fig:cwt_gsf_ERA5}
\end{figure}

Both NAO--GSF cross-wavelet power spectrum and squared wavelet coherence show statistically significant values in the [6--11]-year period range and during 1972--2018 (Figure \ref{fig:xwt_ERA5}b). The peak at decadal timescales is also present both in the single NAO and GSF wavelet (Figure \ref{fig:cwt_nao_ERA5}, Figure \ref{fig:cwt_gsf_ERA5}) and global wavelet (Figure \ref{fig:xwt_ERA5}c) power spectra, providing further evidence of the NAO--GSF interaction. However, the peak in the GSF wavelet power spectrum is statistically significant only from 1985 onwards (Figure \ref{fig:cwt_gsf_ERA5}).

\begin{figure}[ht!]
    \centering
    \includegraphics[width=\textwidth]{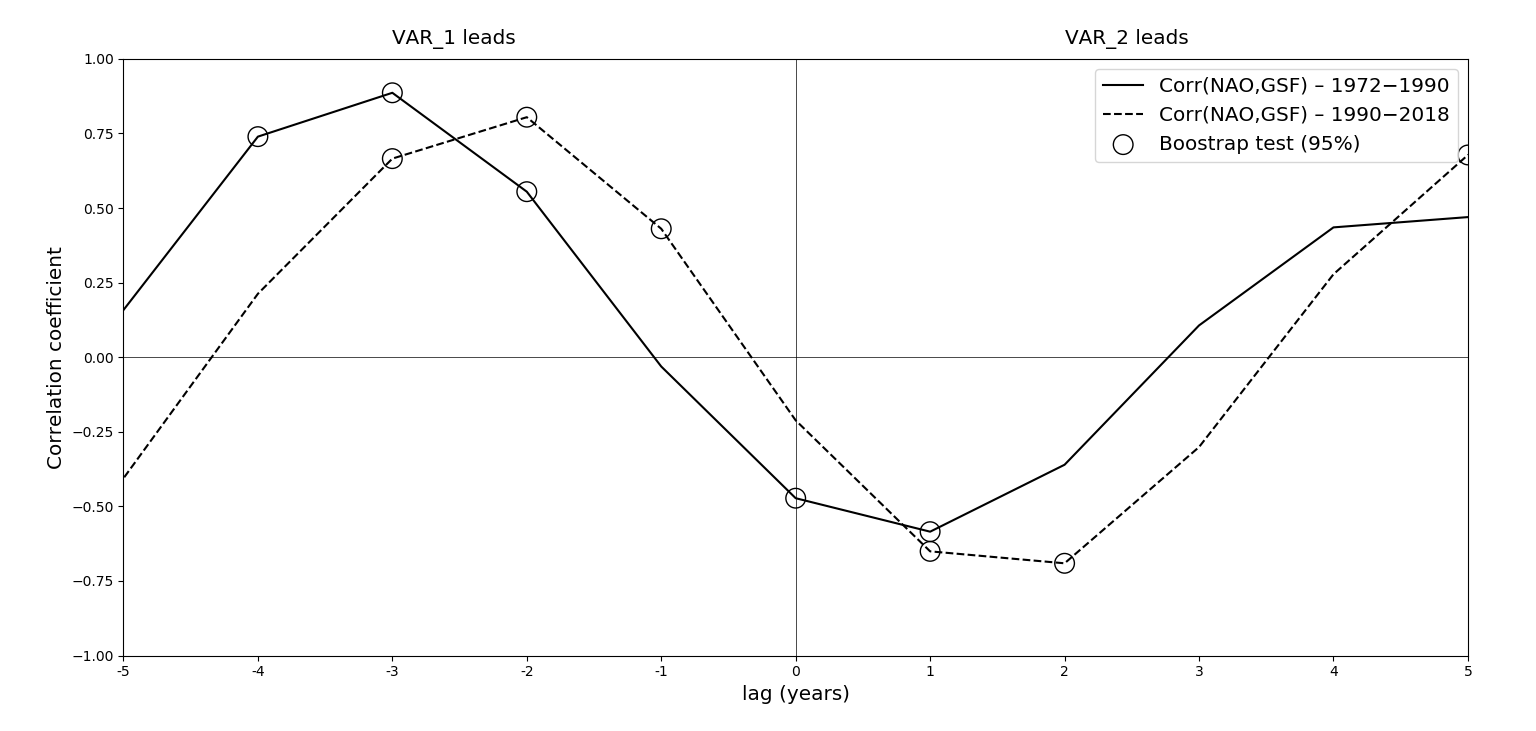}
\caption{Lead-lag cross-correlation between the band-pass filtered NAO and GSF indices during 1972--1990 (solid) and 1990--2018 (dashed). The circles highlight the lead--lag cross-correlations that are statistically significant at the 95\% confidence level.}
\label{fig:nao-gsf}
\end{figure}

In the context of the NAO--GSF decadal interaction during 1972--2018, the spectral features of the cross-wavelet power spectrum are quite homogeneous in time, whereas those of the squared wavelet coherence change around 1990. Indeed, before 1990 the squared wavelet coherence is statistically significant only for periods close to 11 years, with a phase relationship between the NAO and the GSF of about 80$^{\circ}$ on average. In contrast, after 1990 the squared wavelet coherence is statistically significant in the decadal, [6--11]-year range, with a phase relationship between the two time series of about 121$^{\circ}$ on average. These results clearly indicate that the lead--lag relationship between the NAO and the GSF at the decadal timescale changes in time. Specifically, the NAO leads the GSF shifts by about 3 and 2 years, before and after 1990, respectively. Consistently, the lead--lag cross-correlation between the band-pass filtered NAO and GSF indices shows a peak when the former leads the latter by about 3 years during 1972--1990 and by about 2 years during 1990--2018 (Figure \ref{fig:nao-gsf}).

In the last part of the analyzed record (2005--2015), both the NAO--GSF cross-wavelet power spectrum and the squared wavelet coherence show statistically significant values at periods shorter than 6 years, with the two time series closely in anti-phase (Figure \ref{fig:xwt_ERA5}b). This covariability peak is associated with statistically significant values also in the individual wavelet power spectra for the NAO and GSF (Figure \ref{fig:cwt_nao_ERA5}, Figure \ref{fig:cwt_gsf_ERA5}). These results show that the NAO and the GSF covary also at frequencies higher than the decadal one but only for a limited period of time. Consistently, the covariability peak at high frequencies during 2005--2015 is associated with low power in both the NAO and GSF global wavelet power spectra (Figure \ref{fig:cwt_nao_ERA5}c, Figure \ref{fig:cwt_gsf_ERA5}c).

Finally, another peak of covariability emerges at periods longer than 16 years (Figure \ref{fig:xwt_ERA5}b). However, the portion of the time--frequency phase space for periods longer than 16 years is largely affected by edge-effects of the cross-wavelet analysis, so results in this time--frequency region should be considered with caution. Furthermore, this peak in the cross-wavelet power spectrum is associated with a peak in the wavelet power spectrum of the GSF time series, but not for the NAO (Figure \ref{fig:cwt_nao_ERA5}, Figure \ref{fig:cwt_gsf_ERA5}). This suggests that the NAO--GSF covariability at periods longer than 16 years could be an artifact of the strong variability of the GSF alone at these timescales \citep{maraun2004}.

\begin{figure}[ht!]
    \centering
    \includegraphics[width=\textwidth]{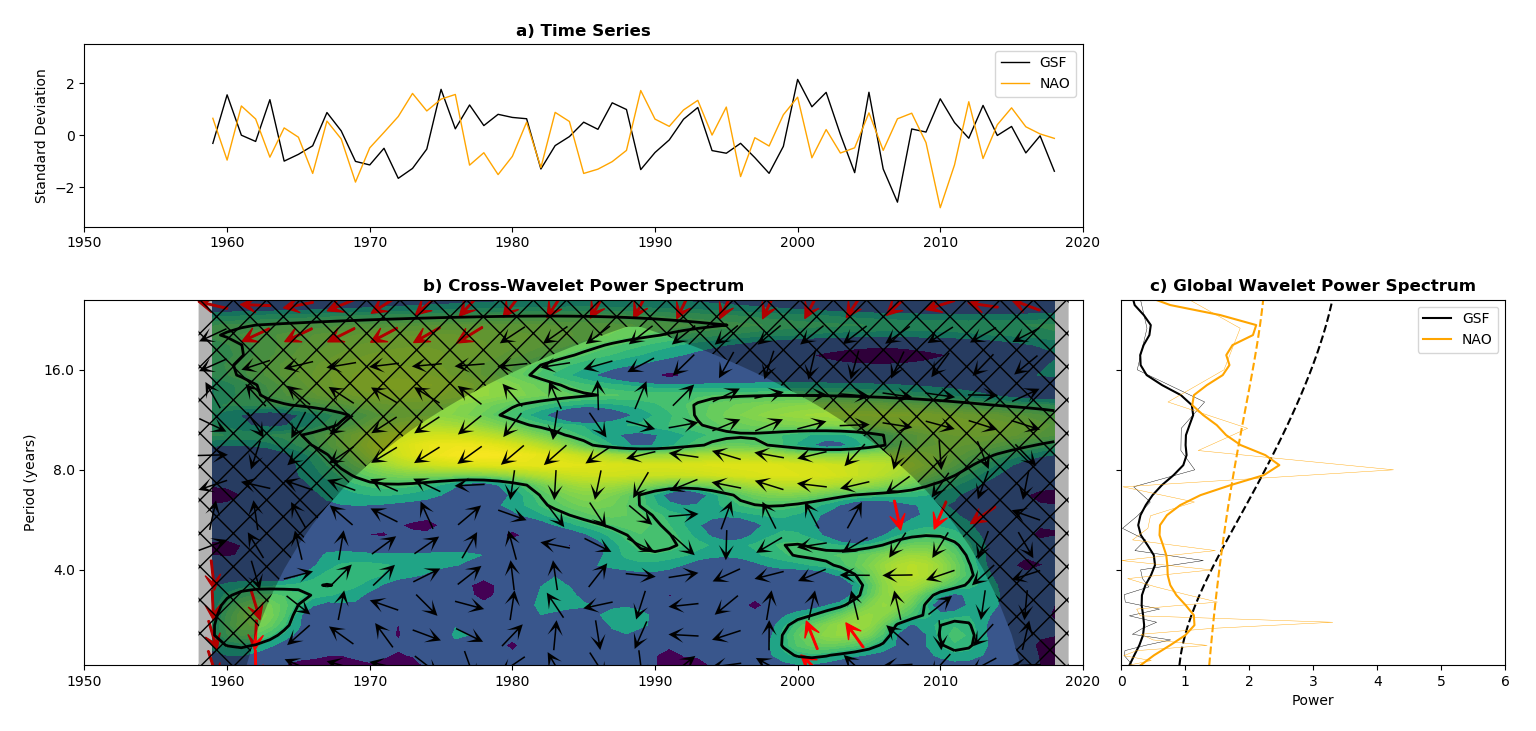}
\caption{a) Detrended and standardized winter GSF (black) and NAO (orange) time series in ORAS5 dataset (1958--2018). b) Cross-wavelet transform of the winter GSF and NAO time series. Thick black contours encircle cross-wavelet power spectrum statistically significant at 90\% level. Phase relationship between the winter GSF and NAO time series is shown as vectors, following the convention in \citet{torrence1999}. In case of in-phase signals, vectors point upwards; in case of anti-phase signals, vectors point downwards. If the GSF leads the NAO, vectors point to the right; if the NAO leads the GSF, vectors point to the left. Thick red vectors indicate phase relationship where squared wavelet coherence is statistically significant at 90\% level \citep{kohyama2021}. c) Global wavelet (bold) and Fourier (thin) power spectrum of detrended and standardized winter GSF (black) and NAO (orange) time series. The bold dashed lines represent the 90\% confidence level of time-averaged red- (black) and white-noise (orange) spectra. The x-axis range in a) and b) is 1950--2020, in order to be comparable with the ERA5 reanalysis dataset (Figure \ref{fig:xwt_ERA5}).}
\label{fig:xwt_ORAS5}
\end{figure}

\begin{figure}[ht!]
    \centering
    \includegraphics[width=\textwidth]{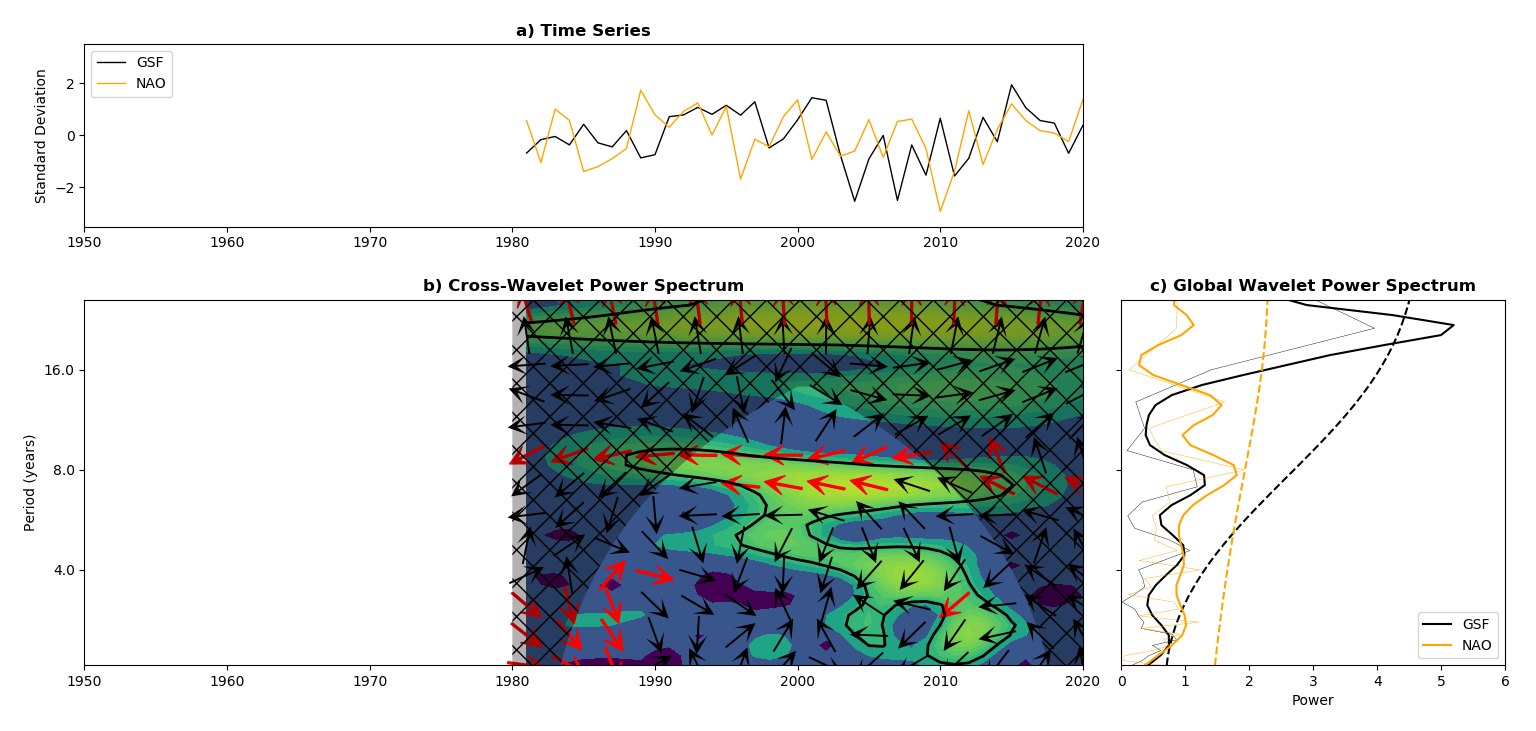}
\caption{a) Detrended and standardized winter GSF (black) and NAO (orange) time series in SODA3.4.2 dataset (1980--2020). b) Cross-wavelet transform of the winter GSF and NAO time series. Thick black contours encircle cross-wavelet power spectrum statistically significant at 90\% level. Phase relationship between the winter GSF and NAO time series is shown as vectors, following the convention in \citet{torrence1999}. In case of in-phase signals, vectors point upwards; in case of anti-phase signals, vectors point downwards. If the GSF leads the NAO, vectors point to the right; if the NAO leads the GSF, vectors point to the left. Thick red vectors indicate phase relationship where squared wavelet coherence is statistically significant at 90\% level \citep{kohyama2021}. c) Global wavelet (bold) and Fourier (thin) power spectrum of detrended and standardized winter GSF (black) and NAO (orange) time series. The bold dashed lines represent the 90\% confidence level of time-averaged red- (black) and white-noise (orange) spectra. The x-axis range in a) and b) is 1950--2020, in order to be comparable with the ERA5 reanalysis dataset (Figure \ref{fig:xwt_ERA5}).}
\label{fig:xwt_SODA}
\end{figure}

The cross-wavelet analysis has been repeated using ORAS5 \citep[1958--2018;][]{zuo2019} and SODA3.4.2 \citep[1980--2020;][]{carton2018} as alternative ocean reanalysis. The general features of the NAO--GSF covariability described above are captured by both datasets (Figure \ref{fig:xwt_ORAS5}, Figure \ref{fig:xwt_SODA}), despite the SST biases over the GS region in the ORAS5 reanalysis and the limited time period covered by the SODA3.4.2 reanalysis. Indeed, both reanalyses show a statistically significant peak in the NAO--GSF cross-wavelet power spectrum at the decadal timescale. SODA3.4.2 also shows a statistically significant peak in the NAO--GSF squared wavelet coherence at the same timescale, with the NAO leading the GSF by about 2 years. In the ORAS5 reanalysis, the NAO--GSF covariability is confined during 1965--2012, in agreement with the results from ERA5 reanalysis. Furthermore, both reanalyses show a peak in the NAO--GSF covariability for periods shorter than 6 years after 2000 (Figure \ref{fig:xwt_ORAS5}, Figure \ref{fig:xwt_SODA}). In the SODA3.4.2 reanalysis, the peak in the NAO--GSF cross-wavelet power spectrum for periods longer than 16 years is also present. Thus, the cross-wavelet analysis applied to ORAS5 and SODA3.4.2 reanalysis datasets provide further evidence of the decadal NAO--GSF covariability.

In order to assess the ocean--atmosphere interaction at the decadal timescale, the data and the selected indices have been band-pass filtered over the [6--11]-year interval. Note that we have only used the time period where the decadal peak is present in the ERA5 dataset (1972--2018) in the analyses described below. 

As mentioned in section \ref{introduction_decadal}, the NAO can force the GSF latitudinal position through the anomalous basin-scale wind-forcing and the buoyancy forcing over the LS \citep{taylor1998b,joyce2000,rossby2000,hameed2004,zhang2006,joyce2019,watelet2020}. However, it is unclear how these mechanisms interact to determine the time lag between the NAO forcing and the GSF shifts. Here, we assess the response of the wind-driven oceanic circulation to the anomalous NAO wind-stress, the excitation of Rossby waves and the response of the westward flow along the Slope Sea to the NAO-related buoyancy forcing over the LS. The assessment is performed over the individual time intervals of 1972--1990 and 1990--2018, separately. The objective is to verify which mechanism can explain the lags between the GSF shifts and the NAO forcing on the decadal timescale before and after 1990.

\begin{figure}[ht!]
    \centering
    \includegraphics[width=\textwidth]{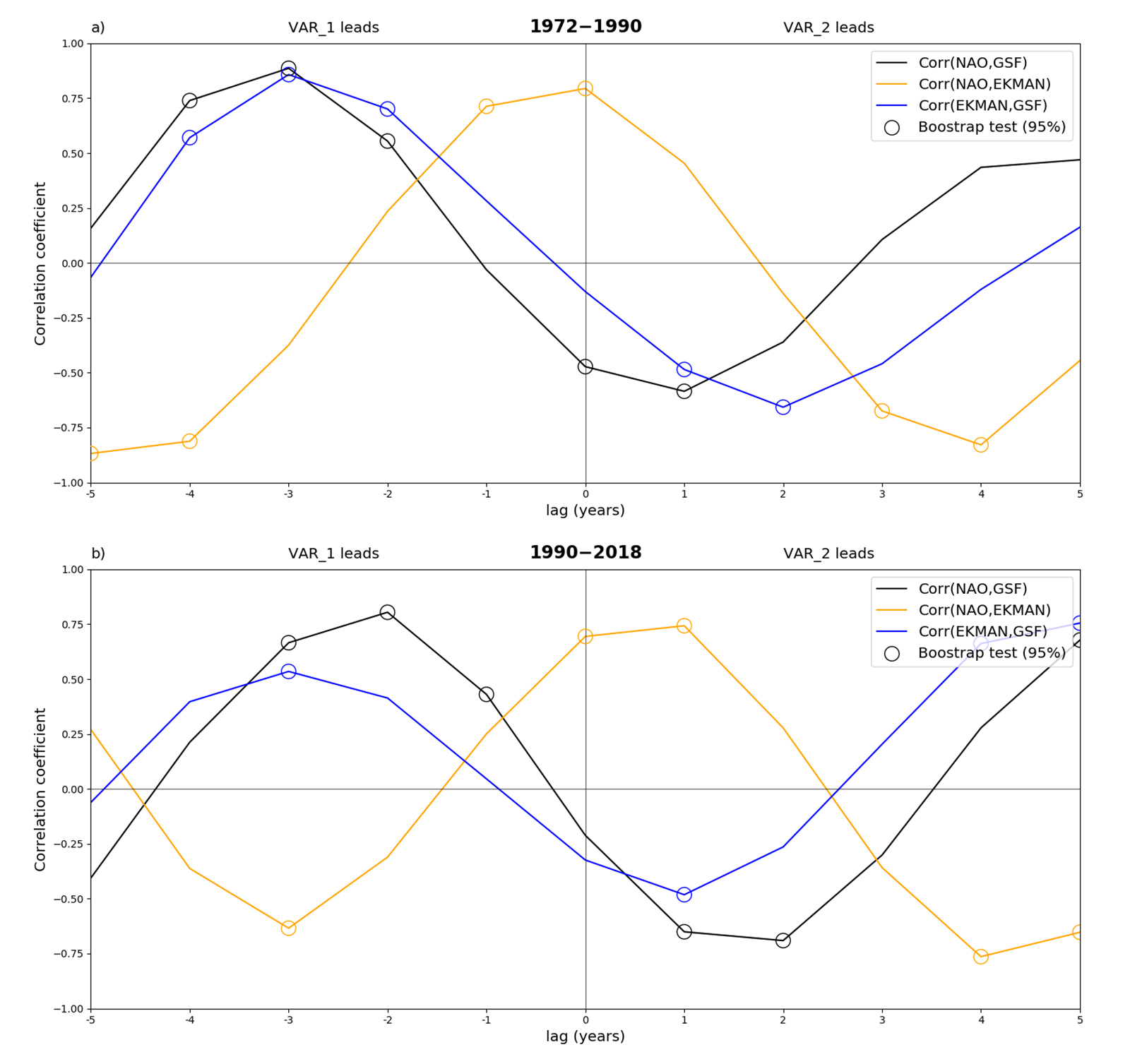}
\caption{Lead-lag cross-correlation between the band-pass filtered NAO, GSF and Ekman indices during 1972--1990 (a) and 1990--2018 (b): NAO--GSF (black), NAO--Ekman (yellow), Ekman--GSF (blue). The first (second) variable leads in the left (right) portion of the plot. The Ekman index has been defined as the meridional Ekman velocity averaged over the GSF area (50$^{\circ}$--68$^{\circ}$W; 35$^{\circ}$--38$^{\circ}$N). The circles highlight the lead-lag cross-correlations that are statistically significant at the 95\% confidence level.
}
\label{fig:ekman}
\end{figure}

\begin{figure}[ht!]
    \centering
    \includegraphics[width=\textwidth]{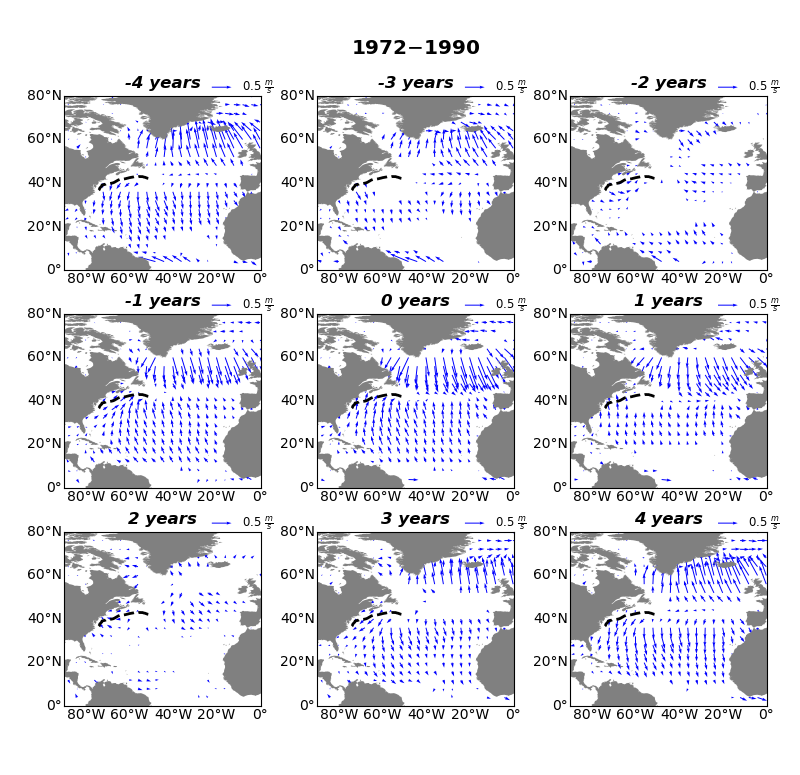}
\caption{Lead--lag linear regression coefficients for band-pass filtered Ekman transport (m s$^{-1}$; arrows) anomalies on the band-pass filtered NAO index in the winter season during 1972--1990. Only regression coefficients for band-pass filtered Ekman transport anomalies that are statistically significant at the 90\% confidence level are shown. The NAO leads (lags) at negative (positive) lags, as in Figure \ref{fig:ekman} (orange line). The winter climatological position of the GSF is indicated by the black dashed line.}
\label{fig:ekman_1972-1990}
\end{figure}

\begin{figure}[ht!]
    \centering
    \includegraphics[width=\textwidth]{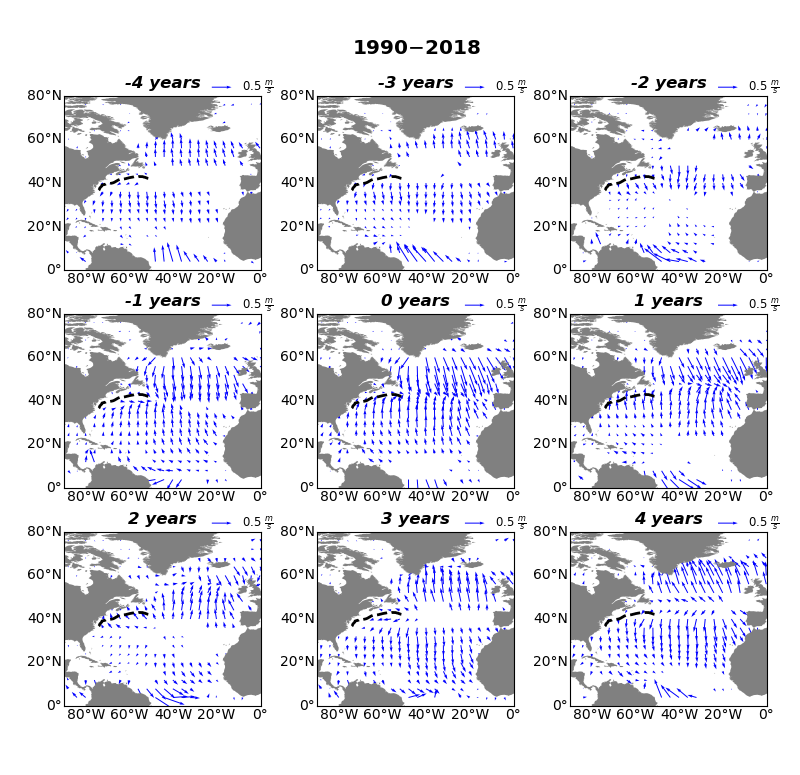}
\caption{Same as Figure \ref{fig:ekman_1972-1990} but during 1990--2018.}
\label{fig:ekman_1990-2018}
\end{figure}

The lead--lag cross-correlations between the band-pass filtered NAO and Ekman indices show maximum values at lag-0 in both intervals (Figure \ref{fig:ekman}). Specifically, a positive (negative) NAO phase induces northward (southward) Ekman transport anomalies over the GS area. The effect of the NAO forcing on the Ekman transport extends over the whole North Atlantic basin, with particularly intense anomalies in the SPG and the subtropical North Atlantic (Figure \ref{fig:ekman_1972-1990}, Figure \ref{fig:ekman_1990-2018}). This is consistent with previous studies showing the instantaneous impact of the NAO-related wind-stress on the wind-driven circulation in the North Atlantic \citep{visbeck2003,deser2010}. Since the wind-driven oceanic circulation adjusts quickly to the wind forcing, the negative peak in the NAO--Ekman transport cross-correlation when the NAO leads the Ekman transport can be understood as an artifact of the periodicity due to the use of the band-pass filtering applied over the decadal range. 

The fast response of the wind-driven oceanic circulation to the NAO forcing also determines that the Ekman transport leads the GSF shifts by the same number of years as the NAO. This is true for the 1972--1990 interval, during which the Ekman transport--GSF and NAO--GSF connections reveal the same lead--lag covariability pattern (Figure \ref{fig:ekman}a). It is not completely so during 1990--2018, when the anomalous Ekman transport anticipates the NAO forcing by 1 year (Figure \ref{fig:ekman}b). This aspect shows that the variability of Ekman transport over the GS area is also influenced by other factors besides the NAO. Apart from that, the Ekman transport--GSF cross-correlation shows that the anomalous oceanic circulation forced by the anomalous wind-stress over the GSF area cannot directly account for the [2--3]-year time lag in the cross-correlation between the GSF and NAO indices.

\begin{figure}[ht!]
    \centering
    \includegraphics[width=\textwidth]{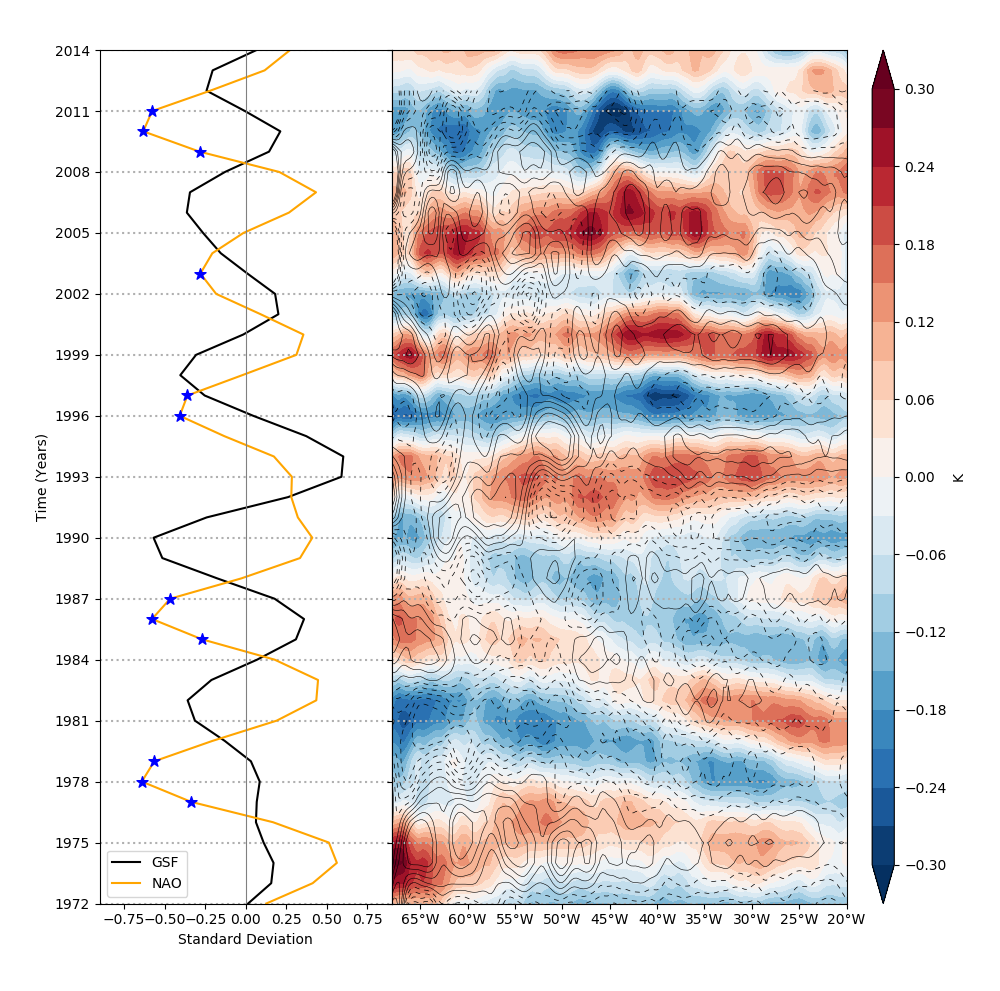}
\caption{Left panel: Band-pass filtered winter GSF (black) and NAO (orange) time series between 1972--2014. Blue stars represent years in which the NAO falls in the lower tercile category. These years are used to perform the composite analysis described in section \ref{discussion_decadal}. Right panel: Hovmöller diagram \citep{hovmoller1949} of the band-pass filtered SST (K; color shading) and SSH  anomalies (cm; black contours; contours every 0.5 cm from -1.75 cm) averaged in the 35$^{\circ}$--38$^{\circ}$N latitudinal range during 1972--2014. It is specified that 6 years are lost at each border of the time series because of the application of the Lanczos filter (see detail in section \ref{data_decadal}). For this reason, the y-axis of the plot ranges during 1972--2014 and not during 1972--2018, that is where the GSF and NAO covary at the decadal timescale (Figure \ref{fig:xwt_ERA5}). The SSH time series are even shorter because the original SSH data are provided until 2018.}
\label{fig:rossby}
\end{figure}

\begin{figure}[ht!]
    \centering
    \includegraphics[width=\textwidth]{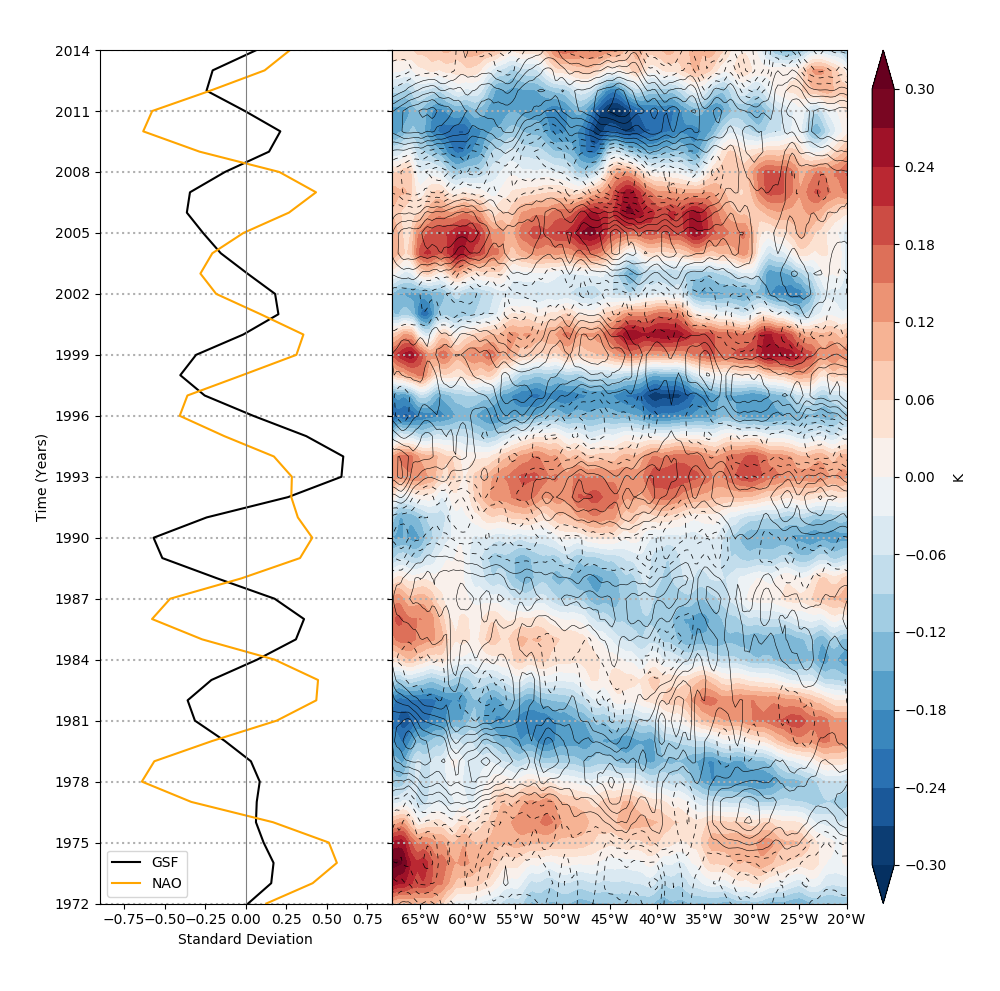}
\caption{Left panel: Band-pass filtered winter GSF (black) and NAO (orange) time series between 1972--2014. Right panel: Hovmöller diagram \citep{hovmoller1949} of the band-pass filtered SST (K; color shading) and wind-stress curl anomalies (10$^{-7}$ N m$^{-3}$; black contours; contours every 0.1 10$^{-7}$ N m$^{-3}$ from -1 10$^{-7}$ N m$^{-3}$) averaged in the 35$^{\circ}$--38$^{\circ}$N and 35$^{\circ}$--37$^{\circ}$N latitudinal range during 1972--2014, respectively. It is specified that 6 years are lost at each border of the time series because of the application of the Lanczos filter (see detail in section \ref{data_decadal}). For this reason, the y-axis of the plot ranges during 1972--2014 and not during 1972--2018, that is where the GSF and NAO covary at the decadal timescale (Figure \ref{fig:xwt_ERA5}).}
\label{fig:rossby_SST_curl}
\end{figure}

\begin{figure}[ht!]
    \centering
    \includegraphics[width=\textwidth]{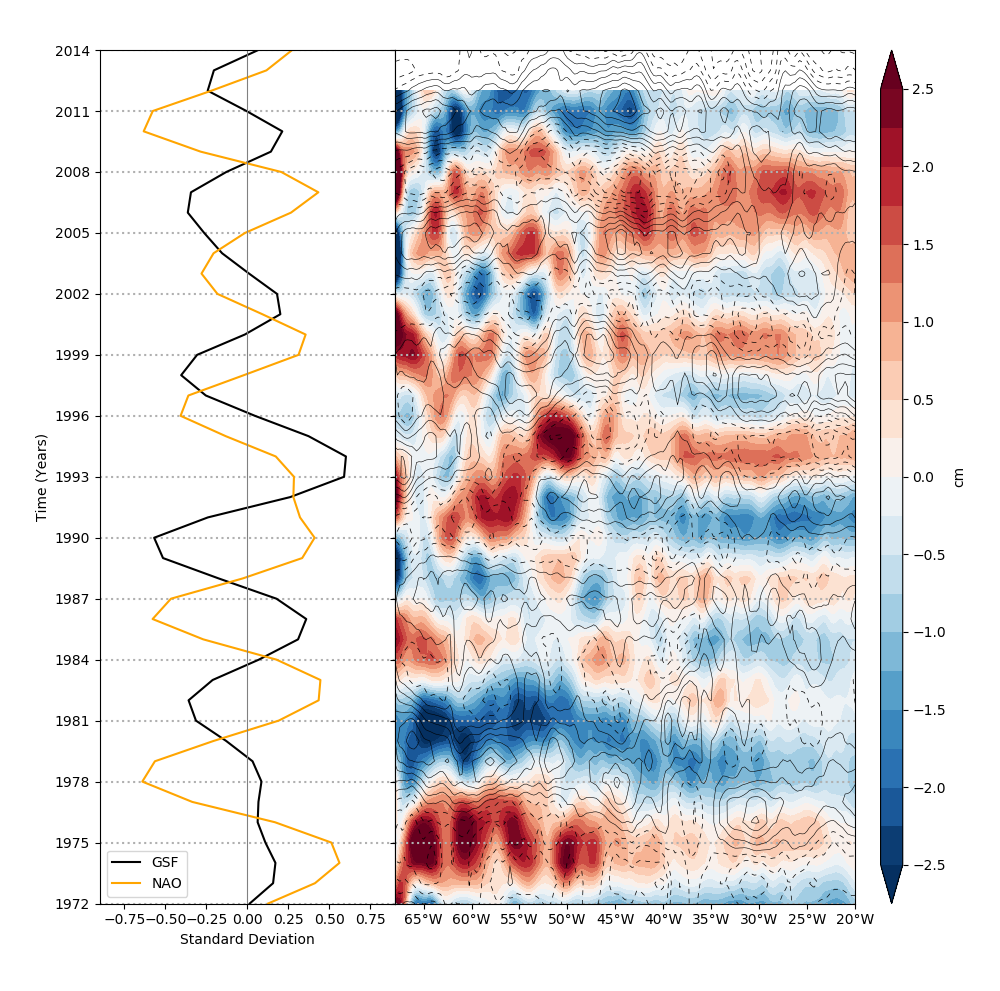}
\caption{Left panel: Band-pass filtered winter GSF (black) and NAO (orange) time series between 1972--2014. Right panel: Hovmöller diagram \citep{hovmoller1949} of the band-pass filtered SSH (cm; color shading) and wind-stress curl anomalies (10$^{-7}$ N m$^{-3}$; black contours; contours every 0.1 10$^{-7}$ N m$^{-3}$ from -1 10$^{-7}$ N m$^{-3}$) averaged in the 35$^{\circ}$--38$^{\circ}$N and 35$^{\circ}$--37$^{\circ}$N latitudinal range during 1972--2014, respectively. It is specified that 6 years are lost at each border of the time series because of the application of the Lanczos filter (see detail in section \ref{data_decadal}). For this reason, the y-axis of the plot ranges during 1972--2014 and not during 1972--2018, that is where the GSF and NAO covary at the decadal timescale (Figure \ref{fig:xwt_ERA5}). The SSH time series are even shorter because the original SSH data are provided until 2018.}
\label{fig:rossby_SSH_curl}
\end{figure}

The Hovmöller diagram \citep{hovmoller1949} of the band-pass filtered SST and SSH anomalies averaged in the 35$^{\circ}$--38$^{\circ}$N latitudinal range shows Rossby wave-like structures propagating from the eastern to western North Atlantic before 1990 (Figure \ref{fig:rossby}). The signal is particularly clear in the SST anomalies, whereas it appears noisier in the SSH anomalies. Positive (negative) SST and SSH anomalies in eastern North Atlantic are associated with negative (positive) wind-stress curl anomalies in the same Atlantic region (20$^{\circ}$--35$^{\circ}$W; Figure \ref{fig:rossby_SST_curl} and Figure \ref{fig:rossby_SSH_curl}). Furthermore, negative (positive) wind-stress curl anomalies in the eastern North Atlantic are associated with the positive (negative) NAO phase. This is consistent with previous studies showing that baroclinic Rossby waves can be excited by the anomalous Ekman convergence/divergence due to large-scale atmospheric circulation anomalies, such as the ones associated with the NAO \citep{anderson1975, sturges1995,sturges1998,fu2002,zhang2016,arthun2020,kowalski2022}. In this context, it has been shown that the momentum transfer between the NAO and the ocean is particularly intense around 30$^{\circ}$W \citep{visbeck1998,esselborn2001}. Consistently, the SST and SSH anomalies show maximum values between 25$^{\circ}$--35$^{\circ}$W in the period where there is evidence of Rossby wave propagation. 

Once excited in the eastern North Atlantic, the positive (negative) SST and SSH anomalies reach the western North Atlantic about 2--3 years later and the GSF shifts northward (southward). Taking into account the distance between 30$^{\circ}$W (as reference for the NAO-related perturbation initiating the Rossby waves) and 59$^{\circ}$W (the center of the longitudinal range used to define the GSF latitudinal position), the [2--3]-year time lag corresponds to a phase speed of about 2.6--4 cm/s. These values are consistent with the phase speeds of baroclinic Rossby waves in extratropical North Atlantic proposed by previous studies \citep[0,93--4,17 cm/s;][]{cipollini1997,cipollini2001,osychny2004,decoetlogon2006,watelet2020}. 

Results above support the idea of the NAO forcing the GSF latitudinal position through Rossby wave excitation, but only during 1978--1990. As specified before, the spectral features of the decadal NAO--GSF covariability in this period are different compared to those from 1990 onwards (Figure \ref{fig:xwt_ERA5}b). It is suggested here that the differences in the spectral features of the decadal NAO--GSF interaction before and after 1990 may be induced by the non-stationarity of the Rossby wave propagation. The evidence of Rossby wave propagation for a limited period of time is consistent with results by \citet{sasaki2011}, showing westward wave propagation only during 1965--1985. After 1990, the SST anomalies show no propagation, with anomalies changing their sign in agreement with the NAO forcing. The reasons behind the absence of Rossby wave propagation after 1990 are unclear.

However, it is interesting to highlight that the NAO index shows a peak of variability on decadal timescale that is stronger during 1970--1990 (Figure \ref{fig:cwt_nao_ERA5}). The changes in the character of the surface atmospheric forcing in time could affect the way the ocean responds to that forcing and thus the Rossby wave propagation. Previous studies have already shown that decadal changes in the NAO-related wind patterns occurring around 1990 can affect the relationship between the NAO and the interannual sea level variability along the coast of North America \citep{kenigson2018,diabate2021}. Furthermore, it should be mentioned that the atmosphere undergoes decadal variability also projecting on other patterns largely orthogonal or/and independent to the NAO, which may also affect the oceanic circulation in the North Atlantic.

Results above also show that, unlike the Ekman transport, the Rossby wave propagation mechanism can be important in setting the time lag between the NAO and the GSF position changes, particularly before 1990. This is supported because the 3-year time lag between the NAO and GSF shifts is about the same as the time required for SST and SSH anomalies generated in the central and eastern North Atlantic to propagate westward across the basin (Figure \ref{fig:rossby}).  

\begin{figure}[ht!]
    \centering
    \includegraphics[width=\textwidth]{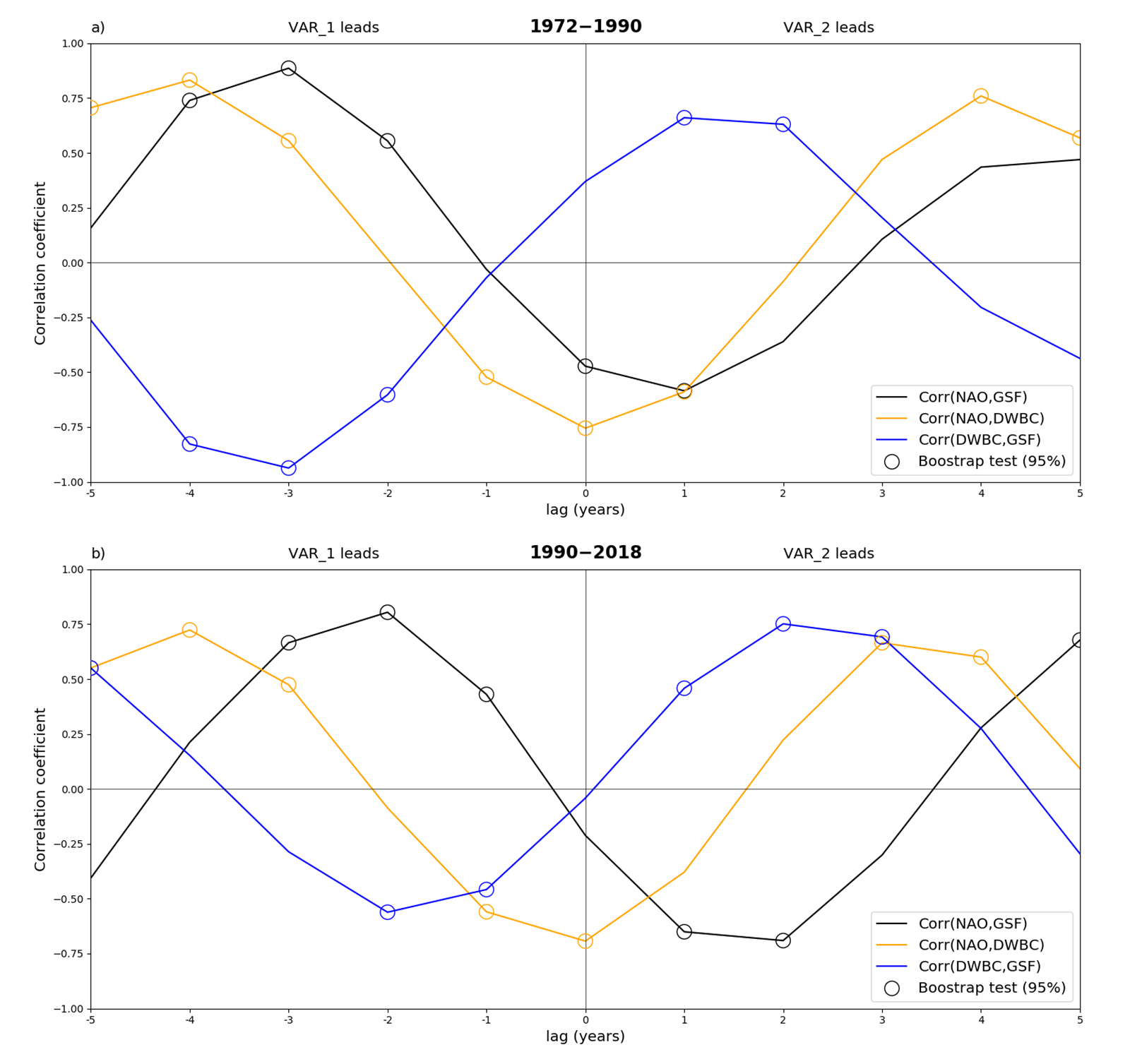}
\caption{Lead--lag cross-correlation between the band-pass filtered NAO, GSF and DWBC indices during 1972--1990 (a) and 1990--2018 (b): NAO--GSF (black), NAO--DWBC (yellow), DWBC--GSF (blue). The first (second) variable leads in the left (right) portion of the plot. The DWBC index has been defined as the total oceanic westward transport in the 43.5$^{\circ}$--45$^{\circ}$N latitudinal range and 1000--3000 m depth (black dashed box). The circles highlight the lead-lag cross-correlations that are statistically significant at the 95\% confidence level.
}
\label{fig:dwbc}
\end{figure}

\begin{figure}[ht!]
    \centering
    \includegraphics[width=\textwidth]{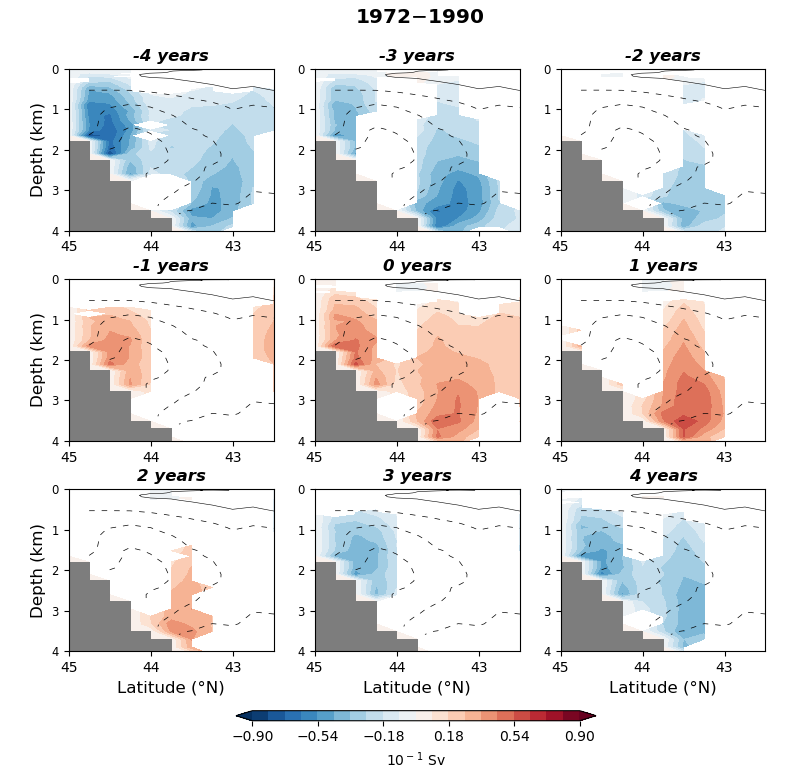}
\caption{Lead--lag linear regression coefficients for band-pass filtered zonal transport (10$^{-1}$ Sv; color shading) anomalies on the band-pass filtered NAO index in the winter season during 1972--1990. The zonal transport anomalies have been averaged in the 50$^{\circ}$--68$^{\circ}$W longitudinal range before to perform the linear regression. Only regression coefficients that are statistically significant at the 90\% confidence level are shown. Grey contours indicate the winter climatology of the zonal oceanic transport every 1 10$^{-1}$ Sv from -3 10$^{-1}$ Sv. The NAO leads (lags) at negative (positive) lags, as in Figure \ref{fig:dwbc} (orange line).
}
\label{fig:dwbc_1972-1990}
\end{figure}

\begin{figure}[ht!]
    \centering
    \includegraphics[width=\textwidth]{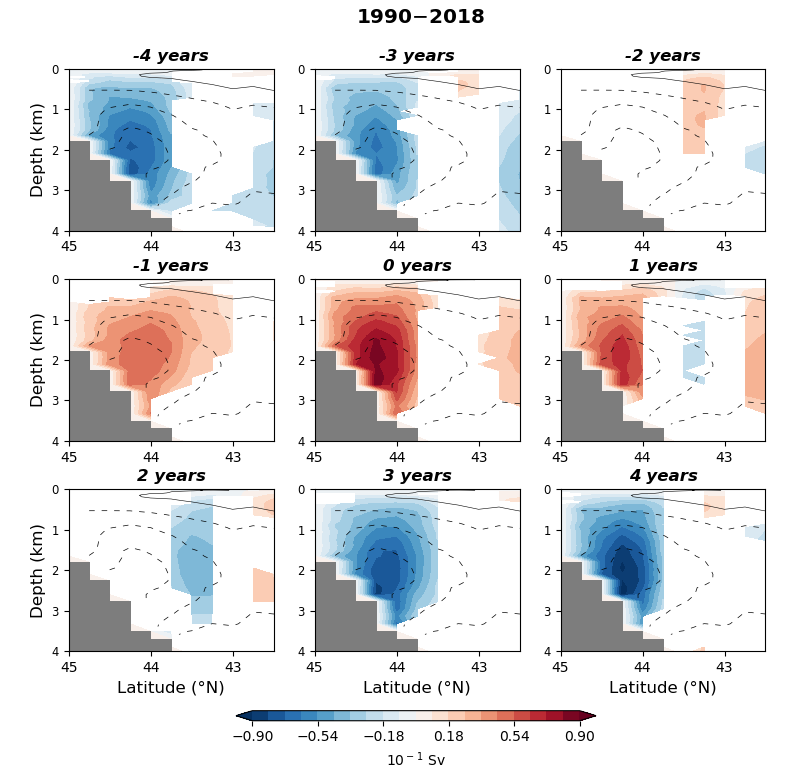}
\caption{Same as Figure \ref{fig:dwbc_1972-1990} but during 1990--2018
}
\label{fig:dwbc_1990-2018}
\end{figure}

\begin{figure}[ht!]
    \centering
    \includegraphics[width=\textwidth]{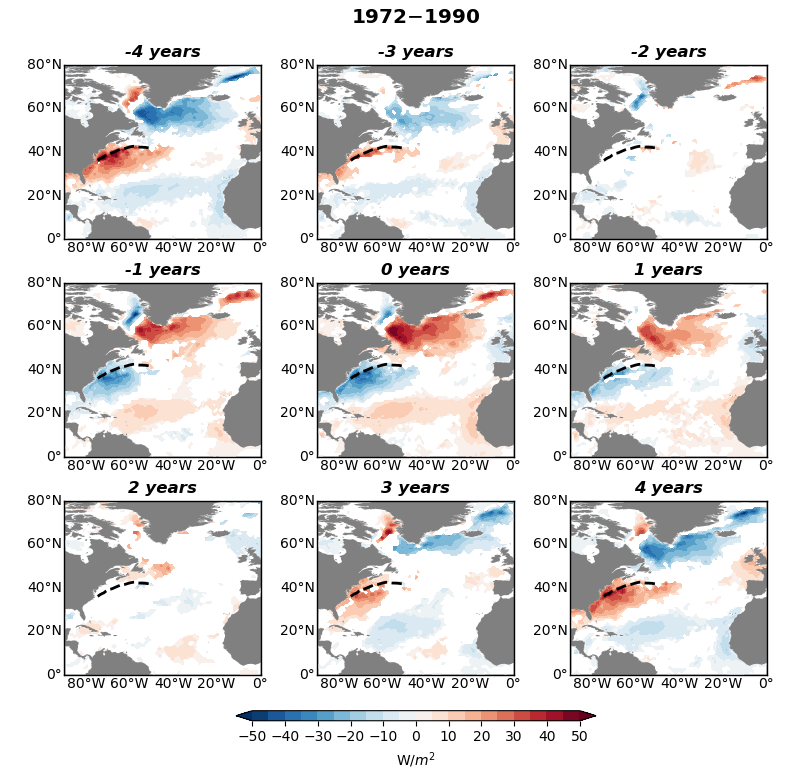}
\caption{Lead--lag linear regression coefficients for band-pass filtered SHF (W m$^{-2}$; color shading) anomalies on the band-pass filtered NAO index in the winter season during 1972--1990. Only regression coefficients that are statistically significant at the 90\% confidence level are shown. The NAO leads (lags) at negative (positive) lags, as in Figure \ref{fig:dwbc}. The winter climatological position of the GSF is indicated by the black dashed line.
}
\label{fig:shf_1972-1990}
\end{figure}

\begin{figure}[ht!]
    \centering
    \includegraphics[width=\textwidth]{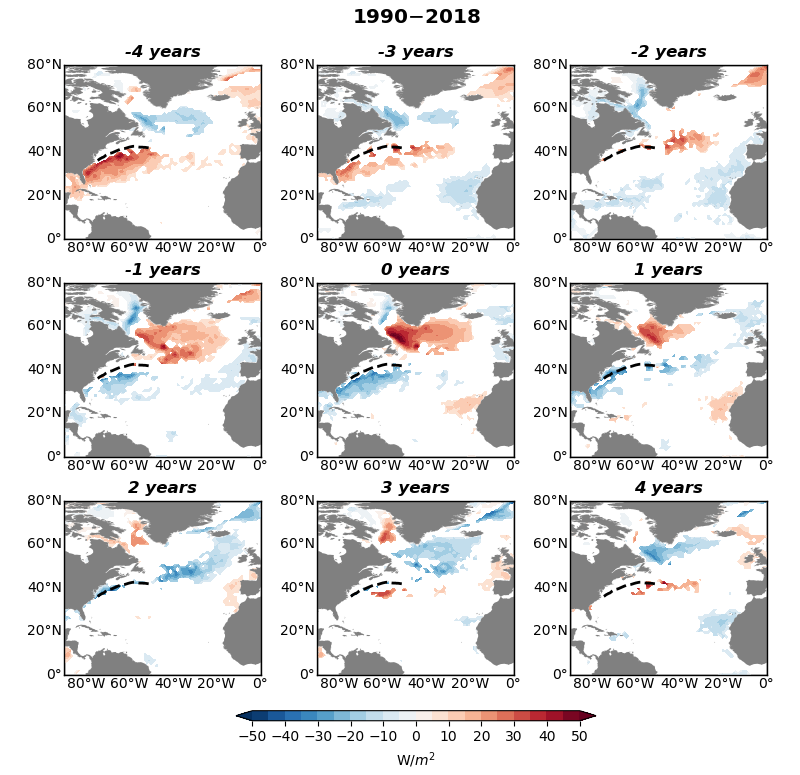}
\caption{Same as Figure \ref{fig:shf_1972-1990} but during 1990--2018.
}
\label{fig:shf_1990-2018}
\end{figure}

Finally, the NAO can influence the GSF path through buoyancy forcing over the LS, affecting the export of the LSW from the LS and thus the southwestward flow along the Slope Sea \citep{joyce2000,rossby2000,hameed2004,zhang2007}. Figure \ref{fig:dwbc} shows that maximum positive correlation between the band-pass filtered NAO and DWBC indices is found when the former leads the latter by 4 years. This is true during both 1972--1990 and 1990--2018. Specifically, the positive (negative) NAO enhances (reduces) the westward volume transport along the Slope Sea after 4 years, with anomalies extending to a 3000 m depth and reaching the greatest amplitude at the DWBC depth (1000--3000 m depth; Figure \ref{fig:dwbc_1972-1990}, Figure \ref{fig:dwbc_1990-2018}). These deep water transport anomalies can be associated with the heat loss (gain) over the LS during a positive (negative) NAO phase (Figure \ref{fig:shf_1972-1990}, Figure \ref{fig:shf_1990-2018}), which is expected to affect the formation (through deep convection) and, ultimately, the export of the LSW via the DWBC. In this context, it is specified that the anomalies in the zonal flow along the Slope Sea are more intense during 1990--2018 (Figure \ref{fig:dwbc_1990-2018}). Furthermore, statistically significant anomalies appear at depths greater than 3000 m during 1972--1990 (Figure \ref{fig:dwbc_1972-1990}). These water masses are generally referred to as Overflow Water, i.e. the densest branch of the DWBC, and previous studies have shown that they are formed in the Greenland and Norwegian Seas \citep{worthington1976,pickart1992}. Therefore, the NAO-related buoyancy forcing over the LS is expected to play no role for the transport anomalies deeper than 3000 m.

The 4-year time lag between the response of the DWBC and the NAO forcing is consistent with the range of values shown in previous observational and modelling studies. \citet{zhang2006} have shown that changes in the deep convection in the LS lead anomalies in the NRG intensity by 4 years, that is the time for the DWBC to travel from the LS to the Grand Banks. In agreement with this, \citet{pena2011} have shown that anomalies in the uppermost layer of the DWBC (500--1000 m depth approximately) take about  4--7 years to reach the GS area. Finally, \citet{georgiou2020} have shown that most of the dense water formed by deep convection in the LS exits the subpolar basin through indirect routes that involve exchanges between the boundary currents and the ocean interior. Such routes are associated with export timescales that are usually between 1--6 years. 

Figure \ref{fig:dwbc} shows a negative peak in the cross-correlation between the DWBC and GSF indices when the former leads the latter by 3 years during 1972--1990 and by 2 years during 1990--2018. This means that the GSF shifts northward (southward) when the westward DWBC transport is reduced (intensified). This is consistent with the fact that the stronger (weaker) DWBC transport induced by the NAO forcing is expected to be associated with a stronger (weaker) bottom vortex stretching along the Slope Sea and thus with a more (less) intense NRG \citep[not shown;][]{zhang2006,zhang2007}. The equivalent barotropic nature of the anomalous westward flow along the Slope Sea (Figure \ref{fig:dwbc_1972-1990}, Figure \ref{fig:dwbc_1990-2018}) is the signature of the changes in the NRG. Thus, the variability of the deep-ocean circulation (DWBC) and associated changes in the NRG also manifests as variability of the upper-ocean circulation \citep[shifts of the GSF;][]{pena2008}. In this context, the fact that the anomalous DWBC transport leads the GSF shifts by 3 years during 1972--1990, and by 2 years during 1990--2018, suggests that the DWBC--GSF relationship is different for the two time periods. Furthermore, it suggests that the anomalous DWBC transport forced by the NAO forcing can directly account for the time lag in the cross-correlation between the NAO and GSF indices only after 1990. Indeed, \citet{zhang2006} have shown that the more (less) intense NRG due to the stronger (weaker) DWBC transport leads the northward (southward) GSF shift by 1 year. Being the DWBC--GSF correlation significant when the former leads the latter by [1--2]-year (and not only by 2 years), the maximum changes in the GSF index are expected to lag the DWBC as well as the NAO index by 2 years. This is not the case before 1990, since the correlation between the GSF index and the DWBC and NAO indices peak when the latters lead the former by [2--4]-year. 

\section{Discussion}
\label{discussion_decadal}

 Results described in the previous section show that, in the ERA5 dataset, the NAO--GSF interaction is non-stationary over the 1950--2020 period. The NAO and the GSF indices covary in the decadal timescales but only during 1972--2018 (Figure \ref{fig:xwt_ERA5}). Over this time period, we have also found that the year 1990 marks a significant discontinuity in the decadal covariability of these two indices. Specifically, the NAO leads the GSF shifts by about 3 years before 1990 and by about 2 years after 1990 (Figure \ref{fig:nao-gsf}). 
It is here noted that other studies have also discussed a discontinuity in the air--sea interaction over North Atlantic occurring around 1990 \citep{andres2013,kenigson2018,diabate2021}. Specifically, the changes in the nature of NAO-related wind-stress during the 20th century have been suggested to cause a decadal shift of the relationship between the NAO and the interannual sea level variability along the eastern coast of North America. These results further confirm the existence of discontinuity in the decadal air--sea interaction over the GS area during the 1990s. Finally, Figure \ref{fig:xwt_ERA5} shows that the NAO and GSF indices covary at multiannual timescales, but only for a limited period of time (2005--2015)
 
Based on these pieces of evidence, we have assessed several mechanisms through which the NAO may be forcing the GSF path on the decadal timescale. Specifically, we have inferred the responses of wind-driven Ekman transport, Rossby waves and the DWBC transport to the NAO through lagged correlation analyses. Results suggest that the response of Ekman transport to the NAO cannot directly account for the lagged response of the GSF to the NAO. Differently, the response of DWBC transport to the NAO forcing and the Rossby wave propagation mechanism appear to be consistent with the time lag between the NAO and the GSF response, but only after and before 1990, respectively. 

\begin{figure}[ht!]
    \centering
    \includegraphics[width=\textwidth]{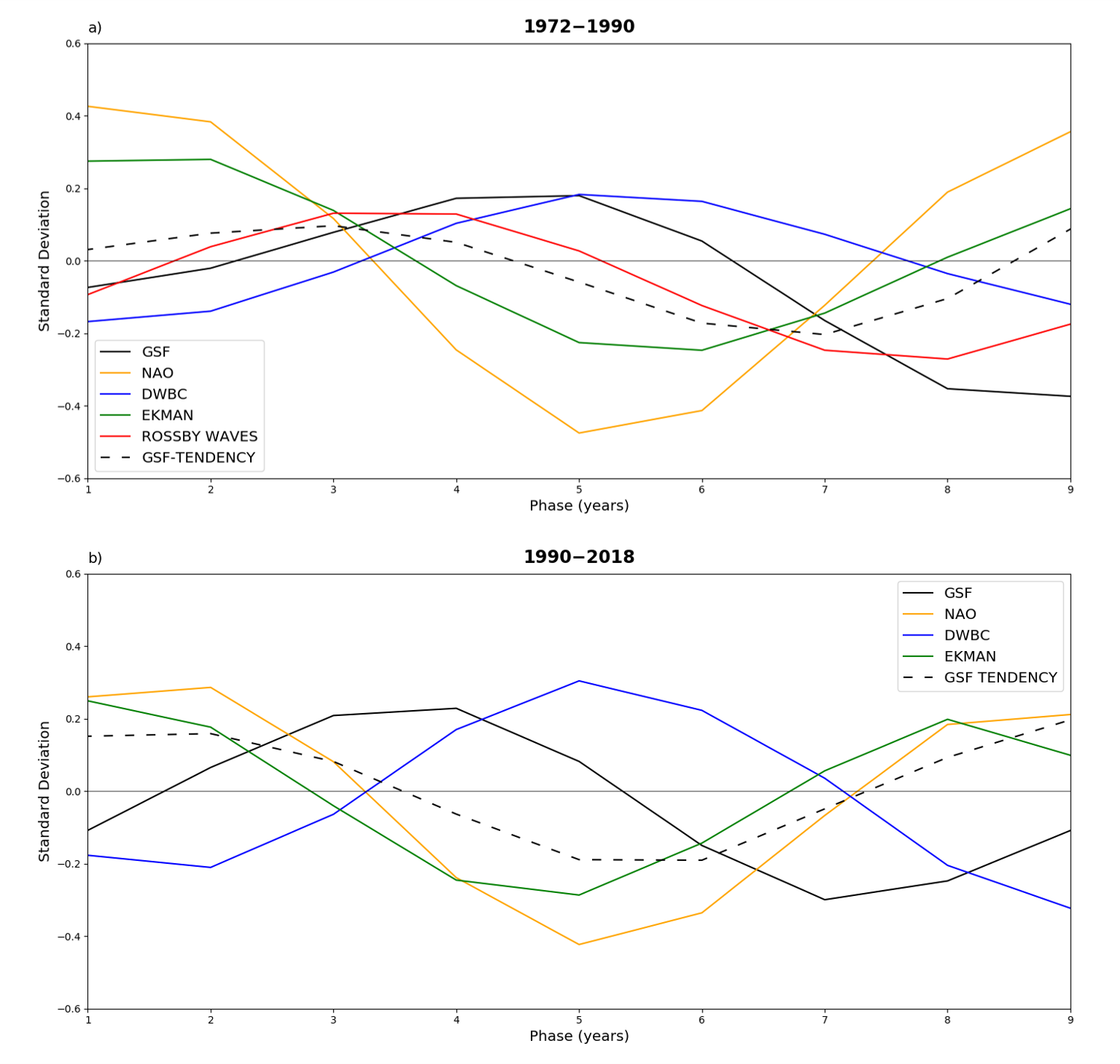}
\caption{a) The average evolution of the GSF shifts (solid black line), GSF tendency (dashed black line), the NAO (orange line), the Ekman index (green line), the DWBC index (blue line) and the Rossby waves index during 1972--1990. b) The average evolution of the GSF shifts (solid black line), GSF tendency (dashed black line), the NAO (orange line), the Ekman index (green line) and the DWBC index (blue line) during 1990--2018. The average evolution of the selected indices during the two time periods has been defined through a lead--lag composite analysis performed for years (blue stars in Figure \ref{fig:rossby}) when the NAO is smaller than its lower tercile.
}
\label{fig:evolution}
\end{figure}

In order to illustrate how these mechanisms interact to set the time lag between the NAO forcing and the GSF response, a lead--lag composite analysis has been performed along the life cycle of the NAO--GSF decadal fluctuation. The analysis has been performed for years (blue stars in Figure \ref{fig:rossby}) when the NAO is smaller than its lower tercile (equal to about -0.22), so as to derive the average evolution of the selected indices associated with the NAO forcing. The average evolution of the selected indices has been obtained by averaging their values year-by-year in a time range of 9 years centered on those below the lower tercile of the NAO. The analysis has been applied for the 1972--1990 and 1990--2018 time periods, separately. In this way, the average evolution of the GSF shifts in response to NAO forcing has been assessed when there is evidence and no evidence of Rossby wave propagation. For the 1972--1990 time period, a Rossby wave index has been defined as the band-pass filtered SST anomalies averaged in the 50$^{\circ}$--60$^{\circ}$W longitudinal range and then normalized. The SST anomalies have been used to define the index instead of the SSH anomalies because they are less noisy closer to the GS region. It is here specified that the size of the selected sample is too small to draw definitive conclusions from the composite analysis. Despite this limitation, the composite analysis shown in Figure \ref{fig:evolution} helps illustrate more clearly the previously identified relationships between the NAO, GSF, Ekman transport, Rossby waves, and DWBC transport. Figure \ref{fig:evolution}a and Figure \ref{fig:evolution}b show the average evolution of the GSF shifts in response to the NAO forcing during 1972--1990 and 1990--2018, respectively.

The composite in Figure \ref{fig:evolution}a shows that a positive NAO phase is associated with positive (northward) Ekman transport anomalies over the GSF region (Figure \ref{fig:ekman}a, Figure \ref{fig:ekman_1972-1990}), which tend to shift the GSF northward (year 1--3). At the same time, negative (or eastward) anomalies in the westward DWBC transport appear along the Slope Sea. These anomalies can be interpreted as the delayed response of the deep oceanic circulation to the previous negative NAO phase \citep{joyce2000}. Indeed, the negative NAO phase is associated with less intense zonal winds and then with downward SHF anomalies over the LS \citep[Figure \ref{fig:shf_1972-1990};][]{eden2001,omrani2022}. The air--sea heat exchanges over the LS weaken the deep convection and the export of the LSW via the DWBC. The changes in the DWBC transit along the Slope Sea 4 years after the NAO peak and the GSF shifts northward (Figure \ref{fig:dwbc}a, Figure \ref{fig:dwbc_1972-1990}). Indeed, the weakened DWBC transport is expected to contract the NRG via the anomalous bottom vortex stretching \citep{zhang2007}. Finally, the GSF is pushed further northward by the Rossby waves excited by the NAO forcing on the eastern North Atlantic and reaching the GSF area after about 2 years (Figure \ref{fig:rossby}; year 3--4). Due to the combined effects of all the mechanisms, the tendency of the GSF index shows a positive trend during the positive NAO phase (year 1--3), reaching its maximum value concurrently with the arrival of the Rossby waves in the GSF area.

At the same time, the NAO reverses the sign and the tendency of the GSF index starts to decrease (year 3--4). However, the GSF still shifts northward and reaches the maximum northern position 1 year later, with a lag of about 3 years respect the positive NAO peak (year 4--5). This is mainly due to the persisting effect of the Rossby waves that counteract the opposite effect of the anomalous Ekman and DWBC transport during the onset of the negative NAO phase. The negative NAO phase is associated with negative (southward) Ekman transport anomalies over the GSF region, which tend to shift the GSF southward (year 4--6). Contemporarily, the positive (westward) anomalies in the DWBC transport induced by the previous positive NAO phase (4 years before) transit along the Slope Sea, pushing further southward the GSF (year 4--6). Finally, the GSF moves further southward because of the sea level anomalies induced by the Rossby waves excited by the negative NAO about 2 years before (year 7--8). Consistently, the tendency of the GSF index shows a negative trend during the negative NAO phase (year 3--7), reaching its minimum value concurrently with the arrival of the Rossby waves in the GSF area. Thanks to their persisting effect, the GSF reaches the maximum southern position 1--2 years later, with a lag of about 3 years respect the negative NAO peak (year 8--9). At this point, the NAO reverses the sign and the cycle starts again.

The phase evolution of the GSF shifts in response to the decadal NAO forcing during 1972--1990 (Figure \ref{fig:evolution}a) is quite comparable with the one during 1990--2018 (Figure \ref{fig:evolution}b). However, the GSF shifts lag the NAO forcing by 3 years during 1972--1990, whereas by 2 years during 1990--2018 (Figure \ref{fig:nao-gsf}, Figure \ref{fig:evolution}). It is here suggested that the differences in the lead--lag NAO--GSF relationship between the two time periods are induced by the non-stationarity of Rossby wave propagation. Indeed, the Rossby waves excited by the NAO forcing on the eastern North Atlantic require about 2--3 years to reach the GS area. This means that the sea level anomalies associated with the wave propagation reach the GSF area when the anomalous Ekman and DWBC transports are in the transition phase from positive to negative values or from negative to positive values (year 3--4; year 7--8). Thus, the tendency of the GSF shifts is extended by about one year. On the other hand, in the absence of Rossby wave propagation, the tendency of the GSF shifts is maximum when the anomalous Ekman and DWBC transport reach their greatest intensity and it immediately reverses once the Ekman and DWBC transport change sign. In this context, the GSF reaches its maximum northern/southern position 2 years after the peak in the GSF tendency, with a lag of about 2 years respect the NAO peak. This fact is interpreted as the combined effects of anomalous Ekman and DWBC transport. Thus, results above suggest that the Rossby wave propagation delays the shift of the front, changing its lag respect the NAO forcing during the two different periods of time.

In the context of the GSF shifts evolution described above, it is specified that the drivers of the decadal NAO variability are outside the scope of present PhD thesis. However, the reversal of the NAO phases (Figure \ref{fig:evolution}; year 3--4 and 7--8) seems to be consistent with the delayed negative feedback on the SST caused by changes in the AMOC, as proposed by previous studies \citep[e.g.][]{reintges2017}. Indeed, the positive NAO phase causes enhanced heat losses over the SPG \citep{eden2001,omrani2022}, which in turn leads to the delayed intensification of the meridional overturning through an increase of the DWBC transport (Figure \ref{fig:dwbc}). The AMOC intensification leads to a warmer SPG (through enhanced meridional heat transport), determining a phase reversal of the NAO in response to the reduced large-scale meridional SST gradient associated with the anomalously warm SPG \citep{marshall2001,delworth2017}. The reverse is true for the negative NAO phase. Furthermore, other processes could play a role in the decadal NAO variability associated with the GSF shifts. For example, there are some hints about the possible role played by the North Pacific sector on the decadal NAO variability. Indeed, large-scale SST and SLP anomalies are found in the North Pacific sector in association with the GSF shifts (not shown). Previous studies have shown that the North Pacific--North Atlantic interaction can account for a great percentage of the variance over the North Atlantic sector \citep{bjerknes1966,honda2001a,honda2001b}. It is speculated that such interaction could have also a role on the decadal NAO variability.

To summarize, results in this part of the PhD thesis suggest that the decadal NAO variability can affect the decadal GSF shifts through changes in deep convection and thus by affecting the DWBC transport (Figure \ref{fig:dwbc}). The peak in the NAO--DWBC cross-correlation when the former leads the latter by 4 years is consistent with the transit time for anomalies in the LS to reach the GS area via the DWBC as suggested by previous observational and modeling studies \citep{pena2011,georgiou2020}. Furthermore, the role of the DWBC transport on the GSF path is suggested by the equivalent barotropic nature of the anomalous zonal flow along the Slope Sea \citep[Figure \ref{fig:dwbc_1972-1990}, Figure \ref{fig:dwbc_1990-2018};][]{pena2008}. However, the role of the LC cannot be excluded. The positive NAO is expected to cause a northward expansion of the STG and a reduction of the cold water sources along the Slope Water \citep{peterson2017,holliday2020,new2021}. In agreement with this, the positive NAO phase is associated with reduced westward flow in the upper 500 m depth. The reverse is true for the negative NAO phase. The fast wind-driven oceanic circulation response to wind forcing could explain why the correlation between the NAO and the anomalous zonal flow at the Slope Water depth peaks at lag-0. It is here specified that the lead--lag cross-correlation between the NAO and the anomalous zonal flow at the Slope Water depth is strictly comparable to Figure \ref{fig:dwbc}, being the zonal flow along the Slope Sea equivalent barotropic. Hence, the changes in the zonal transport along the Slope Sea could be interpreted as the fast response of the LC and the lagged response of the DWBC to the NAO forcing. However, the barotropic nature of the flow along the Slope Sea makes it difficult to differentiate the role of the two currents.

Before closing this section, the possible causes of the non-stationarity in the NAO--GSF relation and associated mechanisms will be discussed. Firstly, we have to consider that the quality of observational data has greatly improved during the last decades. In particular, the absence of the decadal NAO--GSF covariability before 1972 could be an artifact due to a lack of satellite observations during that time period rather than an effective independence between the two time series. The quality of data could also cause the discrepancies in the lead--lag NAO--GSF relationship before and after 1990. The satellite observations provide highly resolved SST data, which are able to properly capture GSF shifts of approximately 50--100 km. This was hardly possible with observational datasets before 1979. However, the limited length of satellite observations can question the spectral features of the NAO--GSF covariability, especially over decadal and multidecadal timescales. In this context, the NAO--GSF covariability described in the current work could be seen as a part of a damped oscillation working on longer timescales, in which periods of weak/non-covariability alternate periods of strong covariability.

Furthermore, it should be taken into account that the GS latitudinal position is potentially affected by a number of phenomena other than the NAO, such as the AMM \citep{hameed2018,wolfe2019}, the AMO \citep{nigam2018}, the AMOC \citep{decoetlogon2006,joyce2010} and the ENSO \citep{perez2014,sanchez2016}. At the same time, the atmosphere undergoes decadal variability also projecting on patterns largely orthogonal or/and independent to the NAO. Thus, the decadal NAO--GSF covariability could be supported or not depending on the slowly varying atmospheric and oceanic background and the eventual phase-locking of the decadal NAO and GSF variability. This aspect could be relevant to explain the dependency of the NAO--GSF lead–lag relationship over time as well as the lack of decadal NAO--GSF covariability before 1972.

Further studies using longer time records of highly resolved SST data and model experiments are highly required for assessing the robustness of the non-stationarity in the NAO--GSF relation and deeper understanding of its possible causes and associated mechanisms.

\section{Chapter conclusions}

The interaction between the NAO and the GSF latitudinal position has been the subject of extensive  investigations. In this context, there are indications of non-stationarity in their interaction but the differences among methodologies used in prior studies make it difficult to draw consistent conclusions. Furthermore, there is a lack of consensus on the key mechanisms underlying the response of the GSF to the NAO. The goal of this part of the PhD thesis was to assess the possible non-stationarity in the NAO--GSF interaction and to help clarify the mechanisms underlying this interaction over the last few decades, using a set of global atmosphere and ocean reanalysis.

Results show that the interaction between the NAO and the GSF latitudinal shifts during the winter season is non-stationary over the 1950--2020 time period. The NAO and the GSF indices covary on the decadal timescales (periods in [6--11]-year range) but only during 1972--2018 (Figure \ref{fig:xwt_ERA5}). A secondary peak in the NAO--GSF covariability emerges on multiannual timescales (periods shorter than 6 years) but only for a limited period of time (2005--2015). The detection of these decadal and multiannual peaks in different observational products provides robustness to this finding (Figure \ref{fig:xwt_ORAS5}, Figure \ref{fig:xwt_SODA}). 

The non-stationarity in the decadal NAO--GSF covariability is also manifested through the dependency of their lead--lag relationship on the analyzed time period. Indeed, the NAO leads the GSF shifts by 3 years during 1972--1990 and by 2 years during 1990--2018 (Figure \ref{fig:nao-gsf}). Note that this may in part explain the discrepancies in the lead--lag relationships between the GSF and the NAO suggested in previous studies (Table \ref{table:NAO-GSF}). Depending on the analysed time period, the lead--lag relationship between the GSF latitudinal position and its drivers may be different.

Overall, the results show that the response of the GSF to the NAO on decadal timescales can be interpreted as the joint effect of several different mechanisms, which are not all stationary across the 1972--2018 period. Before 1990, we interpret the time lag between GSF shifts and NAO to reflect a combination of a quick Ekman response, a lagged response of the DWBC, and the propagation of Rossby waves in response to the NAO (Figure \ref{fig:evolution}a). After 1990, we find an absence of Rossby wave propagation and thus the overall response of the GSF to the NAO is influenced only by the responses of Ekman and DWBC transports to the NAO (Figure \ref{fig:evolution}b). Here it is suggested that the non-stationarity of the Rossby wave propagation causes the time lag between the NAO and the GSF latitudinal position on decadal timescales to differ before and after 1990.
\chapter{\textit{\textbf{Conclusions}}}
\label{conclusions}

\fancyhead[R]{\textit{Chapter \ref{conclusions} - Conclusions}}

The present PhD thesis focused on the ocean--atmosphere interaction associated with the GSF variability, addressing the following scientific questions:
\begin{enumerate}
\item What is the character of the atmospheric response to the interannual GSF meridional shifts and its dependence on the model horizontal resolution?
\item What are the spectral features of the NAO--GSF interaction and the mechanisms through which the NAO forces the GSF meridional shifts on the decadal timescale?
\end{enumerate}

The character of the atmospheric response to the interannual GSF meridional shifts was assessed in the ERA5 reanalysis dataset and in an ensemble of atmosphere-only historical simulations (HighResMIP dataset) forced with observed SSTs (1950--2014). The role of the model horizontal resolution was assessed analyzing models with a nominal horizontal resolution from 25 km to 100 km. We referred to model configurations with a nominal resolution coarser than 50 km as R100 models, whereas to those with a nominal resolution finer or equal to 50 km as R50+ models. It is highlighted that the GSF variability was inspected through the definition of a new SST-based index (see section \ref{GSF_index} for details). Furthermore, the analyses in the vertical--meridional cross-section were developed along an SST front following coordinate system, which represents a novelty in the scientific literature dealing with the oceanic fronts variability and their impact on the atmospheric circulation (see section \ref{coordinate_system} for details).

Results show that the interannual variability in the GSF meridional position is associated with intense SST anomalies straddling its climatological position. Specifically, a northward (southward) shift of the front induces positive (negative) SST anomalies (Figure \ref{fig:sst}). Such SST anomalies induce intense co-located SHF anomalies which represent a thermal forcing for the atmospheric circulation (Figure \ref{fig:shf}).

The local- and large-scale atmospheric response to the SST-induced diabatic heating anomalies is strongly resolution-dependent, with the response in R50+ models resembling the respective observed anomalies. 

Discussing first the local response of the R50+ models, the thermodynamic balance in the lower troposphere is mainly maintained by anomalous meridional heat flux divergence (Figure \ref{fig:budget}f; term IV in equation \ref{heat_budget_equation}) and the anomalous vertical thermal advection (Figure \ref{fig:budget}g; term V in equation \ref{heat_budget_equation}). As expected for baroclinic adjustment, the local intensification of the baroclinic eddy activity is the response to the enhanced near-surface baroclinicity maintained by differential surface heating across the GSF (Figure \ref{fig:t925_grad}). The eddy activity intensifies also north and downstream of the GSF but in this case it is the meridionally differential zonal temperature advection near the east coast of the North American continent that maintains the large-scale baroclinicity (Figure \ref{fig:budget_R50+}). The changes in eddy activity shift northward the eddy-driven jet (Figure \ref{fig:u850}), through the action of the synoptic eddy fluxes (Figure \ref{fig:vt850}, Figure \ref{fig:vv250}). The northward shift of the eddy-driven jet and stormtrack is consistent with large-scale anticyclonic circulation anomalies downstream the GSF (Figure \ref{fig:shf}).

The R100 models show a strongly different local circulation response, with the anomalous diabatic heating mainly balanced by anomalous meridional heat flux divergence, vertical mean advection and meridional mean advection (Figure \ref{fig:budget}b, Figure \ref{fig:budget}c). As in the R50+ models, the local near-surface baroclinicity is enhanced thanks to differential heating across the front. Differently, the large-scale baroclinicity is reduced, with an expected reduction in the eddy activity over the North Atlantic and the southward shift of the eddy-driven jet. The changes in the eddy-driven jet and stormatrack are associated with large-scale cyclonic circulation anomalies.

Interestingly, R50+ models produce a local- as well as a large-scale atmospheric circulation response comparable to the respective observed anomalies. This finding provides important insight into the real character of the atmospheric response to the interannual GSF variability. The improvement in models with a resolution finer than 50 km may be conveyed by a better representation of mesoscale ocean-to-atmosphere forcing that is not resolved in their low-resolution counterparts and/or a more realistic representation of small-scale key atmospheric features (e.g. atmospheric fronts and warm conveyor belts).

The similarity between R50+ models and observations also suggests the possible existence of a positive feedback between ocean and atmosphere. As soon as the NAO forces a GSF shift, the associated SST anomalies tend to force a NAO-like response that strengthens/prolongs the original atmospheric forcing. The existence of a similar positive feedback has been previously suggested by other studies \citep{czaja2002,joyce2019}.

After having assessed the atmospheric response to interannual GSF variability, I moved on analyzing the other direction of ocean--atmosphere interaction, that is the role of the atmospheric forcing on the GSF front variability. In this context, I assessed both the possible existence of non-stationarity in the NAO--GSF interaction and the mechanisms through wich the NAO may be forcing the GSF shifts on decadal timescales. The spectral features of the NAO--GSF interaction in the past decades and the mechanisms underlying this interaction were assessed in a set of global atmosphere and ocean reanalyses.

Results show that the interaction between the NAO and the GSF latitudinal position during the winter season is non-stationary over the 1950--2020 period. This is manifested through the presence of a covariability peak on decadal timescales only during 1972--2018 (Figure \ref{fig:xwt_ERA5}). A secondary peak in the NAO--GSF covariability emerges on multiannual timescales but only during 2005--2015. The detection of the decadal and multiannual peaks in different reanalysis products (ERA5, ORAS5 and SODA3.4.2) provides robustness to these findings (Figure \ref{fig:xwt_ORAS5}, Figure \ref{fig:xwt_SODA}). Finally, the non-stationarity in the decadal NAO--GSF covariability is also shown by the dependency of their lead--lag relationship during 1972--2018. Indeed, the NAO leads the GSF latitudinal shifts by 3 years during 1972--1990 and by 2 years during 1990--2018 (Figure \ref{fig:nao-gsf}).

Overall, the results show that the response of the GSF to the NAO on decadal timescales can be interpreted as the joint effect of several different mechanisms, which are not all stationary across the 1972--2018 period. Before 1990, we interpret the time lag between GSF shifts and NAO to reflect a combination of a quick Ekman response, a lagged response of the DWBC, and the propagation of Rossby waves in response to the NAO (Figure \ref{fig:evolution}a). After 1990, we find an absence of Rossby wave propagation and thus the overall response of the GSF to the NAO is influenced only by the responses of Ekman and DWBC transports to the NAO (Figure \ref{fig:evolution}b). Here it is suggested that the non-stationarity of the Rossby wave propagation causes the time lag between the NAO and the GSF latitudinal position on decadal timescales to differ before and after 1990.

To conclude, this PhD thesis helps to improve our understanding of extratropical climate variability. In fact, it provides important evidence on the role played by small-scale oceanic features, such as the oceanic fronts, on extratropical atmospheric variability.

In particular, results suggest the possible existence of a positive feedback between the GSF and the atmospheric circulation. As soon as the NAO forces a GSF shift, the associated SST anomalies tend to force a NAO-like response that strengthens/prolongs the original atmospheric forcing. This evidence suggests that the atmospheric predictability may be higher than expected, especially on interannual and longer timescales. Given the oceanic state is predictable to some extent, the present results suggest that there are higher chances to properly predict changes in the atmospheric circulation through state-of-the-art high-resolution models.

Furthermore, this PhD thesis shows that the ocean--atmosphere interaction associated with the GSF variability is non-stationary. The non-stationarity in the NAO--GSF covariability has implications for the predictability of the North Atlantic sector. Previous studies have proposed statistical models to predict the GS path based on the atmospheric state some years in advance \citep[e.g.][]{hameed2004,sanchez2016}. However, the relationship between the GS latitudinal position and the atmospheric indices has been generally considered stationary. Thus, results in the present thesis indicate the potential inadequacy of statistical models that do not take into account the non-stationary character of the NAO--GSF interaction. Furthermore, considering the impact that the GSF shifts have on the North Atlantic atmospheric circulation, errors in predicting the GSF latitudinal position can reduce our skills in predicting the atmospheric variability through statistical and dynamical predictions.

\clearpage{}

\fancyhead[R]{\textit{References}}
\addcontentsline{toc}{chapter}{References}
\bibliography{References.bib}

\end{document}